\documentclass[prd,aps,showpacs,epsf,floats,onecolumn]{revtex4}%
\usepackage{amssymb}
\usepackage{amsfonts}
\usepackage{amsmath}
\usepackage{graphicx}
\usepackage{fancyhdr}%
\providecommand{\U}[1]{\protect\rule{.1in}{.1in}}
\begin{document}
\title{\textbf{Bures and Sj\"{o}qvist Metrics over Thermal State Manifolds for Spin
Qubits and Superconducting Flux Qubits}}
\author{\textbf{Carlo Cafaro}$^{1}$ and \textbf{Paul M.\ Alsing}$^{2}$}
\affiliation{$^{1}$SUNY Polytechnic Institute, 12203 Albany, New York, USA}
\affiliation{$^{2}$Air Force Research Laboratory, Information Directorate, 13441 Rome, New
York, USA}

\begin{abstract}
The interplay among differential geometry, statistical physics, and quantum
information science has been increasingly gaining theoretical interest in
recent years.

In this paper, we present an explicit analysis of the Bures and Sj\"{o}qvist
metrics over the manifolds of thermal states for specific spin qubit and the
superconducting flux qubit Hamiltonian models. While the two metrics equally
reduce to the Fubini-Study metric in the asymptotic limiting case of the
inverse temperature approaching infinity for both Hamiltonian models, we
observe that the two metrics are generally different when departing from the
zero-temperature limit. In particular, we discuss this discrepancy in the case
of the superconducting flux Hamiltonian model. We conclude the two metrics
differ in the presence of a nonclassical behavior specified by the
noncommutativity of neighboring mixed quantum states. Such a noncommutativity,
in turn, is quantified by the two metrics in different manners. Finally, we
briefly discuss possible observable consequences of this discrepancy between
the two metrics when using them to predict critical and/or complex behavior of
physical systems of interest in quantum information science.

\end{abstract}

\pacs{Quantum Computation (03.67.Lx), Quantum Information (03.67.Ac), Quantum
Mechanics (03.65.-w), Riemannian Geometry (02.40.Ky), Statistical Mechanics (05.20.-y).}
\maketitle

\fancyhead[R]{\ifnum\value{page}<2\relax\else\thepage\fi}

\thispagestyle{fancy}

\section{Introduction}

Geometry plays a special role in the description and, to a certain extent, in
the understanding of various physical phenomena \cite{pettini07,karol06}. The
concepts of length, area, and volume are ubiquitous in physics and their
meaning can prove quite helpful in explaining physical phenomena from a more
intuitive perspective \cite{cafaroprd22,cafaropre22}. The notions of
\textquotedblleft\emph{longer}\textquotedblright\ and \textquotedblleft%
\emph{shorter}\textquotedblright\ are extensively used in virtually all
disciplines \cite{cafarophysicaa22}. Indeed, geometric formulations of
classical and quantum evolutions along with geometric descriptions of
classical and quantum mechanical aspects of thermal phenomena are becoming
increasingly important in science. Concepts, such as thermodynamic length,
area law, and statistical volumes are omnipresent in geometric thermodynamics,
general relativity, and statistical physics, respectively. The concept of
entropy finds application in essentially any realm of science, from classical
thermodynamics to quantum information science. The notions of
\textquotedblleft\emph{hotter}\textquotedblright\ and \textquotedblleft%
\emph{cooler}\textquotedblright\ are widely used in many fields. Entropy can
be used to provide measures of distinguishability of classical probability
distributions, as well as pure and mixed quantum states. It can also be used
to propose measures of complexity for classical motion, quantum evolution, and
entropic motion on curved statistical manifolds underlying the entropic
dynamics of physical systems for which only partial knowledge of relevant
information can be obtained \cite{cafaroPhD,cafaroCSF,felice18}. Furthermore,
entropy can also be used to express the degree of entanglement in a quantum
state specifying a composite quantum system. For instance, concepts such as
Shannon entropy, von Neumann entropy, and Umegaki relative entropy are
ubiquitous in classical information science, quantum information theory, and
information geometric formulations of mixed quantum state evolutions
\cite{amari}, respectively. In this paper, inspired by the increasing
theoretical interest in the interplay among differential geometry, statistical
physics, and quantum information science
\cite{zanardiprl07,zanardi07,pessoa21,silva21,silva21B,mera22}, we present an
explicit analysis of the Bures \cite{bures69,uhlman76,hubner92} and
Sj\"{o}qvist \cite{erik20} metrics over the manifolds of thermal states for
the spin qubit and the superconducting flux qubit Hamiltonian models. From a
chronological standpoint, the first physical application of the Sj\"{o}qvist
interferometric metric occurs in the original paper by Sj\"{o}qvist himself in
Ref. \cite{erik20}. Here, the author considered his newly proposed
interferometric metric to quantify changes in behavior of a magnetic system in
a thermal state under modifications of temperature and magnetic field
intensity. Keeping the temperature constant while changing the externally
applied magnetic field, the Sj\"{o}qvist interferometric metric was shown to
be physically linked to the magnetic susceptibility. This quantity, in turn,
quantifies how much a material will become magnetized when immersed in a
magnetic field. A second application of the Sj\"{o}qvist interferometric
metric happens in Refs. \cite{silva21,silva21B}. Here, this metric is used to
characterize the finite-temperature phase transitions in the framework of band
insulators. In particular, the authors considered the massive Dirac
Hamiltonian model for a band insulator in two spatial dimensions. The
corresponding Sj\"{o}qvist interferometric metric was calculated and expressed
in terms of two physical parameters, the temperature, and the hopping
parameter. Furthermore, the Sj\"{o}qvist interferometric metric was physically
regarded as an interferometric susceptibility. Interestingly, it was observed
in Refs. \cite{silva21,silva21B} a dramatic difference between the
Sj\"{o}qvist interferometric metric and the Bures metric when studying
topological phase transitions in these types of systems. Specifically, while
the topological phase transition is captured for all temperatures in the case
of the Sj\"{o}qvist interferometric metric, the topological phase transition
is captured only at zero temperature in the case of the Bures metric. Clearly,
the authors leave as an unsolved question the experimental observation of the
singular behavior of the Sj\"{o}qvist interferometric metric in actual
laboratory experiments in Refs. \cite{silva21,silva21B}. A third interesting
work is the one presented in Ref. \cite{mera22}. Here the authors focus on the
zero-temperature aspects of certain quantum systems in their (pure) ground
quantum state. They consider two systems. The first system is described by the
$XY$ anisotropic spin-$1/2$ chain with $N$-sites on a circle in the presence
of an external magnetic field. The second model is the Haldane model, a
two-dimensional condensed-matter lattice model \cite{haldane88}. In the first
case, the two parameters that determine both the Hamiltonian and the parameter
manifold are the anisotropy degree and the magnetic field intensity. In the
second case, instead, the two key parameters become the on-site energy and the
phase in the model being considered. Expressing the so-called quantum metric
in terms of these tunable parameters, they study the thermodynamical limit of
this metric along critical submanifolds of the whole parameter manifold. They
observe a singular (regular) behavior of the metric along normal (tangent)
directions to the critical submanifolds. Therefore, they conclude that tangent
directions to critical manifolds are special. Finally, the authors also point
out that it would be interesting to understand how their findings generalize
to the finite-temperature case where states become mixed. Interestingly,
without reporting any explicit analysis, the authors state they expect the
Bures and the Sj\"{o}qvist metrics to assume different functional forms.

In this paper, inspired by the previously mentioned relevance of comprehending
the physical significance of choosing one metric over another one in such
geometric characterizations of physical aspects of quantum systems, we report
a complete and straightforward analysis of the link between the Sj\"{o}qvist
interferometric metric and the Bures metric for two special classes of
nondegenerate mixed quantum states. Specifically , focusing on manifolds of
thermal states for the spin qubit and the superconducting flux qubit
Hamiltonian models, we observe that while the two metrics both reduce to the
Fubini-Study metric \cite{provost80,wootters81,braunstein94} in the
zero-temperature asymptotic limiting case of the inverse temperature
$\beta\overset{\text{def}}{=}\left(  k_{B}T\right)  ^{-1}$ (with $k_{B}$ being
the Boltzmann constant) approaching infinity for both Hamiltonian models, the
two metrics are generally different. Furthermore, we observe this different
behavior in the case of the superconducting flux Hamiltonian model. More
generally, we note that the two metrics seem to differ when a nonclassical
behavior is present since in this case, the metrics quantify noncommutativity
of neighboring mixed quantum states in different manners. Finally, we briefly
discuss the possible observable consequences of this discrepancy between the
two metrics when using them to predict critical and/or complex behavior of
physical systems of interest. We acknowledge that despite the fact that most
of the preliminary background results presented in this paper are partially
known in the literature, they appear in a scattered fashion throughout several
papers written by researchers working in distinct fields of physics who may
not be necessary aware of each other's findings. For this reason, we present
here an explicit and unified comparative formulation of the Bures and
Sj\"{o}qvist metrics for simple quantum systems in mixed states. In
particular, as mentioned earlier, we illustrate our results on the examples of
thermal state manifolds for spin qubits and superconducting flux qubits. These
applications are original and, to the best of our knowledge, do not appear
anywhere in the literature.

The layout of the rest of this paper is as follows. In Section II, we present
with explicit derivations the expressions of the Bures metric in H\"{u}bner's
\cite{hubner92} and Zanardi's \cite{zanardi07} forms. In Section III, we
present an explicit derivation of the Sj\"{o}qvist interferometric metric
\cite{erik20} between two neighboring nondegenerate density matrices. In
Section IV, we present two Hamiltonian models. The first Hamiltonian model
describes a spin-$1/2$ particle in a uniform and time-independent external
magnetic field oriented along the $z$-axis. The second Hamiltonian model,
instead, specifies a superconducting flux qubit. Bringing these two systems in
thermal equilibrium with a reservoir at finite and non-zero temperature $T$,
we construct the two corresponding parametric families of thermal states. In
Section V, we present an explicit calculation of both the Sj\"{o}qvist and the
Bures metrics for each one of the two distinct families of parametric thermal
states. From our comparative analysis, we find that the two metric coincide
for the first Hamiltonian model (electron in a constant magnetic field along
the $z$-direction), while they differ for the second Hamiltonian model
(superconducting flux qubit). In Section VI, we discuss the effects that arise
from the comparative analysis carried out in Section V concerning the Bures
and Sj\"{o}qvist metrics for spin qubits and superconducting flux qubits
Hamiltonian models introduced in Section IV. Finally, we conclude with our
final remarks along with a summary of our main findings in Section VII.

\section{The Bures metric}

In this section, we present two explicit calculations. In the first
calculation, we carry out a detailed derivation of the Bures metric by
following the original work presented by H\"{u}bner in Ref. \cite{hubner92}.
In the second calculation, we recast the expression of the Bures metric
obtained by H\"{u}bner in a way that is more suitable in the framework of
geometric analyses on thermal states manifolds. Here, we follow the original
work presented by Zanardi and collaborators in Ref. \cite{zanardi07}.

\subsection{The explicit derivation of H\"{u}bner's general expression}

We begin by carrying out an explicit derivation of the Bures metric inspired
by H\"{u}bner \cite{hubner92}. Recall that the squared Bures distance between
two density matrices infinitesimally far apart is given by,%
\begin{equation}
\left[  d_{\mathrm{Bures}}\left(  \rho\text{, }\rho+d\rho\right)  \right]
^{2}=2-2\mathrm{tr}\left[  \rho^{1/2}\left(  \rho+d\rho\right)  \rho
^{1/2}\right]  ^{1/2}\text{.} \label{buri1}%
\end{equation}
To find a useful expression for $\left[  d_{\mathrm{Bures}}\left(  \rho\text{,
}\rho+d\rho\right)  \right]  ^{2}$, we follow the line of reasoning used by
H\"{u}bner in Ref. \cite{hubner92}. Consider an Hermitian matrix $A\left(
t\right)  $ with $t\in%
\mathbb{R}
$ defined as%
\begin{equation}
A\left(  t\right)  \overset{\text{def}}{=}\left[  \rho^{1/2}\left(
\rho+td\rho\right)  \rho^{1/2}\right]  ^{1/2}\text{,} \label{buri2}%
\end{equation}
with $A\left(  t\right)  A\left(  t\right)  =\rho^{1/2}\left(  \rho
+td\rho\right)  \rho^{1/2}$. Note that $A\left(  0\right)  =\rho$ and, for
later use, we assume $\rho$ to be invertible. At this point we observe that
knowledge of the metric tensor $g_{ij}\left(  \rho\right)  $ at $\rho$
requires knowing $\left[  d_{\mathrm{Bures}}\left(  \rho\text{, }\rho
+td\rho\right)  \right]  ^{2}$ up to the second order in $t$. H\"{u}bner's
ansatz (i.e., educated guess) is
\begin{equation}
\left[  d_{\mathrm{Bures}}\left(  \rho\text{, }\rho+td\rho\right)  \right]
^{2}=t^{2}g_{ij}\left(  \rho\right)  d\rho^{i}d\rho^{j}\text{,} \label{buri3}%
\end{equation}
with $\left\{  \rho^{i}\right\}  $ denoting a given set of coordinates on the
manifold of density matrices. From Eq. (\ref{buri3}), we note that
\begin{equation}
g_{ij}\left(  \rho\right)  d\rho^{i}d\rho^{j}=\frac{1}{2}\left(  \frac{d^{2}%
}{dt^{2}}\left[  d_{\mathrm{Bures}}\left(  \rho\text{, }\rho+td\rho\right)
\right]  ^{2}\right)  _{t=0}\text{.} \label{buri4}%
\end{equation}
Observe that using Eq. (\ref{buri2}), the RHS\ of Eq. (\ref{buri4}) becomes%
\begin{align}
\frac{1}{2}\left(  \frac{d^{2}}{dt^{2}}\left[  d_{\mathrm{Bures}}\left(
\rho\text{, }\rho+td\rho\right)  \right]  ^{2}\right)  _{t=0}  &  =\frac{1}%
{2}\frac{d^{2}}{dt^{2}}\left\{  2-2\mathrm{tr}\left[  \rho^{1/2}\left(
\rho+td\rho\right)  \rho^{1/2}\right]  ^{1/2}\right\}  _{t=0}\nonumber\\
&  =\frac{1}{2}\frac{d^{2}}{dt^{2}}\left\{  2-2\mathrm{tr}\left[  A\left(
t\right)  \right]  \right\}  _{t=0}\nonumber\\
&  =-\mathrm{tr}\left[  \ddot{A}\left(  t\right)  \right]  _{t=0}\text{,}%
\end{align}
that is,%
\begin{equation}
\frac{1}{2}\left(  \frac{d^{2}}{dt^{2}}\left[  d_{\mathrm{Bures}}\left(
\rho\text{, }\rho+td\rho\right)  \right]  ^{2}\right)  _{t=0}=-\mathrm{tr}%
\left[  \ddot{A}\left(  t\right)  \right]  _{t=0}\text{.} \label{buri5}%
\end{equation}
From Eqs. (\ref{buri4}) and (\ref{buri5}), we get%
\begin{equation}
g_{ij}\left(  \rho\right)  d\rho^{i}d\rho^{j}=-\mathrm{tr}\left[  \ddot
{A}\left(  t\right)  \right]  _{t=0}\text{.} \label{buri6}%
\end{equation}
Differentiating two times the relation $A\left(  t\right)  A\left(  t\right)
=\rho^{1/2}\left(  \rho+td\rho\right)  \rho^{1/2}$, setting $t=0$ and,
finally, assuming $\rho$ diagonalized in the form%
\begin{equation}
\rho=\sum_{i}\lambda_{i}\left\vert i\right\rangle \left\langle i\right\vert
\text{,}%
\end{equation}
we have
\begin{equation}
\left\{  \frac{d^{2}}{dt^{2}}\left[  A\left(  t\right)  A\left(  t\right)
\right]  \right\}  _{t=0}=\left\{  \frac{d^{2}}{dt^{2}}\left[  \rho
^{1/2}\left(  \rho+td\rho\right)  \rho^{1/2}\right]  \right\}  _{t=0}\text{.}%
\end{equation}
More explicitly, we notice that
\begin{equation}
\frac{d}{dt}\left[  A\left(  t\right)  A\left(  t\right)  \right]  =\dot
{A}\left(  t\right)  A\left(  t\right)  +A\left(  t\right)  \dot{A}\left(
t\right)  \label{buri7}%
\end{equation}
and,%
\begin{equation}
\frac{d}{dt}\left[  \rho^{1/2}\left(  \rho+td\rho\right)  \rho^{1/2}\right]
=\rho^{1/2}d\rho\rho^{1/2}\text{.} \label{buri8}%
\end{equation}
Setting $t=0$, from Eqs. (\ref{buri7}) and (\ref{buri8}) we obtain%
\begin{equation}
\dot{A}\left(  0\right)  A\left(  0\right)  +A\left(  0\right)  \dot{A}\left(
0\right)  =\rho^{1/2}d\rho\rho^{1/2}\text{.} \label{buri8a}%
\end{equation}
After the second differentiation of $A\left(  t\right)  A\left(  t\right)  $
and $\rho^{1/2}\left(  \rho+td\rho\right)  \rho^{1/2}$, we get%
\begin{equation}
\ddot{A}\left(  0\right)  A\left(  0\right)  +2\text{ }\dot{A}\left(
0\right)  \dot{A}\left(  0\right)  +A\left(  0\right)  \ddot{A}\left(
0\right)  =0\text{.} \label{pauli}%
\end{equation}
Multiplying both sides of Eq. (\ref{pauli}) by $A^{-1}\left(  0\right)  $ from
the right and using the cyclicity of the trace operation, we get%
\begin{equation}
\mathrm{tr}\left[  \ddot{A}\left(  0\right)  \right]  =-\mathrm{tr}\left[
A^{-1}\left(  0\right)  \dot{A}\left(  0\right)  ^{2}\right]  \text{.}
\label{buri9}%
\end{equation}
From Eqs. (\ref{buri6}) and (\ref{buri9}), the Bures metric
$ds_{\mathrm{Bures}}^{2}$ becomes%
\begin{align}
ds_{\mathrm{Bures}}^{2}\left(  \rho\text{, }\rho+d\rho\right)   &
=g_{ij}\left(  \rho\right)  d\rho^{i}d\rho^{j}\nonumber\\
&  =-\mathrm{tr}\left[  \ddot{A}\left(  t\right)  \right]  _{t=0}\nonumber\\
&  =\mathrm{tr}\left[  A^{-1}\left(  0\right)  \dot{A}\left(  0\right)
^{2}\right] \nonumber\\
&  =\sum_{i}\left\langle i\left\vert A^{-1}\left(  0\right)  \dot{A}\left(
0\right)  ^{2}\right\vert i\right\rangle \text{,}%
\end{align}
that is,%
\begin{equation}
ds_{\mathrm{Bures}}^{2}\left(  \rho\text{, }\rho+d\rho\right)  =\frac{1}%
{2}\sum_{i,k,l}\left[  \left\langle i\left\vert A^{-1}\left(  0\right)
\right\vert k\right\rangle \left\langle k\left\vert \dot{A}\left(  0\right)
\right\vert l\right\rangle \left\langle l\left\vert \dot{A}\left(  0\right)
\right\vert i\right\rangle +\left\langle i\left\vert A^{-1}\left(  0\right)
\right\vert l\right\rangle \left\langle l\left\vert \dot{A}\left(  0\right)
\right\vert k\right\rangle \left\langle k\left\vert \dot{A}\left(  0\right)
\right\vert i\right\rangle \right]  \text{.} \label{buri9a}%
\end{equation}
Observe that $A\left(  0\right)  \left\vert k\right\rangle =\rho\left\vert
k\right\rangle =\lambda_{k}\left\vert k\right\rangle $ and, therefore,
$A^{-1}\left(  0\right)  \left\vert k\right\rangle =\rho^{-1}\left\vert
k\right\rangle =\lambda_{k}^{-1}\left\vert k\right\rangle $ with $\lambda
_{k}\neq0$ for any $k$. We need to find an expression for $\left\langle
i\left\vert \dot{A}\left(  0\right)  \right\vert j\right\rangle $. From Eq.
(\ref{buri8a}), we get%
\begin{equation}
\left\langle i\left\vert \dot{A}\left(  0\right)  A\left(  0\right)  +A\left(
0\right)  \dot{A}\left(  0\right)  \right\vert j\right\rangle =\left\langle
i\left\vert \rho^{1/2}d\rho\rho^{1/2}\right\vert j\right\rangle \text{.}
\label{buri10}%
\end{equation}
We note that%
\begin{align}
\left\langle i\left\vert \dot{A}\left(  0\right)  A\left(  0\right)  +A\left(
0\right)  \dot{A}\left(  0\right)  \right\vert j\right\rangle  &
=\left\langle i\left\vert \dot{A}\left(  0\right)  \rho+\rho\dot{A}\left(
0\right)  \right\vert j\right\rangle \nonumber\\
&  =%
{\displaystyle\sum\limits_{k}}
\lambda_{k}\left\langle i\left\vert \dot{A}\left(  0\right)  \right\vert
k\right\rangle \left\langle k\left\vert j\right.  \right\rangle +%
{\displaystyle\sum\limits_{k}}
\lambda_{k}\left\langle i\left\vert k\right.  \right\rangle \left\langle
k\left\vert \dot{A}\left(  0\right)  \right\vert j\right\rangle \nonumber\\
&  =\lambda_{j}\left\langle i\left\vert \dot{A}\left(  0\right)  \right\vert
j\right\rangle +\lambda_{i}\left\langle i\left\vert \dot{A}\left(  0\right)
\right\vert j\right\rangle \nonumber\\
&  =\left(  \lambda_{i}+\lambda_{j}\right)  \left\langle i\left\vert \dot
{A}\left(  0\right)  \right\vert j\right\rangle \text{,}%
\end{align}
that is,%
\begin{equation}
\left\langle i\left\vert \dot{A}\left(  0\right)  A\left(  0\right)  +A\left(
0\right)  \dot{A}\left(  0\right)  \right\vert j\right\rangle =\left(
\lambda_{i}+\lambda_{j}\right)  \left\langle i\left\vert \dot{A}\left(
0\right)  \right\vert j\right\rangle \text{.} \label{buri11}%
\end{equation}
Moreover, we observe that
\begin{align}
\left\langle i\left\vert \rho^{1/2}d\rho\rho^{1/2}\right\vert j\right\rangle
&  =\left\langle i\left\vert \left(  \sum_{k}\sqrt{\lambda_{k}}\left\vert
k\right\rangle \left\langle k\right\vert \right)  d\rho\left(  \sum_{m}%
\sqrt{\lambda_{m}}\left\vert m\right\rangle \left\langle m\right\vert \right)
\right\vert \right\rangle \nonumber\\
&  =\sum_{k,m}\sqrt{\lambda_{k}}\sqrt{\lambda_{m}}\left\langle i\left\vert
k\right.  \right\rangle \left\langle k\left\vert d\rho\right\vert
m\right\rangle \left\langle m\left\vert j\right.  \right\rangle \nonumber\\
&  =\sum_{k,m}\sqrt{\lambda_{k}}\sqrt{\lambda_{m}}\delta_{ik}\delta
_{mj}\left\langle k\left\vert d\rho\right\vert m\right\rangle \nonumber\\
&  =\sqrt{\lambda_{i}}\sqrt{\lambda_{j}}\left\langle i\left\vert
d\rho\right\vert j\right\rangle \text{,}%
\end{align}
that is,%
\begin{equation}
\left\langle i\left\vert \rho^{1/2}d\rho\rho^{1/2}\right\vert j\right\rangle
=\sqrt{\lambda_{i}}\sqrt{\lambda_{j}}\left\langle i\left\vert d\rho\right\vert
j\right\rangle \text{.} \label{buri12}%
\end{equation}
Using Eqs. (\ref{buri10}), (\ref{buri11}), and (\ref{buri12}), we have%
\begin{equation}
\left\langle i\left\vert \dot{A}\left(  0\right)  \right\vert j\right\rangle
=\frac{\sqrt{\lambda_{i}}\sqrt{\lambda_{j}}}{\lambda_{i}+\lambda_{j}%
}\left\langle i\left\vert d\rho\right\vert j\right\rangle \text{.}
\label{buri13}%
\end{equation}
Finally, using Eq. (\ref{buri13}), $ds_{\mathrm{Bures}}^{2}$ in Eq.
(\ref{buri9a}) becomes%
\begin{equation}
ds_{\mathrm{Bures}}^{2}\left(  \rho\text{, }\rho+d\rho\right)  =\frac{1}%
{2}\sum_{k,l}\frac{\lambda_{l}+\lambda_{k}}{\left(  \lambda_{l}+\lambda
_{k}\right)  ^{2}}\left\vert \left\langle k\left\vert d\rho\right\vert
l\right\rangle \right\vert ^{2}\text{,}%
\end{equation}
that is, relabelling the dummy indices (i.e., $k\rightarrow i$ and
$l\rightarrow j$),%
\begin{equation}
ds_{\mathrm{Bures}}^{2}\left(  \rho\text{, }\rho+d\rho\right)  =\frac{1}%
{2}\sum_{i,j}\frac{\left\vert \left\langle i\left\vert d\rho\right\vert
j\right\rangle \right\vert ^{2}}{\lambda_{i}+\lambda_{j}}\text{.}
\label{buri14}%
\end{equation}
The derivation of Eq. (\ref{buri14}) ends our revisitation of H\"{u}bner's
original analysis presented in Ref. \cite{hubner92}. Note that in obtaining
Eq. (\ref{buri14}), there is no need to introduce any Hamiltonian \textrm{H}
that might be responsible for the changes from $\rho$ to $\rho+d\rho$. For
this reason, the expression of the Bures metric in Eq. (\ref{buri14}) is said
to be general.

\subsection{The explicit derivation of Zanardi's general expression}

In Ref. \cite{zanardi07}, Zanardi and collaborators provided an alternative
expression of the Bures metric in Eq. (\ref{buri14}). Interestingly, in this
alternative expression, the Bures metric in Eq. (\ref{buri14}) can be
decomposed in its classical and nonclassical parts. To begin, in view of
future geometric investigations in statistical physics, let us use a different
notation and rewrite the Bures metric in Eq. (\ref{buri14}) as
\begin{equation}
ds_{\mathrm{Bures}}^{2}\left(  \rho\text{, }\rho+d\rho\right)
\overset{\text{def}}{=}\frac{1}{2}\sum_{n\text{, }m}\frac{\left\vert
\left\langle m|d\rho|n\right\rangle \right\vert ^{2}}{p_{m}+p_{n}}\text{,}
\label{Bures}%
\end{equation}
with $1\leq m$, $n\leq N$. Let us assume that the quantities $\rho$ and
$d\rho$ in Eq. (\ref{Bures}) are given by,%
\begin{equation}
\rho\overset{\text{def}}{=}\sum_{n}p_{n}\left\vert n\right\rangle \left\langle
n\right\vert \text{,}%
\end{equation}
and%
\begin{equation}
d\rho\overset{\text{def}}{=}\sum_{n}\left[  dp_{n}\left\vert n\right\rangle
\left\langle n\right\vert +p_{n}\left\vert dn\right\rangle \left\langle
n\right\vert +p_{n}\left\vert n\right\rangle \left\langle dn\right\vert
\right]  \text{,} \label{dro}%
\end{equation}
respectively, with $\left\langle n\left\vert m\right.  \right\rangle
=\delta_{n,m}$. Let us use Eqs. (\ref{dro}) and (\ref{Bures}) to find a more
explicit expression for the Bures metric $ds_{\mathrm{Bures}}^{2}\left(
\rho\text{, }\rho+d\rho\right)  $. Observe that the quantity $\left\langle
i\left\vert d\rho\right\vert j\right\rangle $ can be recast as,%
\begin{align}
\left\langle i\left\vert d\rho\right\vert j\right\rangle  &  =\left\langle
i\left\vert \left(  \sum_{n}\left[  dp_{n}\left\vert n\right\rangle
\left\langle n\right\vert +p_{n}\left\vert dn\right\rangle \left\langle
n\right\vert +p_{n}\left\vert n\right\rangle \left\langle dn\right\vert
\right]  \right)  \right\vert j\right\rangle \nonumber\\
&  =\sum_{n}dp_{n}\left\langle i|n\right\rangle \left\langle n|j\right\rangle
+p_{n}\left\langle i|dn\right\rangle \left\langle n|j\right\rangle
+p_{n}\left\langle i|n\right\rangle \left\langle dn|j\right\rangle \nonumber\\
&  =\sum_{n}dp_{n}\delta_{in}\delta_{nj}+p_{n}\left\langle i|dn\right\rangle
\delta_{nj}+p_{n}\delta_{in}\left\langle dn|j\right\rangle \nonumber\\
&  =dp_{i}\delta_{ij}+p_{j}\left\langle i|dj\right\rangle +p_{i}\left\langle
di|j\right\rangle \text{,}%
\end{align}
that is,%
\begin{equation}
\left\langle i\left\vert d\rho\right\vert j\right\rangle =dp_{i}\delta
_{ij}+p_{j}\left\langle i|dj\right\rangle +p_{i}\left\langle di|j\right\rangle
\text{.} \label{1}%
\end{equation}
Note that the orthonormality condition $\left\langle i|j\right\rangle
=\delta_{ij}$ implies $\left\langle di|j\right\rangle +\left\langle
i|dj\right\rangle =0$, that is%
\begin{equation}
\left\langle di|j\right\rangle =-\left\langle i|dj\right\rangle \text{.}
\label{2}%
\end{equation}
Using Eq. (\ref{2}), $\left\langle i\left\vert d\rho\right\vert j\right\rangle
$ in Eq. (\ref{1}) becomes%
\begin{equation}
\left\langle i\left\vert d\rho\right\vert j\right\rangle =dp_{i}\delta
_{ij}+p_{j}\left\langle i|dj\right\rangle -p_{i}\left\langle i|dj\right\rangle
=\delta_{ij}dp_{i}+\left(  p_{j}-p_{i}\right)  \left\langle i|dj\right\rangle
\text{,}\nonumber
\end{equation}
that is,%
\begin{equation}
\left\langle i\left\vert d\rho\right\vert j\right\rangle =\delta_{ij}%
dp_{i}+\left(  p_{j}-p_{i}\right)  \left\langle i|dj\right\rangle \text{.}
\label{3}%
\end{equation}
Making use of Eq. (\ref{3}), we can now find an explicit expression of the
quantity $\left\vert \left\langle m|d\rho|n\right\rangle \right\vert ^{2}$ in
Eq. (\ref{Bures}). Indeed, from Eq. (\ref{3}) we have%
\begin{align}
\left\vert \left\langle m|d\rho|n\right\rangle \right\vert ^{2}  &
=\left\langle m|d\rho|n\right\rangle \left\langle m|d\rho|n\right\rangle
^{\ast}\nonumber\\
&  =\left\langle m|d\rho|n\right\rangle \left\langle n|d\rho|m\right\rangle
\nonumber\\
&  =\left[  \delta_{mn}dp_{m}+\left(  p_{n}-p_{m}\right)  \left\langle
m|dn\right\rangle \right]  \left[  \delta_{nm}dp_{n}+\left(  p_{m}%
-p_{n}\right)  \left\langle n|dm\right\rangle \right] \nonumber\\
&  =\delta_{mn}dp_{m}\delta_{nm}dp_{n}+\delta_{mn}dp_{m}\left(  p_{m}%
-p_{n}\right)  \left\langle n|dm\right\rangle +\left(  p_{n}-p_{m}\right)
\left\langle m|dn\right\rangle \delta_{nm}dp_{n}+\nonumber\\
&  +\left(  p_{n}-p_{m}\right)  \left\langle m|dn\right\rangle \left(
p_{m}-p_{n}\right)  \left\langle n|dm\right\rangle \nonumber\\
&  =\delta_{nm}dp_{n}^{2}-\left(  p_{n}-p_{m}\right)  ^{2}\left\langle
m|dn\right\rangle \left\langle n|dm\right\rangle \text{,} \label{4}%
\end{align}
that is,%
\begin{equation}
\left\vert \left\langle m|d\rho|n\right\rangle \right\vert ^{2}=\delta
_{nm}dp_{n}^{2}-\left(  p_{n}-p_{m}\right)  ^{2}\left\langle m|dn\right\rangle
\left\langle n|dm\right\rangle \text{.} \label{5}%
\end{equation}
Using Eq. (\ref{2}), Eq. (\ref{5}) reduces to%
\begin{align}
\left\vert \left\langle m|d\rho|n\right\rangle \right\vert ^{2}  &
=\delta_{nm}dp_{n}^{2}+\left(  p_{n}-p_{m}\right)  ^{2}\left\langle
dm|n\right\rangle \left\langle n|dm\right\rangle \nonumber\\
&  =\delta_{nm}dp_{n}^{2}+\left(  p_{n}-p_{m}\right)  ^{2}\left\vert
\left\langle n|dm\right\rangle \right\vert ^{2}\text{,}%
\end{align}
that is,%
\begin{equation}
\left\vert \left\langle m|d\rho|n\right\rangle \right\vert ^{2}=\delta
_{nm}dp_{n}^{2}+\left(  p_{n}-p_{m}\right)  ^{2}\left\vert \left\langle
n|dm\right\rangle \right\vert ^{2}\text{.} \label{6}%
\end{equation}
Finally, substituting Eq. (\ref{6}) into Eq. (\ref{Bures}),
$ds_{\mathrm{Bures}}^{2}\left(  \rho\text{, }\rho+d\rho\right)  $ becomes%
\begin{align}
ds_{\mathrm{Bures}}^{2}\left(  \rho\text{, }\rho+d\rho\right)   &  =\frac
{1}{2}\sum_{n\text{, }m}\frac{\delta_{nm}dp_{n}^{2}+\left(  p_{n}%
-p_{m}\right)  ^{2}\left\vert \left\langle n|dm\right\rangle \right\vert ^{2}%
}{p_{m}+p_{n}}\nonumber\\
&  =\frac{1}{4}\sum_{n}\frac{dp_{n}^{2}}{p_{n}}+\frac{1}{2}\sum_{n\neq
m}\left\vert \left\langle n|dm\right\rangle \right\vert ^{2}\frac{\left(
p_{n}-p_{m}\right)  ^{2}}{p_{m}+p_{n}}\text{,}%
\end{align}
that is,%
\begin{equation}
ds_{\mathrm{Bures}}^{2}\left(  \rho\text{, }\rho+d\rho\right)  =\frac{1}%
{4}\sum_{n}\frac{dp_{n}^{2}}{p_{n}}+\frac{1}{2}\sum_{n\neq m}\left\vert
\left\langle n|dm\right\rangle \right\vert ^{2}\frac{\left(  p_{n}%
-p_{m}\right)  ^{2}}{p_{m}+p_{n}}\text{.} \label{7}%
\end{equation}
Eq. (\ref{7}) is the explicit expression of $ds_{\mathrm{Bures}}^{2}\left(
\rho\text{, }\rho+d\rho\right)  $ we were searching for. As a side remark,
note that if both $\left\vert n\right\rangle $ and $\left\vert m\right\rangle
\in\ker\left(  \rho\right)  $, we have that $\left\langle n|d\rho
|m\right\rangle =0$. Indeed, from $\rho\left\vert m\right\rangle =\left\vert
0\right\rangle $, we have $d\rho\left\vert m\right\rangle +\rho\left\vert
dm\right\rangle =\left\vert 0\right\rangle $. Therefore, we have $\left\langle
n|d\rho|m\right\rangle +\left\langle n|\rho|dm\right\rangle =\left\langle
n|0\right\rangle $, that is, $\left\langle n|d\rho|m\right\rangle =0$ since
$\left\langle n|0\right\rangle =0$ and $\left\langle n|\rho|dm\right\rangle
=0$. The expression of the Bures metric in Eq. (\ref{7}) can be regarded as
given by two contribution, a classical and a nonclassical term. The first term
in $ds_{\mathrm{Bures}}^{2}\left(  \rho\text{, }\rho+d\rho\right)  $ in Eq.
(\ref{7}) is the classical one and is represented by the classical Fisher-Rao
information metric between the two probability distributions $\left\{
p_{n}\right\}  _{1\leq n\leq N}$ and $\left\{  p_{n}+dp_{n}\right\}  _{1\leq
n\leq N}$. The second term is the nonclassical one and emerges from the
noncommutativity of the density matrices $\rho$ and $\rho+d\rho$ (i.e.,
$\left[  \rho\text{, }\rho+d\rho\right]  =\left[  \rho\text{, }d\rho\right]
\neq0$, in general). When $\left[  \rho\text{, }\rho+d\rho\right]  =0$, the
problem becomes classical and the Bures metric reduces to the classical
Fisher-Rao metric.

\subsection{The explicit derivation of Zanardi's expression for thermal
states}

In what follows, we specialize on the functional form of $ds_{\mathrm{Bures}%
}^{2}\left(  \rho\text{, }\rho+d\rho\right)  $ in Eq. (\ref{7}) for thermal
states. Specifically, let us focus on mixed quantum states $\rho\left(
\beta\text{, }\lambda\right)  $ of the form,%
\begin{equation}
\rho\left(  \beta\text{, }\lambda\right)  \overset{\text{def}}{=}%
\frac{e^{-\beta\mathrm{H}\left(  \lambda\right)  }}{\mathcal{Z}}%
=\frac{e^{-\beta\mathrm{H}\left(  \lambda\right)  }}{\mathrm{tr}\left(
e^{-\beta\mathrm{H}\left(  \lambda\right)  }\right)  }\text{,} \label{ro}%
\end{equation}
with $\mathcal{Z}\overset{\text{def}}{=}\mathrm{tr}\left(  e^{-\beta
\mathrm{H}\left(  \lambda\right)  }\right)  $ denoting the partition function
of the system. The Hamiltonian $\mathrm{H}$ in Eq. (\ref{ro}) depends on a set
of parameters $\left\{  \lambda\right\}  $ and is such that $\mathrm{H}%
\left\vert n\right\rangle =E_{n}\left\vert n\right\rangle $ or, equivalently%
\begin{equation}
\mathrm{H}=\sum_{n}E_{n}\left\vert n\right\rangle \left\langle n\right\vert
\text{,} \label{spectral}%
\end{equation}
with $1\leq n\leq N$. Using the spectral decomposition of $\mathrm{H}$ in Eq.
(\ref{spectral}), $\rho\left(  \beta\text{, }\lambda\right)  $ in Eq.
(\ref{ro}) can be recast as%
\begin{equation}
\rho\left(  \beta\text{, }\lambda\right)  =\sum_{n}p_{n}\left\vert
n\right\rangle \left\langle n\right\vert =\sum_{n}\frac{e^{-\beta E_{n}}%
}{\mathcal{Z}}\left\vert n\right\rangle \left\langle n\right\vert \text{,}
\label{rouse}%
\end{equation}
with $p_{n}\overset{\text{def}}{=}e^{-\beta E_{n}}/\mathcal{Z}$. Note that the
$\lambda$-dependence of $\rho$ in Eq. (\ref{rouse}) appears, in general, in
both $E_{n}=E_{n}\left(  \lambda\right)  $ and $\left\vert n\right\rangle
=\left\vert n\left(  \lambda\right)  \right\rangle $. Before presenting the
general case where both $\beta$ and the set of $\left\{  \lambda\right\}  $
can change, we focus on the sub-case where $\beta$ is kept constant while
$\left\{  \lambda\right\}  $ is allowed to change.

\subsubsection{Case: $\beta$-constant and $\lambda$-nonconstant}

In what follows, assuming that $\beta$ is fixed, we wish to find the
expression of $ds_{\mathrm{Bures}}^{2}\left(  \rho\text{, }\rho+d\rho\right)
$ in Eq. (\ref{7}) when $\rho$ is given as in Eq. (\ref{rouse}). \ Clearly, we
note that the two key quantities that we need to find are $dp_{i}^{2}$ and
$\left\langle i|dj\right\rangle $. Let us start with the latter. From
$\mathrm{H}\left\vert j\right\rangle =E_{j}\left\vert j\right\rangle $, we
have $d\mathrm{H}\left\vert j\right\rangle +\mathrm{H}\left\vert
dj\right\rangle =dE_{j}\left\vert j\right\rangle +E_{j}\left\vert
dj\right\rangle $.\ Assuming $i\neq j$, we have%
\begin{equation}
\left\langle i\right\vert d\mathrm{H}\left\vert j\right\rangle +\left\langle
i\right\vert \mathrm{H}\left\vert dj\right\rangle =\left\langle i\right\vert
dE_{j}\left\vert j\right\rangle +\left\langle i\right\vert E_{j}\left\vert
dj\right\rangle =dE_{j}\delta_{ij}+E_{j}\left\langle i|dj\right\rangle
=E_{j}\left\langle i|dj\right\rangle \text{,}%
\end{equation}
that is,%
\begin{equation}
\left\langle i\right\vert d\mathrm{H}\left\vert j\right\rangle +\left\langle
i\right\vert \mathrm{H}\left\vert dj\right\rangle =E_{j}\left\langle
i|dj\right\rangle \text{.} \label{8}%
\end{equation}
Observe that,%
\begin{equation}
\left\langle i\right\vert \mathrm{H}\left\vert dj\right\rangle =\left\langle
dj\right\vert \mathrm{H}^{\dagger}\left\vert i\right\rangle ^{\ast
}=\left\langle dj\right\vert \mathrm{H}\left\vert i\right\rangle ^{\ast}%
=E_{i}\left\langle dj|i\right\rangle ^{\ast}=E_{i}\left\langle
i|dj\right\rangle \text{.} \label{9}%
\end{equation}
Substituting Eq. (\ref{9}) into Eq. (\ref{8}), we get
\begin{equation}
\left\langle i\right\vert d\mathrm{H}\left\vert j\right\rangle +E_{i}%
\left\langle i|dj\right\rangle =E_{j}\left\langle i|dj\right\rangle \text{,}%
\end{equation}
that is,%
\begin{equation}
\left\langle i|dj\right\rangle =\frac{\left\langle i\right\vert d\mathrm{H}%
\left\vert j\right\rangle }{E_{j}-E_{i}}\text{.} \label{10}%
\end{equation}
Eq. (\ref{10}) is the first piece of relevant information we were looking for.
Let us not focus on calculating $dp_{i}^{2}$. From $p_{i}\overset{\text{def}%
}{=}e^{-\beta E_{i}}/\mathcal{Z}$, we get%
\begin{align}
dp_{i}  &  =d\left(  \frac{e^{-\beta E_{i}}}{\mathcal{Z}}\right) \nonumber\\
&  =\frac{1}{\mathcal{Z}}d\left(  e^{-\beta E_{i}}\right)  +e^{-\beta E_{i}%
}d\left(  \frac{1}{\mathcal{Z}}\right) \nonumber\\
&  =\frac{1}{\mathcal{Z}}\frac{d}{dE_{i}}\left(  e^{-\beta E_{i}}\right)
dE_{i}+e^{-\beta E_{i}}\left(  -\frac{1}{\mathcal{Z}^{2}}d\mathcal{Z}\right)
\nonumber\\
&  =-\beta\frac{e^{-\beta E_{i}}}{\mathcal{Z}}dE_{i}-\frac{e^{-\beta E_{i}}%
}{\mathcal{Z}}\frac{d\mathcal{Z}}{\mathcal{Z}}\nonumber\\
&  =-\beta\frac{e^{-\beta E_{i}}}{\mathcal{Z}}dE_{i}-\frac{e^{-\beta E_{i}}%
}{\mathcal{Z}}\sum_{j}\left(  \frac{d\mathcal{Z}}{dE_{j}}\frac{dE_{j}%
}{\mathcal{Z}}\right)  \text{,}%
\end{align}
that is,%
\begin{equation}
dp_{i}=-\beta\frac{e^{-\beta E_{i}}}{\mathcal{Z}}dE_{i}-\frac{e^{-\beta E_{i}%
}}{\mathcal{Z}}\sum_{j}\left(  \frac{d\mathcal{Z}}{dE_{j}}\frac{dE_{j}%
}{\mathcal{Z}}\right)  \text{.} \label{11}%
\end{equation}
At this point, note that%
\begin{align}
\sum_{j}\frac{d\mathcal{Z}}{dE_{j}}\frac{dE_{j}}{\mathcal{Z}}  &  =\sum
_{j}\frac{d}{dE_{j}}\left(  \sum_{k}e^{-\beta E_{k}}\right)  \frac{dE_{j}%
}{\mathcal{Z}}\nonumber\\
&  =-\beta\sum_{j}\frac{e^{-\beta E_{j}}}{\mathcal{Z}}dE_{j}\nonumber\\
&  =-\beta\sum_{j}p_{j}dE_{j}\text{.} \label{12}%
\end{align}
Substituting Eq. (\ref{12}) into Eq. (\ref{11}), we get%
\begin{equation}
dp_{i}=-\beta p_{i}dE_{i}+\beta p_{i}\sum_{j}p_{j}dE_{j}=-\beta p_{i}\left[
dE_{i}-\sum_{j}p_{j}dE_{j}\right]  \text{.} \label{13}%
\end{equation}
Eq. (\ref{13}) is the second piece of relevant information we were looking
for. We can now calculate $ds_{\mathrm{Bures}}^{2}\left(  \rho\text{, }%
\rho+d\rho\right)  $ in Eq. (\ref{7}) by means of Eqs. (\ref{10}) and
(\ref{13}). We obtain,%
\begin{align}
\frac{1}{4}\sum_{i}\frac{dp_{i}^{2}}{p_{i}}  &  =\frac{1}{4}\sum_{i}\beta
^{2}\frac{p_{i}^{2}}{p_{i}}\left[  dE_{i}-\sum_{j}p_{j}dE_{j}\right]
^{2}\nonumber\\
&  =\frac{\beta^{2}}{4}\sum_{i}p_{i}\left[  dE_{i}-\left\langle
dE\right\rangle _{\beta}\right]  ^{2}\nonumber\\
&  =\frac{\beta^{2}}{4}\left(  \left\langle dE^{2}\right\rangle _{\beta
}-\left\langle dE\right\rangle _{\beta}^{2}\right) \nonumber\\
&  =\frac{\beta^{2}}{4}\left(  \left\langle d\mathrm{H}_{d}^{2}\right\rangle
_{\beta}-\left\langle d\mathrm{H}_{d}\right\rangle _{\beta}^{2}\right)
\text{,}%
\end{align}
that is,%
\begin{equation}
\frac{1}{4}\sum_{i}\frac{dp_{i}^{2}}{p_{i}}=\frac{\beta^{2}}{4}\left(
\left\langle d\mathrm{H}_{d}^{2}\right\rangle _{\beta}-\left\langle
d\mathrm{H}_{d}\right\rangle _{\beta}^{2}\right)  \text{.} \label{14}%
\end{equation}
The quantity $d\mathrm{H}_{d}$ in Eq. (\ref{14}) is defined as%
\begin{equation}
d\mathrm{H}_{d}\overset{\text{def}}{=}\sum_{j}dE_{j}\left\vert j\right\rangle
\left\langle j\right\vert \text{,}%
\end{equation}
and is different from $d\mathrm{H}$. For clarity, we also observe that%
\begin{equation}
\left\langle d\mathrm{H}_{d}\right\rangle _{\beta}\overset{\text{def}}{=}%
\sum_{i}p_{i}dE_{i}\text{, and }\left\langle d\mathrm{H}_{d}^{2}\right\rangle
_{\beta}\overset{\text{def}}{=}\sum_{i}p_{i}dE_{i}^{2}\text{.}%
\end{equation}
Finally, using Eqs. (\ref{10}) and (\ref{14}), we get%
\begin{equation}
ds_{\mathrm{Bures}}^{2}\left(  \rho\text{, }\rho+d\rho\right)  =\frac
{\beta^{2}}{4}\left(  \left\langle d\mathrm{H}_{d}^{2}\right\rangle _{\beta
}-\left\langle d\mathrm{H}_{d}\right\rangle _{\beta}^{2}\right)  +\frac{1}%
{2}\sum_{n\neq m}\left\vert \frac{\left\langle n|d\mathrm{H}|m\right\rangle
}{E_{n}-E_{m}}\right\vert ^{2}\frac{\left(  e^{-\beta E_{n}}-e^{-\beta E_{m}%
}\right)  ^{2}}{\mathcal{Z}\left(  e^{-\beta E_{n}}+e^{-\beta E_{m}}\right)
}\text{.} \label{bures1}%
\end{equation}
The Bures metric $ds_{\mathrm{Bures}}^{2}\left(  \rho\text{, }\rho
+d\rho\right)  $ in Eq. (\ref{bures1}) is the Bures metric in Eq. (\ref{7})
between two mixed thermal states $\rho\left(  \beta\text{, }\lambda\right)  $
and $\left(  \rho+d\rho\right)  \left(  \beta\text{, }\lambda\right)  $ when
only changes in $\lambda$ are permitted.

\subsubsection{Case: $\beta$-nonconstant and $\lambda$-nonconstant}

In what follows, we consider the general case where both $\beta$ and the set
of $\left\{  \lambda\right\}  $ can change. The sub-case where $\beta$ changes
while the set of $\left\{  \lambda\right\}  $ is kept constant is then
obtained as a special case. For simplicity, let us assume we have two
parameters, $\beta$ and a single parameter $\lambda$ that we denote with $h$
(a magnetic field intensity, for instance). In this two-dimensional parametric
case, we generally have that
\begin{equation}
ds_{\mathrm{Bures}}^{2}\left(  \beta\text{, }h\right)  =\left(
\begin{array}
[c]{cc}%
d\beta & dh
\end{array}
\right)  \left(
\begin{array}
[c]{cc}%
g_{\beta\beta} & g_{\beta h}\\
g_{h\beta} & g_{hh}%
\end{array}
\right)  \left(
\begin{array}
[c]{c}%
d\beta\\
dh
\end{array}
\right)  =g_{\beta\beta}d\beta^{2}+g_{hh}dh^{2}+2g_{\beta h}d\beta dh\text{,}
\label{15}%
\end{equation}
where we used the fact that $g_{h\beta}=g_{\beta h}$. From Eq. (\ref{15}), we
note that%
\begin{equation}
ds_{\mathrm{Bures}}^{2}\left(  \beta\text{, }h\right)  =\left\{
\begin{array}
[c]{c}%
g_{\beta\beta}\left(  \beta\text{, }h\right)  d\beta^{2}\text{, if
}h=\text{\textrm{const}.}\\
g_{hh}\left(  \beta\text{, }h\right)  dh^{2}\text{, if }\beta
=\text{\textrm{const}.}\\
g_{\beta\beta}\left(  \beta\text{, }h\right)  d\beta^{2}+g_{hh}\left(
\beta\text{, }h\right)  dh^{2}+2g_{\beta h}\left(  \beta\text{, }h\right)
d\beta dh\text{, if }\beta\neq\text{\textrm{const}. and }h\neq
\text{\textrm{const}.}%
\end{array}
\right.  \text{.}%
\end{equation}
Recalling $ds_{\mathrm{Bures}}^{2}\left(  \rho\text{, }\rho+d\rho\right)  $ in
Eq. (\ref{7}), we start by calculating the expression of $dp_{n}$ with
$p_{n}=p_{n}\left(  h\text{, }\beta\right)  \overset{\text{def}}{=}e^{-\beta
E_{n}}/\mathcal{Z}$. We observe that $dp_{n}$ can be written as,%
\begin{equation}
dp_{n}=\frac{\partial p_{n}}{\partial h}dh+\frac{\partial p_{n}}{\partial
\beta}d\beta\text{,} \label{util1}%
\end{equation}
where $\partial p_{n}/\partial\beta$ is given by,%
\begin{align}
\frac{\partial p_{n}}{\partial\beta}  &  =\frac{\partial}{\partial\beta
}\left(  \frac{e^{-\beta E_{n}}}{\mathcal{Z}}\right) \nonumber\\
&  =\frac{1}{\mathcal{Z}}\frac{\partial}{\partial\beta}\left(  e^{-\beta
E_{n}}\right)  +e^{-\beta E_{n}}\frac{\partial}{\partial\beta}\left(  \frac
{1}{\mathcal{Z}}\right) \nonumber\\
&  =-\frac{E_{n}}{\mathcal{Z}}e^{-\beta E_{n}}+e^{-\beta E_{n}}\frac{\partial
}{\partial\mathcal{Z}}\left(  \frac{1}{\mathcal{Z}}\right)  \frac
{\partial\mathcal{Z}}{\partial\beta}\nonumber\\
&  =-p_{n}E_{n}-\frac{e^{-\beta E_{n}}}{\mathcal{Z}}\frac{1}{\mathcal{Z}}%
\frac{\partial\mathcal{Z}}{\partial\beta}\nonumber\\
&  =-p_{n}E_{n}-p_{n}\frac{\partial\ln\mathcal{Z}}{\partial\beta}\nonumber\\
&  =-p_{n}E_{n}+p_{n}\frac{1}{\mathcal{Z}}\sum_{n}E_{n}e^{-\beta E_{n}%
}\nonumber\\
&  =-p_{n}E_{n}+p_{n}\sum_{n}p_{n}E_{n}\nonumber\\
&  =-p_{n}E_{n}+p_{n}\left\langle \mathrm{H}\right\rangle \text{,}%
\end{align}
that is,%
\begin{equation}
\frac{\partial p_{n}}{\partial\beta}d\beta=p_{n}\left[  \left\langle
\mathrm{H}\right\rangle -E_{n}\right]  d\beta\text{.} \label{util2}%
\end{equation}
Note that the expectation value $\left\langle \mathrm{H}\right\rangle $ in Eq.
(\ref{util2}) is defined as $\left\langle \mathrm{H}\right\rangle $
$\overset{\text{def}}{=}\sum_{n}p_{n}E_{n}$. From Eq. (\ref{13}), we also have%
\begin{equation}
\frac{\partial p_{n}}{\partial h}dh=-\beta p_{n}\left[  \frac{\partial E_{n}%
}{\partial h}-\sum_{j}p_{j}\frac{\partial E_{j}}{\partial h}\right]  dh=\beta
p_{n}\left[  \left\langle \left(  \partial_{h}\mathrm{H}\right)
_{d}\right\rangle -\partial_{h}E_{n}\right]  dh\text{,} \label{util3}%
\end{equation}
where $\left\langle \left(  \partial_{h}\mathrm{H}\right)  _{d}\right\rangle $
is defined as%
\begin{equation}
\left\langle \left(  \partial_{h}\mathrm{H}\right)  _{d}\right\rangle
\overset{\text{def}}{=}\sum_{j}p_{j}\partial_{h}E_{j}\text{.} \label{masini}%
\end{equation}
Using Eqs. (\ref{util1}), (\ref{util2}), and (\ref{util3}), we wish to
calculate the term $\left(  1/4\right)  \sum_{n}dp_{n}^{2}/p_{n}$ in
$ds_{\mathrm{Bures}}^{2}\left(  \rho\text{, }\rho+d\rho\right)  $ in Eq.
(\ref{7}). Let us begin by observing that%
\begin{equation}
dp_{n}^{2}=\left(  \frac{\partial p_{n}}{\partial h}dh+\frac{\partial p_{n}%
}{\partial\beta}d\beta\right)  ^{2}=\left(  \partial_{h}p_{n}dh\right)
^{2}+\left(  \partial_{\beta}p_{n}d\beta\right)  ^{2}+2\partial_{\beta}%
p_{n}\partial_{h}p_{n}d\beta dh\text{.}%
\end{equation}
Therefore, we get%
\begin{align}
\frac{1}{4}\sum_{n}\frac{dp_{n}^{2}}{p_{n}}  &  =\frac{1}{4}\sum_{n}%
\frac{\left(  \partial_{h}p_{n}dh\right)  ^{2}+\left(  \partial_{\beta}%
p_{n}d\beta\right)  ^{2}+2\partial_{\beta}p_{n}\partial_{h}p_{n}d\beta
dh}{p_{n}}\nonumber\\
&  =\frac{1}{4}\sum_{n}\frac{\left(  \partial_{h}p_{n}\right)  ^{2}}{p_{n}%
}dh^{2}+\frac{1}{4}\sum_{n}\frac{\left(  \partial_{\beta}p_{n}\right)  ^{2}%
}{p_{n}}d\beta^{2}+\frac{1}{4}\sum_{n}\frac{2\partial_{\beta}p_{n}\partial
_{h}p_{n}}{p_{n}}d\beta dh\text{.}%
\end{align}
First, note that%
\begin{equation}
\frac{1}{4}\sum_{n}\frac{\left(  \partial_{\beta}p_{n}\right)  ^{2}}{p_{n}%
}d\beta^{2}=\frac{1}{4}\left[  \left\langle \mathrm{H}^{2}\right\rangle
-\left\langle \mathrm{H}\right\rangle ^{2}\right]  d\beta^{2}\text{,}
\label{A}%
\end{equation}
where $\left\langle \mathrm{H}\right\rangle $ and $\left\langle \mathrm{H}%
^{2}\right\rangle $ are defined as%
\begin{equation}
\left\langle \mathrm{H}\right\rangle \overset{\text{def}}{=}\sum_{i}p_{i}%
E_{i}\text{, and }\left\langle \mathrm{H}^{2}\right\rangle \overset{\text{def}%
}{=}\sum_{i}p_{i}E_{i}^{2}\text{,}%
\end{equation}
respectively. Indeed, using Eq. (\ref{util2}), we have%
\begin{align}
\sum_{n}\frac{\left(  \partial_{\beta}p_{n}\right)  ^{2}}{p_{n}}  &  =\sum
_{n}\frac{p_{n}^{2}\left[  \left\langle \mathrm{H}\right\rangle -E_{n}\right]
^{2}}{p_{n}}\nonumber\\
&  =\sum_{n}\frac{p_{n}^{2}\left\langle \mathrm{H}\right\rangle ^{2}+p_{n}%
^{2}E_{n}^{2}-2\left\langle \mathrm{H}\right\rangle p_{n}^{2}E_{n}}{p_{n}%
}\nonumber\\
&  =\left\langle \mathrm{H}\right\rangle ^{2}+\left\langle \mathrm{H}%
^{2}\right\rangle -2\left\langle \mathrm{H}\right\rangle ^{2}\nonumber\\
&  =\left\langle \mathrm{H}^{2}\right\rangle -\left\langle \mathrm{H}%
\right\rangle ^{2}\text{.}%
\end{align}
Second, observe that%
\begin{equation}
\frac{1}{4}\sum_{n}\frac{\left(  \partial_{h}p_{n}\right)  ^{2}}{p_{n}}%
dh^{2}=\frac{1}{4}\beta^{2}\left\{  \left\langle \left[  \left(  \partial
_{h}\mathrm{H}\right)  _{d}\right]  ^{2}\right\rangle -\left\langle \left(
\partial_{h}\mathrm{H}\right)  _{d}\right\rangle ^{2}\right\}  dh^{2}\text{,}
\label{B}%
\end{equation}
where $\left\langle \left(  \partial_{h}\mathrm{H}\right)  _{d}\right\rangle $
is given in Eq. (\ref{masini}) and $\left\langle \left[  \left(  \partial
_{h}\mathrm{H}\right)  _{d}\right]  ^{2}\right\rangle $ is defined as%
\begin{equation}
\left\langle \left[  \left(  \partial_{h}\mathrm{H}\right)  _{d}\right]
^{2}\right\rangle \overset{\text{def}}{=}\sum_{i}p_{i}\left(  \partial
_{h}E_{i}\right)  ^{2}\text{.}%
\end{equation}
Indeed, using Eq. (\ref{util3}), we have%
\begin{align}
\sum_{n}\frac{\left(  \partial_{h}p_{n}\right)  ^{2}}{p_{n}}  &  =\sum
_{n}\frac{\left(  \beta p_{n}\right)  ^{2}\left[  \left\langle \left(
\partial_{h}\mathrm{H}\right)  _{d}\right\rangle -\partial_{h}E_{n}\right]
^{2}}{p_{n}}\nonumber\\
&  =\beta^{2}\sum_{n}\left[  p_{n}\left\langle \left(  \partial_{h}%
\mathrm{H}\right)  _{d}\right\rangle ^{2}+p_{n}\left(  \partial_{h}%
E_{n}\right)  ^{2}-2p_{n}\left\langle \left(  \partial_{h}\mathrm{H}\right)
_{d}\right\rangle \partial_{h}E_{n}\right] \nonumber\\
&  =\beta^{2}\left\{  \left\langle \left(  \partial_{h}\mathrm{H}\right)
_{d}\right\rangle ^{2}+\left\langle \left[  \left(  \partial_{h}%
\mathrm{H}\right)  _{d}\right]  ^{2}\right\rangle -2\left\langle \left(
\partial_{h}\mathrm{H}\right)  _{d}\right\rangle ^{2}\right\} \nonumber\\
&  =\beta^{2}\left\{  \left\langle \left[  \left(  \partial_{h}\mathrm{H}%
\right)  _{d}\right]  ^{2}\right\rangle -\left\langle \left(  \partial
_{h}\mathrm{H}\right)  _{d}\right\rangle ^{2}\right\}  \text{.}%
\end{align}
Third, we note that%
\begin{equation}
\frac{1}{4}\sum_{n}\frac{2\partial_{\beta}p_{n}\partial_{h}p_{n}}{p_{n}}d\beta
dh=\frac{1}{4}2\beta\left[  \left\langle \mathrm{H}\left(  \partial
_{h}\mathrm{H}\right)  _{d}\right\rangle -\left\langle \mathrm{H}\right\rangle
\left\langle \left(  \partial_{h}\mathrm{H}\right)  _{d}\right\rangle \right]
d\beta dh\text{.} \label{C}%
\end{equation}
Indeed, using Eqs. (\ref{util2}) and (\ref{util3}), we get%
\begin{align}
\sum_{n}\frac{2\partial_{\beta}p_{n}\partial_{h}p_{n}}{p_{n}}  &  =\sum
_{n}\frac{2p_{n}\left[  \left\langle \mathrm{H}\right\rangle -E_{n}\right]
\beta p_{n}\left[  \left\langle \left(  \partial_{h}\mathrm{H}\right)
_{d}\right\rangle -\partial_{h}E_{n}\right]  }{p_{n}}\nonumber\\
&  =\sum_{n}2\beta\left[  \left\langle \mathrm{H}\right\rangle -E_{n}\right]
\left[  \left\langle \left(  \partial_{h}\mathrm{H}\right)  _{d}\right\rangle
-\partial_{h}E_{n}\right]  p_{n}\nonumber\\
&  =\sum_{n}2\beta\left[  \left\langle \mathrm{H}\right\rangle \left\langle
\left(  \partial_{h}\mathrm{H}\right)  _{d}\right\rangle -\left\langle
\mathrm{H}\right\rangle \partial_{h}E_{n}-E_{n}\left\langle \left(
\partial_{h}\mathrm{H}\right)  _{d}\right\rangle +E_{n}\partial_{h}%
E_{n}\right]  p_{n}\nonumber\\
&  =2\beta\left[  \left\langle \mathrm{H}\right\rangle \left\langle \left(
\partial_{h}\mathrm{H}\right)  _{d}\right\rangle -\left\langle \mathrm{H}%
\right\rangle \left\langle \left(  \partial_{h}\mathrm{H}\right)
_{d}\right\rangle -\left\langle \mathrm{H}\right\rangle \left\langle \left(
\partial_{h}\mathrm{H}\right)  _{d}\right\rangle +\left\langle \mathrm{H}%
\left(  \partial_{h}\mathrm{H}\right)  _{d}\right\rangle \right] \nonumber\\
&  =2\beta\left[  \left\langle \mathrm{H}\left(  \partial_{h}\mathrm{H}%
\right)  _{d}\right\rangle -\left\langle \mathrm{H}\right\rangle \left\langle
\left(  \partial_{h}\mathrm{H}\right)  _{d}\right\rangle \right]  \text{,}%
\end{align}
where $\left\langle \mathrm{H}\left(  \partial_{h}\mathrm{H}\right)
_{d}\right\rangle $ is defined as%
\begin{equation}
\left\langle \mathrm{H}\left(  \partial_{h}\mathrm{H}\right)  _{d}%
\right\rangle \overset{\text{def}}{=}\sum_{i}p_{i}E_{i}\partial_{h}%
E_{i}\text{.}%
\end{equation}
Finally, employing Eqs. (\ref{A}), (\ref{B}), and (\ref{C}), the most general
expression of $ds_{\mathrm{Bures}}^{2}\left(  \rho\text{, }\rho+d\rho\right)
$ in Eq. (\ref{7}) between two mixed thermal states $\rho\left(  \beta\text{,
}h\right)  $ and $\left(  \rho+d\rho\right)  \left(  \beta\text{, }h\right)  $
when either changes in the parameter $\beta$ or $h$ are allotted becomes%
\begin{align}
ds_{\mathrm{Bures}}^{2}\left(  \rho\text{, }\rho+d\rho\right)   &  =\frac
{1}{4}\left[  \left\langle \mathrm{H}^{2}\right\rangle -\left\langle
\mathrm{H}\right\rangle ^{2}\right]  d\beta^{2}\nonumber\\
&  +\frac{1}{4}\left\{  \beta^{2}\left\{  \left\langle \left[  \left(
\partial_{h}\mathrm{H}\right)  _{d}\right]  ^{2}\right\rangle -\left\langle
\left(  \partial_{h}\mathrm{H}\right)  _{d}\right\rangle ^{2}\right\}
+2\sum_{n\neq m}\left\vert \frac{\left\langle n|\partial_{h}\mathrm{H}%
|m\right\rangle }{E_{n}-E_{m}}\right\vert ^{2}\frac{\left(  e^{-\beta E_{n}%
}-e^{-\beta E_{m}}\right)  ^{2}}{\mathcal{Z}\left(  e^{-\beta E_{n}}+e^{-\beta
E_{m}}\right)  }\right\}  dh^{2}+\nonumber\\
&  +\frac{1}{4}\left\{  2\beta\left[  \left\langle \mathrm{H}\left(
\partial_{h}\mathrm{H}\right)  _{d}\right\rangle -\left\langle \mathrm{H}%
\right\rangle \left\langle \left(  \partial_{h}\mathrm{H}\right)
_{d}\right\rangle \right]  \right\}  d\beta dh\text{.} \label{general}%
\end{align}
Note that\textbf{ }$ds_{\mathrm{Bures}}^{2}\left(  \rho\text{, }\rho
+d\rho\right)  $\textbf{ }is the sum of two contributions, the classical
Fisher-Rao information metric contribution and\textbf{ }the non-classical
metric contribution expressed in the summation term in the right-hand-side of
Eq. (\ref{general}). For later convenience, we also remark that the quadratic
term\textbf{ }$\left\vert \left\langle n|\partial_{h}\mathrm{H}|m\right\rangle
\right\vert ^{2}$\textbf{ }in the summation term in the right-hand-side of Eq.
(\ref{general}) is invariant under change of sign of the Hamiltonian of the
system.\textbf{ }Clearly, from Eq. (\ref{general}) we find that
$ds_{\mathrm{Bures}}^{2}\left(  \rho\text{, }\rho+d\rho\right)  =(1/4)\left[
\left\langle \mathrm{H}^{2}\right\rangle -\left\langle \mathrm{H}\right\rangle
^{2}\right]  d\beta^{2}$ when $h=$\textrm{const}. and only $\beta$ can change.
If $\beta=$\textrm{const}., $ds_{\mathrm{Bures}}^{2}\left(  \rho\text{, }%
\rho+d\rho\right)  $ in Eq. (\ref{general}) reduces to Eq. (\ref{bures1}). The
explicit derivation of Eq. (\ref{general}) ends our calculation of the Bures
metric between neighboring thermal states undergoing temperature and/or
magnetic field intensity changes as originally presented by Zanardi and
collaborators in Ref. \cite{zanardi07}.

\section{The Sj\"{o}qvist metric}

In this section, we introduce the Sj\"{o}qvist metric \cite{erik20} for
nondegerante mixed states with an explicit derivation. Assume to consider two
rank-$N$ neighboring nondegenerate density operators $\rho\left(  t\right)  $
and $\rho\left(  t+dt\right)  $ linked by means of a smooth path $t\mapsto
\rho\left(  t\right)  $ specifying the evolution of a given quantum system.
The nondegeneracy property implies that the phase of the eigenvectors
represents the gauge freedom in the spectral decomposition of the density
operators. As a consequence, there exists a one-to-one correspondence between
the set of two orthogonal rays $\left\{  e^{i\phi_{k}\left(  t\right)
}\left\vert e_{k}\left(  t\right)  \right\rangle :0\leq\phi_{k}\left(
t\right)  <2\pi\right\}  _{1\leq k\leq N}$ that specify the spectral
decomposition along the path $t\mapsto\rho\left(  t\right)  $ and the rank-$N$
nondegenerate density operator $\rho\left(  t\right)  $. Obviously, if some
nonzero eigenvalue of $\rho\left(  t\right)  $ is degenerate, this
correspondence would no longer exist. We present next the explicit derivation
of the Sj\"{o}qvist metric.

\subsection{The explicit derivation}

Consider two neighboring states $\rho\left(  t\right)  $ and $\rho\left(
t+dt\right)  $ with spectral decompositions given by $\mathcal{B}\left(
t\right)  =\left\{  \sqrt{p_{k}\left(  t\right)  }e^{if_{k}\left(  t\right)
}\left\vert n_{k}\left(  t\right)  \right\rangle \right\}  _{1\leq k\leq N}$
and $\mathcal{B}\left(  t+dt\right)  =\left\{  \sqrt{p_{k}\left(  t+dt\right)
}e^{if_{k}\left(  t+dt\right)  }\left\vert n_{k}\left(  t+dt\right)
\right\rangle \right\}  _{1\leq k\leq N}$, respectively. The quantity $N$
denotes the rank of the nondegenerate density operator $\rho\left(  t\right)
$. Consider the infinitesimal distance $d^{2}\left(  t\text{, }t+dt\right)  $
between $\rho\left(  t\right)  $ and $\rho\left(  t+dt\right)  $ defined as%
\begin{equation}
d^{2}\left(  t\text{, }t+dt\right)  \overset{\text{def}}{=}\sum_{k=1}%
^{N}\left\Vert \sqrt{p_{k}\left(  t\right)  }e^{if_{k}\left(  t\right)
}\left\vert n_{k}\left(  t\right)  \right\rangle -\sqrt{p_{k}\left(
t+dt\right)  }e^{if_{k}\left(  t+dt\right)  }\left\vert n_{k}\left(
t+dt\right)  \right\rangle \right\Vert ^{2}\text{.} \label{distb}%
\end{equation}
The Sj\"{o}qvist metric is defined as the minimum of $d^{2}\left(  t\text{,
}t+dt\right)  $ in Eq. (\ref{distb}). Note that the squared norm term
$\left\Vert \sqrt{p_{k}\left(  t\right)  }e^{if_{k}\left(  t\right)
}\left\vert n_{k}\left(  t\right)  \right\rangle -\sqrt{p_{k}\left(
t+dt\right)  }e^{if_{k}\left(  t+dt\right)  }\left\vert n_{k}\left(
t+dt\right)  \right\rangle \right\Vert ^{2}$ can be written as
\begin{align}
&  \left(  \sqrt{p_{k}\left(  t\right)  }e^{-if_{k}\left(  t\right)
}\left\langle n_{k}\left(  t\right)  \right\vert -\sqrt{p_{k}\left(
t+dt\right)  }e^{-if_{k}\left(  t+dt\right)  }\left\langle n_{k}\left(
t+dt\right)  \right\vert \right) \nonumber\\
&  \left(  \sqrt{p_{k}\left(  t\right)  }e^{if_{k}\left(  t\right)
}\left\vert n_{k}\left(  t\right)  \right\rangle -\sqrt{p_{k}\left(
t+dt\right)  }e^{if_{k}\left(  t+dt\right)  }\left\vert n_{k}\left(
t+dt\right)  \right\rangle \right) \nonumber\\
&  =p_{k}\left(  t\right)  +p_{k}\left(  t+dt\right)  -\sqrt{p_{k}\left(
t\right)  p_{k}\left(  t+dt\right)  }e^{i\left[  f_{k}\left(  t+dt\right)
-f_{k}\left(  t\right)  \right]  }\left\langle n_{k}\left(  t\right)
|n_{k}\left(  t+dt\right)  \right\rangle +\nonumber\\
&  -\sqrt{p_{k}\left(  t\right)  p_{k}\left(  t+dt\right)  }e^{-i\left[
f_{k}\left(  t+dt\right)  -f_{k}\left(  t\right)  \right]  }\left\langle
n_{k}\left(  t+dt\right)  |n_{k}\left(  t\right)  \right\rangle \text{,}%
\end{align}
with $e^{i\left[  f_{k}\left(  t+dt\right)  -f_{k}\left(  t\right)  \right]
}\left\langle n_{k}\left(  t\right)  |n_{k}\left(  t+dt\right)  \right\rangle
+e^{-i\left[  f_{k}\left(  t+dt\right)  -f_{k}\left(  t\right)  \right]
}\left\langle n_{k}\left(  t+dt\right)  |n_{k}\left(  t\right)  \right\rangle
$ equal to%
\begin{align}
&  2\operatorname{Re}\left\{  e^{i\left[  f_{k}\left(  t+dt\right)
-f_{k}\left(  t\right)  \right]  }\left\langle n_{k}\left(  t\right)
|n_{k}\left(  t+dt\right)  \right\rangle \right\} \nonumber\\
&  =2\operatorname{Re}\left\{  e^{i\left[  f_{k}\left(  t+dt\right)
-f_{k}\left(  t\right)  \right]  }\left\vert \left\langle n_{k}\left(
t\right)  |n_{k}\left(  t+dt\right)  \right\rangle \right\vert e^{i\arg\left(
\left\langle n_{k}\left(  t\right)  |n_{k}\left(  t+dt\right)  \right\rangle
\right)  }\right\} \nonumber\\
&  =2\left\vert \left\langle n_{k}\left(  t\right)  |n_{k}\left(  t+dt\right)
\right\rangle \right\vert \operatorname{Re}\left\{  e^{i\left[  f_{k}\left(
t+dt\right)  -f_{k}\left(  t\right)  \right]  }e^{i\arg\left(  \left\langle
n_{k}\left(  t\right)  |n_{k}\left(  t+dt\right)  \right\rangle \right)
}\right\} \nonumber\\
&  =2\left\vert \left\langle n_{k}\left(  t\right)  |n_{k}\left(  t+dt\right)
\right\rangle \right\vert \cos\left[  f_{k}\left(  t+dt\right)  -f_{k}\left(
t\right)  +\arg\left(  \left\langle n_{k}\left(  t\right)  |n_{k}\left(
t+dt\right)  \right\rangle \right)  \right]  \text{.}%
\end{align}
Note that $f_{k}\left(  t+dt\right)  =f_{k}\left(  t\right)  +\dot{f}%
_{k}\left(  t\right)  dt+O\left(  dt^{2}\right)  $ and $\left\vert
n_{k}\left(  t+dt\right)  \right\rangle =\left\vert n_{k}\left(  t\right)
\right\rangle +\left\vert \dot{n}_{k}\left(  t\right)  \right\rangle
dt+O\left(  dt^{2}\right)  $. Therefore, we have
\begin{equation}
\cos\left[  f_{k}\left(  t+dt\right)  -f_{k}\left(  t\right)  +\arg\left(
\left\langle n_{k}\left(  t\right)  |n_{k}\left(  t+dt\right)  \right\rangle
\right)  \right]  =\cos\left\{  \dot{f}_{k}\left(  t\right)  dt+\arg\left[
1+\left\langle n_{k}\left(  t\right)  |\dot{n}_{k}\left(  t\right)
\right\rangle dt\right]  +O\left(  dt^{2}\right)  \right\}  \text{.}%
\end{equation}
Setting $\dot{f}_{k}\left(  t\right)  dt+\arg\left[  1+\left\langle
n_{k}\left(  t\right)  |\dot{n}_{k}\left(  t\right)  \right\rangle dt\right]
+O\left(  dt^{2}\right)  \overset{\text{def}}{=}\lambda_{k}\left(  t\text{,
}t+dt\right)  $, the infinitesimal distance $d^{2}\left(  t\text{,
}t+dt\right)  $ becomes%
\begin{equation}
d^{2}\left(  t\text{, }t+dt\right)  =2-2\sum_{k=1}^{N}\sqrt{p_{k}\left(
t\right)  p_{k}\left(  t+dt\right)  }\left\vert \left\langle n_{k}\left(
t\right)  |n_{k}\left(  t+dt\right)  \right\rangle \right\vert \cos\left[
\lambda_{k}\left(  t\text{, }t+dt\right)  \right]  \text{.}%
\end{equation}
Then, the Sj\"{o}qvist metric $ds_{\mathrm{Sj\ddot{o}qvist}}^{2}$ is the
minimum of $d^{2}\left(  t\text{, }t+dt\right)  $, $d_{\min}^{2}\left(
t\text{, }t+dt\right)  $, and is obtained when $\lambda_{k}\left(  t\text{,
}t+dt\right)  $ equals zero for any $1\leq k\leq N$. Its expression is given
by,%
\begin{equation}
ds_{\mathrm{Sj\ddot{o}qvist}}^{2}=2-2\sum_{k=1}^{N}\sqrt{p_{k}\left(
t\right)  p_{k}\left(  t+dt\right)  }\left\vert \left\langle n_{k}\left(
t\right)  |n_{k}\left(  t+dt\right)  \right\rangle \right\vert \text{.}
\label{zero1}%
\end{equation}
It is worthwhile emphasizing that the minimum of $d^{2}\left(  t\text{,
}t+dt\right)  $ is achieved by selecting phases $\left\{  f_{k}\left(
t\right)  \text{, }f_{k}\left(  t+dt\right)  \right\}  $ such that%
\begin{equation}
\dot{f}_{k}\left(  t\right)  dt+\arg\left[  1+\left\langle n_{k}\left(
t\right)  \left\vert \dot{n}_{k}\left(  t\right)  \right.  \right\rangle
dt+O\left(  dt^{2}\right)  \right]  =0 \label{condo1}%
\end{equation}
Observing that $e^{\left\langle n_{k}\left(  t\right)  \left\vert \dot{n}%
_{k}\left(  t\right)  \right.  \right\rangle dt}=1+\left\langle n_{k}\left(
t\right)  \left\vert \dot{n}_{k}\left(  t\right)  \right.  \right\rangle
dt+O\left(  dt^{2}\right)  $ is such that $\arg\left[  e^{\left\langle
n_{k}\left(  t\right)  \left\vert \dot{n}_{k}\left(  t\right)  \right.
\right\rangle dt}\right]  =-i\left\langle n_{k}\left(  t\right)  \left\vert
\dot{n}_{k}\left(  t\right)  \right.  \right\rangle dt$, Eq. (\ref{condo1})
can be rewritten to the first order in $dt$ as%
\begin{equation}
\dot{f}_{k}\left(  t\right)  -i\left\langle n_{k}\left(  t\right)  \left\vert
\dot{n}_{k}\left(  t\right)  \right.  \right\rangle =0\text{.} \label{condo2}%
\end{equation}
Eq. (\ref{condo2}) denotes the parallel transport condition $\left\langle
\psi_{k}\left(  t\right)  \left\vert \psi_{k}\left(  t\right)  \right.
\right\rangle =0$ where $\left\vert \psi_{k}\left(  t\right)  \right\rangle
\overset{\text{def}}{=}e^{if_{k}\left(  t\right)  }\left\vert n_{k}\left(
t\right)  \right\rangle $ is associated with individual pure state paths in
the chosen ensemble that defines the mixed state $\rho\left(  t\right)  $
\cite{aharonov87}. To find a more useful expression of $ds_{\mathrm{Sj\ddot
{o}qvist}}^{2}$, let us start by observing that,%
\begin{equation}
\sqrt{p_{k}\left(  t\right)  p_{k}\left(  t+dt\right)  }=p_{k}\left(
t\right)  \sqrt{1+\frac{dp_{k}\left(  t\right)  }{p_{k}\left(  t\right)  }%
}=p_{k}+\frac{1}{2}\dot{p}_{k}dt-\frac{1}{8}\frac{\dot{p}_{k}^{2}}{p_{k}%
}dt^{2}+O\left(  dt^{2}\right)  \text{.} \label{zero2}%
\end{equation}
Furthermore, to the second order in $dt$, the state $\left\vert n_{k}\left(
t+dt\right)  \right\rangle $ can be written as%
\begin{equation}
\left\vert n_{k}\left(  t+dt\right)  \right\rangle =\left\vert n_{k}\left(
t\right)  \right\rangle +\left\vert \dot{n}_{k}\left(  t\right)  \right\rangle
dt+\frac{1}{2}\left\vert \ddot{n}_{k}\left(  t\right)  \right\rangle
dt^{2}+O\left(  dt^{2}\right)  \text{.} \label{oro1}%
\end{equation}
Therefore, to the second order in $dt$, the quantum overlap $\left\langle
n_{k}\left(  t\right)  |n_{k}\left(  t+dt\right)  \right\rangle $ becomes
\begin{equation}
\left\langle n_{k}\left(  t\right)  |n_{k}\left(  t+dt\right)  \right\rangle
=\left\langle n_{k}\left(  t\right)  |n_{k}\left(  t\right)  \right\rangle
+\left\langle n_{k}\left(  t\right)  |\dot{n}_{k}\left(  t\right)
\right\rangle dt+\frac{1}{2}\left\langle n_{k}\left(  t\right)  |\ddot{n}%
_{k}\left(  t\right)  \right\rangle dt^{2}+O\left(  dt^{2}\right)
\end{equation}
Let us focus now on calculating $\left\vert \left\langle n_{k}\left(
t\right)  |n_{k}\left(  t+dt\right)  \right\rangle \right\vert =\sqrt
{\left\vert \left\langle n_{k}\left(  t\right)  |n_{k}\left(  t+dt\right)
\right\rangle \right\vert ^{2}}$, where%
\begin{equation}
\left\vert \left\langle n_{k}\left(  t\right)  |n_{k}\left(  t+dt\right)
\right\rangle \right\vert ^{2}=\left\langle n_{k}\left(  t\right)
|n_{k}\left(  t+dt\right)  \right\rangle \left\langle n_{k}\left(
t+dt\right)  |n_{k}\left(  t\right)  \right\rangle \text{.} \label{oro2}%
\end{equation}
Using Eq. (\ref{oro1}), Eq. (\ref{oro2}) becomes%
\begin{align}
\left\langle n_{k}\left(  t\right)  |n_{k}\left(  t+dt\right)  \right\rangle
\left\langle n_{k}\left(  t+dt\right)  |n_{k}\left(  t\right)  \right\rangle
&  \approx\left[  \left\langle n_{k}\left(  t\right)  |n_{k}\left(  t\right)
\right\rangle +\left\langle n_{k}\left(  t\right)  |\dot{n}_{k}\left(
t\right)  \right\rangle dt+\frac{1}{2}\left\langle n_{k}\left(  t\right)
|\ddot{n}_{k}\left(  t\right)  \right\rangle dt^{2}\right]  \cdot\nonumber\\
&  \cdot\left[  \left\langle n_{k}\left(  t\right)  |n_{k}\left(  t\right)
\right\rangle +\left\langle \dot{n}_{k}\left(  t\right)  |n_{k}\left(
t\right)  \right\rangle dt+\frac{1}{2}\left\langle \ddot{n}_{k}\left(
t\right)  |n_{k}\left(  t\right)  \right\rangle dt^{2}\right] \nonumber\\
&  =\left[  1+\left\langle n_{k}|\dot{n}_{k}\right\rangle dt+\frac{1}%
{2}\left\langle n_{k}|\ddot{n}_{k}\right\rangle dt^{2}\right]  \left[
1+\left\langle \dot{n}_{k}|n_{k}\right\rangle dt+\frac{1}{2}\left\langle
\ddot{n}_{k}|n_{k}\right\rangle dt^{2}\right] \nonumber\\
&  \approx1+\left\langle \dot{n}_{k}|n_{k}\right\rangle dt+\frac{1}%
{2}\left\langle \ddot{n}_{k}|n_{k}\right\rangle dt^{2}+\left\langle n_{k}%
|\dot{n}_{k}\right\rangle dt+\left\langle n_{k}|\dot{n}_{k}\right\rangle
\left\langle \dot{n}_{k}|n_{k}\right\rangle dt^{2}+\frac{1}{2}\left\langle
n_{k}|\ddot{n}_{k}\right\rangle dt^{2}\nonumber\\
&  =1+\left[  \left\langle \dot{n}_{k}|n_{k}\right\rangle +\left\langle
n_{k}|\dot{n}_{k}\right\rangle \right]  dt+\left\langle n_{k}|\dot{n}%
_{k}\right\rangle \left\langle \dot{n}_{k}|n_{k}\right\rangle dt^{2}+\frac
{1}{2}\left[  \left\langle n_{k}|\ddot{n}_{k}\right\rangle +\left\langle
\ddot{n}_{k}|n_{k}\right\rangle \right]  dt^{2}\nonumber\\
&  =1+\left\langle n_{k}|\dot{n}_{k}\right\rangle \left\langle \dot{n}%
_{k}|n_{k}\right\rangle dt^{2}-\left\langle \dot{n}_{k}|\dot{n}_{k}%
\right\rangle dt^{2}\text{,}%
\end{align}
that is,%
\begin{equation}
\left\langle n_{k}\left(  t\right)  |n_{k}\left(  t+dt\right)  \right\rangle
\left\langle n_{k}\left(  t+dt\right)  |n_{k}\left(  t\right)  \right\rangle
=1+\left\langle n_{k}|\dot{n}_{k}\right\rangle \left\langle \dot{n}_{k}%
|n_{k}\right\rangle dt^{2}-\left\langle \dot{n}_{k}|\dot{n}_{k}\right\rangle
dt^{2}+O\left(  dt^{2}\right)  \text{,} \label{zero3}%
\end{equation}
since $\left\langle n_{k}|n_{k}\right\rangle =1$ implies $\left\langle \dot
{n}_{k}|n_{k}\right\rangle +\left\langle n_{k}|\dot{n}_{k}\right\rangle =0$
and $\left\langle n_{k}|\ddot{n}_{k}\right\rangle +\left\langle \ddot{n}%
_{k}|n_{k}\right\rangle =-2\left\langle \dot{n}_{k}|\dot{n}_{k}\right\rangle
$. Finally, using Eqs. (\ref{zero2}) and (\ref{zero3}) along with noting that
$\sum_{k}\dot{p}_{k}=0$, the Sj\"{o}qvist metric $ds_{\mathrm{Sj\ddot{o}%
qvist}}^{2}$ in Eq. (\ref{zero1}) becomes%
\begin{align}
ds_{\mathrm{Sj\ddot{o}qvist}}^{2}  &  \approx2-2\sum_{k=1}^{N}\left(
p_{k}+\frac{1}{2}\dot{p}_{k}dt-\frac{1}{8}\frac{\dot{p}_{k}^{2}}{p_{k}}%
dt^{2}\right)  \left(  1+\frac{1}{2}\left\langle n_{k}|\dot{n}_{k}%
\right\rangle \left\langle \dot{n}_{k}|n_{k}\right\rangle dt^{2}-\frac{1}%
{2}\left\langle \dot{n}_{k}|\dot{n}_{k}\right\rangle dt^{2}\right) \nonumber\\
&  \approx2-2\sum_{k=1}^{N}p_{k}-\sum_{k=1}^{N}p_{k}\left\langle n_{k}|\dot
{n}_{k}\right\rangle \left\langle \dot{n}_{k}|n_{k}\right\rangle dt^{2}%
+\sum_{k=1}^{N}p_{k}\left\langle \dot{n}_{k}|\dot{n}_{k}\right\rangle
dt^{2}+\frac{1}{4}\sum_{k=1}^{N}\frac{\dot{p}_{k}^{2}}{p_{k}}dt^{2}\text{,}%
\end{align}
that is,%
\begin{align}
ds_{\mathrm{Sj\ddot{o}qvist}}^{2}  &  \approx\frac{1}{4}\sum_{k=1}^{N}%
\frac{\dot{p}_{k}^{2}}{p_{k}}dt^{2}+\sum_{k=1}^{N}p_{k}\left[  \left\langle
\dot{n}_{k}|\dot{n}_{k}\right\rangle -\left\langle n_{k}|\dot{n}%
_{k}\right\rangle \left\langle \dot{n}_{k}|n_{k}\right\rangle \right]
dt^{2}\nonumber\\
&  \approx\frac{1}{4}\sum_{k=1}^{N}\frac{\dot{p}_{k}^{2}}{p_{k}}dt^{2}%
+\sum_{k=1}^{N}p_{k}\left[  \left\langle \dot{n}_{k}|\left(  \mathrm{I}%
-\left\vert n_{k}\right\rangle \left\langle n_{k}\right\vert \right)  |\dot
{n}_{k}\right\rangle \right]  dt^{2}\text{,} \label{89}%
\end{align}
where \textrm{I} in Eq. (\ref{89}) denotes the identity operator on the
$N$-dimensional Hilbert space. Finally, neglecting terms that are smaller than
$O\left(  dt^{2}\right)  $ in Eq. (\ref{zero1}) and defining $ds_{k}%
^{2}\overset{\text{def}}{=}\left[  \left\langle \dot{n}_{k}|\left(
\mathrm{I}-\left\vert n_{k}\right\rangle \left\langle n_{k}\right\vert
\right)  |\dot{n}_{k}\right\rangle \right]  dt^{2}$, the expression of the
Sj\"{o}qvist metric will be formally taken to be%
\begin{equation}
ds_{\mathrm{Sj\ddot{o}qvist}}^{2}\overset{\text{def}}{=}\frac{1}{4}\sum
_{k=1}^{N}\frac{\dot{p}_{k}^{2}}{p_{k}}dt^{2}+\sum_{k=1}^{N}p_{k}ds_{k}%
^{2}\text{.} \label{vetta}%
\end{equation}
The derivation of Eq. (\ref{vetta}) concludes our explicit calculation of the
Sj\"{o}qvist metric for nondegerante mixed states. Interestingly, note that
$ds_{k}^{2}\overset{\text{def}}{=}\left\langle \dot{n}_{k}\left\vert \left(
\mathrm{I}-\left\vert n_{k}\right\rangle \left\langle n_{k}\right\vert
\right)  \right\vert \dot{n}_{k}\right\rangle dt^{2}$ in Eq. (\ref{vetta}) can
be written as $ds_{k}^{2}=\left\langle \nabla n_{k}\left\vert \nabla
n_{k}\right.  \right\rangle $ with $\left\vert \nabla n_{k}\right\rangle
\overset{\text{def}}{=}\mathrm{P}_{\bot}^{\left(  k\right)  }\left\vert
\dot{n}_{k}\right\rangle $ being the covariant derivative of $\left\vert
n_{k}\right\rangle $ and $\mathrm{P}_{\bot}^{\left(  k\right)  }%
\overset{\text{def}}{=}\mathrm{I}-\left\vert n_{k}\right\rangle \left\langle
n_{k}\right\vert $ denoting the projector onto states perpendicular to
$\left\vert n_{k}\right\rangle $.\textbf{ }In analogy to the Bures metric case
(see the comment right below Eq. (\ref{general})), we stress for later
convenience that the quadratic term\textbf{ }$ds_{k}^{2}$\textbf{ }does not
change under change of sign of the Hamiltonian of the system. The expression
of the Sj\"{o}qvist metric in Eq. (\ref{vetta}) can be viewed as expressed by
two contributions, a classical and a nonclassical term. The first term in Eq.
(\ref{vetta}) is the classical one and is represented by the classical
Fisher-Rao information metric between the two probability distributions
$\left\{  p_{k}\right\}  _{1\leq k\leq N}$ and $\left\{  p_{k}+dp_{k}\right\}
_{1\leq k\leq N}$. The second term is the nonclassical one and is represented
by a weighted average of pure state Fubini-Study metrics along directions
specified by state vectors $\left\{  \left\vert n_{k}\right\rangle \right\}
_{1\leq k\leq N}$. We are now ready to introduce our Hamiltonian models.

\section{The Hamiltonian Models}

In this section, we present two Hamiltonian models. The first Hamiltonian
model specifies a spin-$1/2$ particle in a uniform and time-independent
external magnetic field oriented along the $z$-axis. The second Hamiltonian
model, instead, describes a superconducting flux qubit. Finally, we construct
the two corresponding parametric families of thermal states by bringing these
two systems in thermal equilibrium with a reservoir at finite and non-zero
temperature $T$.\begin{figure}[t]
\centering
\includegraphics[width=0.75\textwidth] {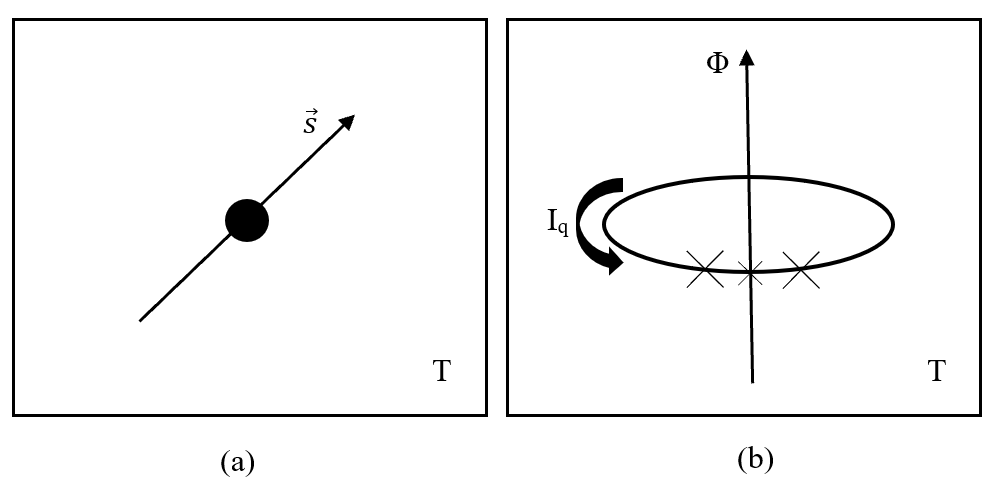}\caption{Schematic depiction of a
spin qubit (a) and a superconducting flux qubit (b) in thermal equilibrium
with a reservoir at non-zero temperature T. The spin qubit in (a) has opposite
orientations of the spin along the quantization axis as its two states. The
superconducting flux qubit in (b), instead, has circulating currents of
opposite sign as its two states.}%
\end{figure}

\subsection{Spin-1/2 qubit Hamiltonian}

Consider a spin-$1/2$ particle represented by an electron of $m$, charge $-e$
with $e\geq0$ immersed in an external magnetic field $\vec{B}\left(  t\right)
$. From a quantum-mechanical perspective, the Hamiltonian of this system can
be described the Hermitian operator \textrm{H}$\left(  t\right)  $\textrm{
}given by $\mathrm{H}\left(  t\right)  \overset{\text{def}}{=}-\vec{\mu
}\mathbf{\cdot}\vec{B}\left(  t\right)  $ \cite{sakurai}, with $\vec{\mu}$
denoting the electron magnetic moment operator. The quantity $\vec{\mu}$ is
defined as $\vec{\mu}\overset{\text{def}}{=}-\left(  e/m\right)  \vec{s}$ with
$\vec{s}\overset{\text{def}}{=}\left(  \hslash/2\right)  \vec{\sigma}$ being
the spin operator. Clearly, $\hslash\overset{\text{def}}{=}h/(2\pi)$ is the
reduced Planck constant and $\vec{\sigma}\overset{\text{def}}{=}\left(
\sigma_{x}\text{, }\sigma_{y}\text{, }\sigma_{z}\right)  $ is the usual Pauli
spin vector operator. Assuming a time-independent magnetic field along the
$z$-direction given by $\vec{B}\left(  t\right)  =B_{0}\hat{z}$ and
introducing the frequency $\omega\overset{\text{def}}{=}(e/m)B_{0}$, the
spin-$1/2$ qubit (SQ) Hamiltonian becomes
\begin{equation}
\mathrm{H}_{\mathrm{SQ}}\left(  \omega\right)  \overset{\text{def}}{=}%
\frac{\hslash\omega}{2}\sigma_{z}\text{,} \label{spinH}%
\end{equation}
where $\sigma_{z}\overset{\text{def}}{=}\left\vert \uparrow\right\rangle
\left\langle \uparrow\right\vert -\left\vert \downarrow\right\rangle
\left\langle \downarrow\right\vert $ with $\left\vert \uparrow\right\rangle $
and $\left\vert \downarrow\right\rangle $ denoting the spin-up and the
spin-down quantum states, respectively. Observe that with the sign convention
used for\textbf{ }$\mathrm{H}_{\mathrm{SQ}}\left(  \omega\right)  $\textbf{
}in Eq. (\ref{spinH}), we have that\textbf{ }$\left\vert \downarrow
\right\rangle $\textbf{ (}$\left\vert \uparrow\right\rangle $\textbf{)
}denotes the ground (excited) state of the system with energy\textbf{
}$-\hslash\omega/2$\textbf{ (}$+\hslash\omega/2$\textbf{).}

\subsection{Superconducting flux qubit Hamiltonian}

It is known that a qubit is a two-level (or, a two-state) quantum system and,
moreover, it is possible to realize the two levels in a number of ways. For
example, the two-levels can be regarded as the spin-up and spin-down of an
electron, or as the vertical and horizontal polarization of a single photon.
Interestingly, the two-levels of a qubit can be also realized as the
supercurrent flowing in an anti-clockwise and clockwise directions in a
superconducting loop \cite{clarke08,devoret13}. A flux qubit is a
superconducting loop interrupted by one or three Josephson junctions (i.e., a
dissipationless device with a nonlinear inductance). An arbitrary flux qubit
can be described as a superposition of two persistent current basis states.
The two quantum states are total magnetic flux $\Phi$ pointing up $\left\vert
\uparrow\right\rangle $ and $\Phi$ pointing down $\left\langle \downarrow
\right\vert $. Alternatively, as previously mentioned, the two-levels of the
quantum system can be described as the supercurrent $I_{q}$ circulating in the
loop anti-clockwise and $I_{q}$ circulating clockwise. The Hamiltonian of a
superconducting flux qubit (\textrm{SFQ}) in persistent current basis
$\left\{  \left\vert \uparrow\right\rangle \text{, }\left\langle
\downarrow\right\vert \right\}  $ is given by
\cite{chiorescu03,pekola07,paauw09,pekola16},%
\begin{equation}
\mathrm{H}_{\mathrm{SFQ}}\left(  \Delta\text{, }\epsilon\right)
\overset{\text{def}}{=}-\frac{\hslash}{2}\left(  \Delta\sigma_{x}%
+\epsilon\sigma_{z}\right)  \text{.} \label{superH}%
\end{equation}
In Eq. (\ref{superH}), $\hslash\overset{\text{def}}{=}h/\left(  2\pi\right)  $
is the reduced Planck constant, while $\sigma_{x}$ and $\sigma_{z}$ are Pauli
matrices. Furthermore, $\hslash\epsilon\overset{\text{def}}{=}2I_{q}\left(
\Phi_{e}-\frac{\Phi_{0}}{2}\right)  $ is the magnetic energy bias defined in
terms of the supercurrent $I_{q}$, the externally applied magnetic flux
$\Phi_{e}$, and the magnetic flux quantum $\Phi_{0}\overset{\text{def}%
}{=}h/\left(  2e\right)  $ with $e$ being the absolute value of the electron
charge. Finally, $\hslash\Delta$ is the energy gap at the degeneracy point
specified by the relation $\Phi_{e}=\Phi_{0}/2$ (i.e., $\epsilon=0$) and
represents the minimum splitting of the energy levels of the ground state
$\left\vert g\right\rangle $ and the first excited state $\left\vert
e\right\rangle $ of the superconducting qubit. At the gap, the coherence
properties of the qubit are optimal. Away from the degeneracy point,
$\epsilon\neq0$ and the energy-level splitting becomes $\hslash\nu
\overset{\text{def}}{=}\hslash\sqrt{\epsilon^{2}+\Delta^{2}}$, with $\nu$
being the transition angular frequency of the qubit. The energy level
splitting $\hslash\Delta$ depends on the critical current of the three
Josephson junctions and their capacitance \cite{paauw09}. For flux qubits one
has $\Delta\sim E_{C}/E_{J}$ with $E_{C}$ and $E_{J}$ denoting the Cooper pair
charging energy and the Josephson coupling energy \cite{pekola16},
respectively. In summary, a flux qubit can be represented by a double-well
potential whose shape (symmetrical versus asymmetrical) can be tuned with the
externally applied magnetic flux $\Phi_{e}$. When $\Phi_{e}=\Phi_{0}/2$, the
double-well is symmetric, the energy eigenstates (i.e., ground state and first
excited states $\left\vert g\right\rangle $ and $\left\vert e\right\rangle $,
respectively) are symmetric (i.e., $\left\vert g\right\rangle
\overset{\text{def}}{=}\left[  \left\vert \uparrow\right\rangle +\left\vert
\downarrow\right\rangle \right]  /\sqrt{2}$) and antisymmetric (i.e.,
$\left\vert e\right\rangle \overset{\text{def}}{=}\left[  \left\vert
\uparrow\right\rangle -\left\vert \downarrow\right\rangle \right]  /\sqrt{2}$)
superpositions of the two states $\left\vert \uparrow\right\rangle $ and
$\left\vert \downarrow\right\rangle $ and, finally, the splitting of the
energy levels of $\left\vert g\right\rangle $ and $\left\vert e\right\rangle $
is $\Delta$. Instead, when $\Phi_{e}\neq\Phi_{0}/2$, the double-well is not
symmetric, the energy eigenstates are arbitrary superpositions of the basis
states $\left\vert \uparrow\right\rangle $ and $\left\vert \downarrow
\right\rangle $ (i.e., $\alpha\left\vert \uparrow\right\rangle \pm
\beta\left\vert \downarrow\right\rangle $ with $\left\vert \alpha\right\vert
^{2}+\left\vert \beta\right\vert ^{2}=1$) and, finally, the energy gap becomes
$\hslash\nu\overset{\text{def}}{=}\hslash\sqrt{\epsilon^{2}+\Delta^{2}}$. For
more details on the theory underlying superconducting flux qubits, we refer to
Ref. \cite{clarke08}.

The transition from (isolated) physical systems specified by pure states
evolving according to the Hamiltonians in Eqs. (\ref{spinH})\ and
(\ref{superH}) to the same (open) physical systems described by mixed quantum
states can be explained as follows. Assume a quantum system specified by an
Hamiltonian $\mathrm{H}$ is in thermal equilibrium with a reservoir at
non-zero temperature $T$. Then, following the principles of quantum
statistical mechanics \cite{huang87}, the system has temperature $T$ and its
state is described by a thermal state \cite{strocchi08} specified by a density
matrix $\rho$ given by,%
\begin{equation}
\rho\overset{\text{def}}{=}\frac{e^{-\beta\mathrm{H}}}{\mathrm{tr}\left(
e^{-\beta\mathrm{H}}\right)  }\text{.} \label{densityma}%
\end{equation}
In Eq. (\ref{densityma}), $\beta\overset{\text{def}}{=}\left(  k_{B}T\right)
^{-1}$ denotes the so-called inverse temperature, while $k_{B}$ is the
Boltzmann constant. In what follows, we shall consider two families of mixed
quantum thermal states given by%
\begin{equation}
\rho_{\mathrm{SQ}}\left(  \beta\text{, }\omega\right)  \overset{\text{def}%
}{=}\frac{e^{-\beta\mathrm{H}_{\mathrm{SQ}}\left(  \omega\right)  }%
}{\mathrm{tr}\left(  e^{-\beta\mathrm{H}_{\mathrm{SQ}}\left(  \omega\right)
}\right)  }\text{ and, }\rho_{\mathrm{SFQ}}\left(  \beta\text{, }%
\epsilon\right)  \overset{\text{def}}{=}\frac{e^{-\beta\mathrm{H}%
_{\mathrm{SFQ}}\left(  \epsilon\right)  }}{\mathrm{tr}\left(  e^{-\beta
\mathrm{H}_{\mathrm{SFQ}}\left(  \epsilon\right)  }\right)  }\text{.}
\label{densita}%
\end{equation}
Note that in $\rho_{\mathrm{SFQ}}\left(  \beta\text{, }\epsilon\right)  $ in
Eq. (\ref{densita}), we assume that the parameter $\Delta$ is fixed. For a
work on how to tune the energy gap $\Delta$ in a flux qubit from an
experimental standpoint, we refer to Ref. \cite{paauw09}. In Fig. $1$, we
present a schematic depiction of of a spin qubit and a superconducting flux
qubit in thermal equilibrium with a reservoir at non-zero temperature $T$.

\section{Applications}

In this section, we calculate both the Sj\"{o}qvist and the Bures metrics for
each one of the two distinct families of parametric thermal states mentioned
in the previous section. From our comparative investigation, we find that the
two metric coincide for the first Hamiltonian model (electron in a constant
magnetic field along the $z$-direction), while they differ for the second
Hamiltonian model (superconducting flux qubit).

\subsection{Spin qubits}

Let us consider a system with an Hamiltonian described by $\mathrm{H}%
_{\mathrm{SQ}}\left(  \omega\right)  \overset{\text{def}}{=}\left(
\hslash\omega/2\right)  \sigma_{z}$ in Eq. (\ref{spinH}). Observe that
$\mathrm{H}_{\mathrm{SQ}}\left(  \omega\right)  $ can be recast as%
\begin{equation}
\mathrm{H}_{\mathrm{SQ}}\left(  \omega\right)  =\sum_{n=0}^{1}E_{n}\left\vert
n\right\rangle \left\langle n\right\vert =\frac{\hslash\omega}{2}\left\vert
0\right\rangle \left\langle 0\right\vert -\frac{\hslash\omega}{2}\left\vert
1\right\rangle \left\langle 1\right\vert \text{,} \label{h1}%
\end{equation}
where $E_{0}\overset{\text{def}}{=}\hslash\omega/2$, $E_{1}\overset{\text{def}%
}{=}-\hslash\omega/2$, and $\left\{  \left\vert n\right\rangle \right\}
\overset{\text{def}}{=}\left\{  \left\vert 0\right\rangle =\left\vert
\uparrow\right\rangle \text{, }\left\vert 1\right\rangle =\left\vert
\downarrow\right\rangle \right\}  $. For clarity, note that\textbf{
}$\left\vert 1\right\rangle =\left\vert \downarrow\right\rangle $\textbf{
(}$\left\vert 0\right\rangle =\left\vert \uparrow\right\rangle $\textbf{)
}denotes here the ground (excited) state corresponding to the lowest (highest)
energy level with\textbf{ }$E_{1}\overset{\text{def}}{=}-\hslash\omega/2$
($E_{0}\overset{\text{def}}{=}\hslash\omega/2$). Observe that the thermal
state $\rho_{\mathrm{SQ}}$ emerging from the Hamiltonian $\mathrm{H}%
_{\mathrm{SQ}}$ in Eq. (\ref{h1}) can be written as%
\begin{equation}
\rho_{\mathrm{SQ}}=\rho_{\mathrm{SQ}}\left(  \beta\text{, }\omega\right)
\overset{\text{def}}{=}\frac{e^{-\beta\mathrm{H}_{\mathrm{SQ}}\left(
\omega\right)  }}{\mathrm{tr}\left(  e^{-\beta\mathrm{H}_{\mathrm{SQ}}\left(
\omega\right)  }\right)  }\text{.} \label{mike1}%
\end{equation}
The thermal state $\rho_{\mathrm{SQ}}\left(  \beta\text{, }\omega\right)  $ in
Eq. (\ref{mike1}) can be rewritten as,%
\begin{align}
\rho_{\mathrm{SQ}}\left(  \beta\text{, }\omega\right)   &  =\frac
{e^{-\beta\frac{\hslash\omega}{2}\sigma_{z}}}{\mathrm{tr}\left(
e^{-\beta\frac{\hslash\omega}{2}\sigma_{z}}\right)  }\nonumber\\
&  =\frac{\left(
\begin{array}
[c]{cc}%
e^{-\beta\frac{\hslash\omega}{2}} & 0\\
0 & e^{\beta\frac{\hslash\omega}{2}}%
\end{array}
\right)  }{e^{-\beta\frac{\hslash\omega}{2}}+e^{\beta\frac{\hslash\omega}{2}}%
}\nonumber\\
&  =\frac{\left(
\begin{array}
[c]{cc}%
\cosh\left(  \beta\frac{\hslash\omega}{2}\right)  -\sinh\left(  \beta
\frac{\hslash\omega}{2}\right)  & 0\\
0 & \cosh\left(  \beta\frac{\hslash\omega}{2}\right)  +\sinh\left(  \beta
\frac{\hslash\omega}{2}\right)
\end{array}
\right)  }{2\cosh\left(  \beta\frac{\hslash\omega}{2}\right)  }\nonumber\\
&  =\frac{1}{2}\left(
\begin{array}
[c]{cc}%
1-\tanh\left(  \beta\frac{\hslash\omega}{2}\right)  & 0\\
0 & 1+\tanh\left(  \beta\frac{\hslash\omega}{2}\right)
\end{array}
\right) \nonumber\\
&  =\frac{1}{2}\left[  \mathrm{I}-\tanh\left(  \beta\frac{\hslash\omega}%
{2}\right)  \sigma_{z}\right]  \text{,}%
\end{align}
that is,%
\begin{equation}
\rho_{\mathrm{SQ}}\left(  \beta\text{, }\omega\right)  =\frac{1}{2}\left[
\mathrm{I}-\tanh\left(  \beta\frac{\hslash\omega}{2}\right)  \sigma
_{z}\right]  \text{.} \label{ro1}%
\end{equation}
In what follows, we shall use $\rho_{\mathrm{SQ}}\left(  \beta\text{, }%
\omega\right)  $ in Eq. (\ref{ro1}) to calculate the Bures and the
Sj\"{o}qvist metrics.

\subsubsection{The Bures metric}

We begin by noticing that $ds_{\mathrm{Bures}}^{2}$ in Eq. (\ref{general})
becomes in our case%
\begin{align}
ds_{\mathrm{Bures}}^{2}  &  =\frac{1}{4}\left[  \left\langle \mathrm{H}%
^{2}\right\rangle -\left\langle \mathrm{H}\right\rangle ^{2}\right]
d\beta^{2}\nonumber\\
&  +\frac{1}{4}\left\{  \beta^{2}\left\{  \left\langle \left[  \left(
\partial_{\omega}\mathrm{H}\right)  _{d}\right]  ^{2}\right\rangle
-\left\langle \left(  \partial_{\omega}\mathrm{H}\right)  _{d}\right\rangle
^{2}\right\}  +2\sum_{n\neq m}\left\vert \frac{\left\langle n|\partial
_{\omega}\mathrm{H}|m\right\rangle }{E_{n}-E_{m}}\right\vert ^{2}\frac{\left(
e^{-\beta E_{n}}-e^{-\beta E_{m}}\right)  ^{2}}{\mathcal{Z}\cdot\left(
e^{-\beta E_{n}}+e^{-\beta E_{m}}\right)  }\right\}  d\omega^{2}+\nonumber\\
&  +\frac{1}{4}\left\{  2\beta\left[  \left\langle \mathrm{H}\left(
\partial_{\omega}\mathrm{H}\right)  _{d}\right\rangle -\left\langle
\mathrm{H}\right\rangle \left\langle \left(  \partial_{\omega}\mathrm{H}%
\right)  _{d}\right\rangle \right]  \right\}  d\beta d\omega\text{,}
\label{general1}%
\end{align}
where, for simplicity of notation, we denote $\mathrm{H}_{\mathrm{SQ}}\left(
\omega\right)  $ in Eq. (\ref{h1}) with $\mathrm{H}$. To calculate
$ds_{\mathrm{Bures}}^{2}$ in Eq. (\ref{general}), we perform three distinct
calculations. Specifically, we compute the metric tensor components
$g_{\beta\beta}$, $2g_{\beta\omega}$, and $g_{\omega\omega}$ defined as
\begin{equation}
g_{\beta\beta}\left(  \beta\text{, }\omega\right)  \overset{\text{def}%
}{=}\frac{1}{4}\left[  \left\langle \mathrm{H}^{2}\right\rangle -\left\langle
\mathrm{H}\right\rangle ^{2}\right]  \text{, }2g_{\beta\omega}\left(
\beta\text{, }\omega\right)  \overset{\text{def}}{=}\frac{1}{4}\left\{
2\beta\left[  \left\langle \mathrm{H}\left(  \partial_{\omega}\mathrm{H}%
\right)  _{d}\right\rangle -\left\langle \mathrm{H}\right\rangle \left\langle
\left(  \partial_{\omega}\mathrm{H}\right)  _{d}\right\rangle \right]
\right\}  \text{, } \label{secondo}%
\end{equation}
and,%
\begin{equation}
g_{\omega\omega}\left(  \beta\text{, }\omega\right)  \overset{\text{def}%
}{=}\frac{1}{4}\left\{  \beta^{2}\left\{  \left\langle \left[  \left(
\partial_{\omega}\mathrm{H}\right)  _{d}\right]  ^{2}\right\rangle
-\left\langle \left(  \partial_{\omega}\mathrm{H}\right)  _{d}\right\rangle
^{2}\right\}  +2\sum_{n\neq m}\left\vert \frac{\left\langle n|\partial
_{\omega}\mathrm{H}|m\right\rangle }{E_{n}-E_{m}}\right\vert ^{2}\frac{\left(
e^{-\beta E_{n}}-e^{-\beta E_{m}}\right)  ^{2}}{\mathcal{Z}\cdot\left(
e^{-\beta E_{n}}+e^{-\beta E_{m}}\right)  }\right\}  \text{,} \label{terzo}%
\end{equation}
respectively.

\paragraph{First sub-calculation}

Let us begin with calculating $(1/4)\left[  \left\langle \mathrm{H}%
^{2}\right\rangle -\left\langle \mathrm{H}\right\rangle ^{2}\right]
d\beta^{2}$. Observe that the expectation value $\left\langle \mathrm{H}%
^{2}\right\rangle $ of $\mathrm{H}^{2}$ is given by,
\begin{align}
\left\langle \mathrm{H}^{2}\right\rangle  &  =\mathrm{tr}\left(
\mathrm{H}^{2}\rho\right)  =\sum_{i=0}^{1}p_{i}E_{i}=p_{0}E_{0}+p_{1}%
E_{1}\nonumber\\
&  =\frac{e^{-\beta E_{0}}}{\mathcal{Z}}\left(  \frac{\hslash\omega}%
{2}\right)  ^{2}+\frac{e^{-\beta E_{1}}}{\mathcal{Z}}\left(  -\frac
{\hslash\omega}{2}\right)  ^{2}\nonumber\\
& \nonumber\\
&  =\frac{\hslash^{2}\omega^{2}}{4}\left(  \frac{e^{-\beta\frac{\hslash\omega
}{2}}}{\mathcal{Z}}+\frac{e^{\beta\frac{\hslash\omega}{2}}}{\mathcal{Z}%
}\right) \nonumber\\
&  =\frac{\hslash^{2}\omega^{2}}{4}\text{,}%
\end{align}
that is,%
\begin{equation}
\left\langle \mathrm{H}^{2}\right\rangle =\frac{\hslash^{2}\omega^{2}}%
{4}\text{,} \label{a}%
\end{equation}
where the partition function is $\mathcal{Z}\overset{\text{def}}{=}%
e^{-\beta\frac{\hslash\omega}{2}}+e^{\beta\frac{\hslash\omega}{2}}%
=2\cosh\left(  \beta\frac{\hslash\omega}{2}\right)  $. Furthermore, we note
that the expectation value $\left\langle \mathrm{H}\right\rangle $ of the
Hamiltonian is
\begin{align}
\left\langle \mathrm{H}\right\rangle  &  =\mathrm{tr}\left(  \rho
\mathrm{H}\right)  =\sum_{i=0}^{1}p_{i}E_{i}=p_{0}E_{0}+p_{1}E_{1}%
=\frac{e^{-\beta E_{0}}}{\mathcal{Z}}\frac{\hslash\omega}{2}-\frac{e^{-\beta
E_{1}}}{\mathcal{Z}}\frac{\hslash\omega}{2}\nonumber\\
&  =\frac{e^{-\beta E_{0}}-e^{-\beta E_{1}}}{\mathcal{Z}}\frac{\hslash\omega
}{2}=\frac{e^{-\beta\frac{\hslash\omega}{2}}-e^{\beta\frac{\hslash\omega}{2}}%
}{e^{-\beta\frac{\hslash\omega}{2}}+e^{\beta\frac{\hslash\omega}{2}}}%
\frac{\hslash\omega}{2}=-\frac{2\sinh\left(  \beta\frac{\hslash\omega}%
{2}\right)  }{2\cosh\left(  \beta\frac{\hslash\omega}{2}\right)  }%
\frac{\hslash\omega}{2}=-\frac{\hslash\omega}{2}\tanh\left(  \beta
\frac{\hslash\omega}{2}\right)  \text{,}%
\end{align}
that is,%
\begin{equation}
\left\langle \mathrm{H}\right\rangle =-\frac{\hslash\omega}{2}\tanh\left(
\beta\frac{\hslash\omega}{2}\right)  \text{.} \label{b}%
\end{equation}
Therefore, using Eqs. (\ref{a}) and (\ref{b}), $(1/4)\left[  \left\langle
\mathrm{H}^{2}\right\rangle -\left\langle \mathrm{H}\right\rangle ^{2}\right]
d\beta^{2}$ becomes%
\begin{equation}
g_{\beta\beta}\left(  \beta\text{, }\omega\right)  d\beta^{2}%
\overset{\text{def}}{=}\frac{1}{4}\left[  \left\langle \mathrm{H}%
^{2}\right\rangle -\left\langle \mathrm{H}\right\rangle ^{2}\right]
d\beta^{2}=\frac{\hslash^{2}}{16}\omega^{2}\left[  1-\tanh^{2}\left(
\beta\frac{\hslash\omega}{2}\right)  \right]  d\beta^{2}\text{.} \label{A1}%
\end{equation}
For completeness, we remark that\textbf{ }$1-\tanh^{2}\left[  \beta\left(
\hslash\omega/2\right)  \right]  $\textbf{ }in Eq. (\ref{A1}) can also be
expressed as\textbf{ }$1$\textbf{/}$\cosh^{2}\left[  \beta\left(
\hslash\omega/2\right)  \right]  $. The calculation of $g_{\beta\beta}\left(
\beta\text{, }\omega\right)  $ in Eq. (\ref{A1}) ends our first sub-calculation.

\paragraph{Second sub-calculation}

Let us focus on the second term in Eq. (\ref{secondo}). We start by noting
that $\left\langle \mathrm{H}\left(  \partial_{\omega}\mathrm{H}\right)
_{d}\right\rangle $ is given by%
\begin{align}
\left\langle \mathrm{H}\left(  \partial_{\omega}\mathrm{H}\right)
_{d}\right\rangle  &  =\sum_{i=0}^{1}p_{i}E_{i}\partial_{\omega}E_{i}%
=p_{0}E_{0}\partial_{\omega}E_{0}+p_{1}E_{1}\partial_{\omega}E_{1}\nonumber\\
&  =p_{0}\frac{\hslash\omega}{2}\partial_{\omega}\left(  \frac{\hslash\omega
}{2}\right)  +p_{1}\left(  -\frac{\hslash\omega}{2}\right)  \partial_{\omega
}\left(  -\frac{\hslash\omega}{2}\right) \nonumber\\
&  =\frac{\hslash^{2}}{4}\omega p_{0}+\frac{\hslash^{2}}{4}\omega p_{1}%
=\frac{\hslash^{2}}{4}\omega\left(  p_{0}+p_{1}\right)  =\frac{\hslash^{2}}%
{4}\omega\text{,}%
\end{align}
that is,%
\begin{equation}
\left\langle \mathrm{H}\left(  \partial_{\omega}\mathrm{H}\right)
_{d}\right\rangle =\frac{\hslash^{2}}{4}\omega\text{.} \label{c}%
\end{equation}
Moreover, $\left\langle \left(  \partial_{\omega}\mathrm{H}\right)
_{d}\right\rangle $ can be expressed as%
\begin{align}
\left\langle \left(  \partial_{\omega}\mathrm{H}\right)  _{d}\right\rangle  &
=\sum_{i=0}^{1}p_{i}\partial_{\omega}E_{i}=p_{0}\partial_{\omega}E_{0}%
+p_{1}\partial_{\omega}E_{1}=\frac{\hslash}{2}p_{0}-\frac{\hslash}{2}%
p_{1}\nonumber\\
&  =\frac{\hslash}{2}\frac{e^{-\beta E_{0}}-e^{-\beta E_{1}}}{\mathcal{Z}%
}=\frac{\hslash}{2}\frac{e^{-\beta\frac{\hslash\omega}{2}}-e^{\beta
\frac{\hslash\omega}{2}}}{e^{-\beta\frac{\hslash\omega}{2}}+e^{\beta
\frac{\hslash\omega}{2}}}=-\frac{\hslash}{2}\frac{2\sinh\left(  \beta
\frac{\hslash\omega}{2}\right)  }{2\cosh\left(  \beta\frac{\hslash\omega}%
{2}\right)  }\nonumber\\
&  =-\frac{\hslash}{2}\tanh\left(  \beta\frac{\hslash\omega}{2}\right)
\text{,}%
\end{align}
that is,
\begin{equation}
\left\langle \left(  \partial_{\omega}\mathrm{H}\right)  _{d}\right\rangle
=-\frac{\hslash}{2}\tanh\left(  \beta\frac{\hslash\omega}{2}\right)  \text{.}
\label{d}%
\end{equation}
Therefore, using Eqs. (\ref{b}), (\ref{c}), and (\ref{d}), we obtain%
\begin{equation}
2g_{\beta\omega}\left(  \beta\text{, }\omega\right)  d\beta d\omega
\overset{\text{def}}{=}\frac{1}{4}\left\{  2\beta\left[  \left\langle
\mathrm{H}\left(  \partial_{\omega}\mathrm{H}\right)  _{d}\right\rangle
-\left\langle \mathrm{H}\right\rangle \left\langle \left(  \partial_{\omega
}\mathrm{H}\right)  _{d}\right\rangle \right]  \right\}  d\beta d\omega
=\frac{\hslash^{2}}{8}\beta\omega\left[  1-\tanh^{2}\left(  \beta\frac
{\hslash\omega}{2}\right)  \right]  d\beta d\omega\text{.} \label{B1}%
\end{equation}
The calculation of $2g_{\beta\omega}\left(  \beta\text{, }\omega\right)  $ in
Eq. (\ref{B1}) ends our second sub-calculation.

\paragraph{Third sub-calculation}

Let us now calculate the term in Eq. (\ref{terzo}). Recall from Eq. (\ref{d})
that $\left\langle \left(  \partial_{\omega}\mathrm{H}\right)  _{d}%
\right\rangle =-\frac{\hslash}{2}\tanh\left(  \beta\frac{\hslash\omega}%
{2}\right)  $. Therefore, we have%
\begin{equation}
\left\langle \left(  \partial_{\omega}\mathrm{H}\right)  _{d}\right\rangle
^{2}=\frac{\hslash^{2}}{4}\tanh^{2}\left(  \beta\frac{\hslash\omega}%
{2}\right)  \text{.} \label{f}%
\end{equation}
Moreover, we note that $\left\langle \left[  \left(  \partial_{\omega
}\mathrm{H}\right)  _{d}\right]  ^{2}\right\rangle $ can be rewritten as%
\begin{align}
\left\langle \left[  \left(  \partial_{\omega}\mathrm{H}\right)  _{d}\right]
^{2}\right\rangle  &  =\sum_{i=0}^{1}p_{i}\left(  \partial_{\omega}%
E_{i}\right)  ^{2}=p_{0}\left(  \partial_{\omega}E_{0}\right)  ^{2}%
+p_{1}\left(  \partial_{\omega}E_{1}\right)  ^{2}\nonumber\\
&  =p_{0}\left[  \partial_{\omega}\left(  -\frac{\hslash\omega}{2}\right)
\right]  ^{2}+p_{1}\left[  \partial_{\omega}\left(  \frac{\hslash\omega}%
{2}\right)  \right]  ^{2}\nonumber\\
&  =\frac{\hslash^{2}}{4}p_{0}+\frac{\hslash^{2}}{4}p_{1}=\frac{\hslash^{2}%
}{4}\left(  p_{0}+p_{1}\right)  =\frac{\hslash^{2}}{4}\text{,}%
\end{align}
that is,%
\begin{equation}
\left\langle \left[  \left(  \partial_{\omega}\mathrm{H}\right)  _{d}\right]
^{2}\right\rangle =\frac{\hslash^{2}}{4}\text{.} \label{g}%
\end{equation}
Finally, note that%
\begin{align}
2\sum_{n\neq m}\left\vert \frac{\left\langle n|\partial_{\omega}%
\mathrm{H}|m\right\rangle }{E_{n}-E_{m}}\right\vert ^{2}\frac{\left(
e^{-\beta E_{n}}-e^{-\beta E_{m}}\right)  ^{2}}{\mathcal{Z}\cdot\left(
e^{-\beta E_{n}}+e^{-\beta E_{m}}\right)  }  &  =\frac{2}{\mathcal{Z}%
}\left\vert \frac{\left\langle 0|\partial_{\omega}\mathrm{H}|1\right\rangle
}{E_{0}-E_{1}}\right\vert ^{2}\frac{e^{-\beta E_{0}}-e^{-\beta E_{1}}%
}{e^{-\beta E_{0}}+e^{-\beta E_{1}}}+\frac{2}{\mathcal{Z}}\left\vert
\frac{\left\langle 1|\partial_{\omega}\mathrm{H}|0\right\rangle }{E_{1}-E_{0}%
}\right\vert ^{2}\frac{e^{-\beta E_{1}}-e^{-\beta E_{0}}}{e^{-\beta E_{1}%
}+e^{-\beta E_{0}}}\nonumber\\
&  =0\text{,} \label{h}%
\end{align}
since $\left\langle 0|\partial_{\omega}\mathrm{H}|1\right\rangle =\left\langle
1|\partial_{\omega}\mathrm{H}|0\right\rangle =0$ as a consequence of the fact
that $\mathrm{H}=\mathrm{H}_{\mathrm{SQ}}\left(  \omega\right)  $ in Eq.
(\ref{spinH}) is diagonal. Therefore, using Eqs. (\ref{f}), (\ref{g}), and
(\ref{h}), we finally get that Eq. (\ref{terzo}) becomes%
\begin{equation}
g_{\omega\omega}\left(  \beta\text{, }\omega\right)  d\omega^{2}=\frac
{\hslash^{2}}{16}\beta^{2}\left[  1-\tanh^{2}\left(  \beta\frac{\hslash\omega
}{2}\right)  \right]  d\omega^{2}\text{.} \label{C1}%
\end{equation}
The calculation of $g_{\omega\omega}\left(  \beta\text{, }\omega\right)  $ in
Eq. (\ref{C1}) ends our third sub-calculation.

In conclusion, exploiting Eqs. (\ref{A1}), (\ref{B1}), and (\ref{C1}), the
Bures metric $ds_{\mathrm{Bures}}^{2}=g_{\beta\beta}\left(  \beta\text{,
}\omega\right)  d\beta^{2}+g_{\omega\omega}\left(  \beta\text{, }%
\omega\right)  d\omega^{2}+2g_{\beta\omega}\left(  \beta\text{, }%
\omega\right)  d\beta d\omega$ in Eq. (\ref{general1}) becomes%
\begin{equation}
ds_{\mathrm{Bures}}^{2}=\frac{\hslash^{2}\omega^{2}}{16}\left[  1-\tanh
^{2}\left(  \beta\frac{\hslash\omega}{2}\right)  \right]  d\beta^{2}%
+\frac{\hslash^{2}\beta^{2}}{16}\left[  1-\tanh^{2}\left(  \beta\frac
{\hslash\omega}{2}\right)  \right]  d\omega^{2}+\frac{\hslash^{2}\beta\omega
}{8}\left[  1-\tanh^{2}\left(  \beta\frac{\hslash\omega}{2}\right)  \right]
d\beta d\omega\text{.} \label{07}%
\end{equation}
Using Einstein's summation convention, $ds_{\mathrm{Bures}}^{2}=g_{ij}%
^{\left(  \mathrm{Bures}\right)  }\left(  \beta\text{, }\omega\right)
d\theta^{i}d\theta^{j}$ with $\theta^{1}\overset{\text{def}}{=}\beta$ and
$\theta^{2}\overset{\text{def}}{=}\omega$. Finally, using Eq. (\ref{07}), the
Bures metric metric tensor $g_{ij}^{\left(  \mathrm{Bures}\right)  }\left(
\beta\text{, }\omega\right)  $ becomes%
\begin{equation}
g_{ij}^{\left(  \mathrm{Bures}\right)  }\left(  \beta\text{, }\omega\right)
=\frac{\hslash^{2}}{16}\left[  1-\tanh^{2}\left(  \beta\frac{\hslash\omega}%
{2}\right)  \right]  \left(
\begin{array}
[c]{cc}%
\omega^{2} & \beta\omega\\
\beta\omega & \beta^{2}%
\end{array}
\right)  \text{,} \label{f2}%
\end{equation}
with $1\leq i$, $j\leq2$. Note that $g_{ij}^{\left(  \mathrm{Bures}\right)
}\left(  \beta\text{, }\omega\right)  $\textbf{ }in Eq. (\ref{f2}) equals the
classical Fisher-Rao metric since there is no non-classical contribution in
this case. The derivation of $g_{ij}^{\left(  \mathrm{Bures}\right)  }\left(
\beta\text{, }\omega\right)  $ in Eq. (\ref{f2}) ends our calculation of the
Bures metric tensor for spin qubits. Interestingly, we observe that setting
$k_{B}=1$, $\beta=t^{-1}$, and $\omega_{z}=t$, our Eq. (\ref{f2}) reduces to
the last relation obtained by Zanardi and collaborators in Ref.
\cite{zanardi07}.

\subsubsection{The Sj\"{o}qvist metric}

Given the expression of $\rho_{\mathrm{SQ}}\left(  \beta\text{, }%
\omega\right)  $ in\ Eq. (\ref{ro1}), we can proceed with the calculation of
the Sj\"{o}qvist metric given by%
\begin{equation}
ds_{\mathrm{Sj\ddot{o}qvist}}^{2}=\frac{1}{4}\sum_{k=0}^{1}\frac{dp_{k}^{2}%
}{p_{k}}+\sum_{k=0}^{1}p_{k}\left\langle dn_{k}|\left(  \mathrm{I}-\left\vert
n_{k}\right\rangle \left\langle n_{k}\right\vert \right)  |dn_{k}\right\rangle
\text{.} \label{smetric}%
\end{equation}
In our case, we note that the probabilities $p_{0}$ and $p_{1}$ are given by
\begin{equation}
p_{0}=p_{0}\left(  \beta\text{, }\omega\right)  \overset{\text{def}}{=}%
\frac{1-\tanh\left(  \beta\frac{\hslash\omega}{2}\right)  }{2}\text{, and
}p_{1}=p_{1}\left(  \beta\text{, }\omega\right)  \overset{\text{def}}{=}%
\frac{1+\tanh\left(  \beta\frac{\hslash\omega}{2}\right)  }{2}\text{,}
\label{prob}%
\end{equation}
respectively. Furthermore, the states $\left\vert n_{0}\right\rangle $ and
$\left\vert n_{1}\right\rangle $ are
\begin{equation}
\left\vert n_{0}\right\rangle \overset{\text{def}}{=}\left\vert 0\right\rangle
\text{, and }\left\vert n_{1}\right\rangle \overset{\text{def}}{=}\left\vert
1\right\rangle \text{.} \label{stati}%
\end{equation}
Observe that since $n_{k}=n_{k}\left(  \beta\text{, }\omega\right)  $, we have
that $dn_{k}\overset{\text{def}}{=}\frac{\partial n_{k}}{\partial\beta}%
d\beta+\frac{\partial n_{k}}{\partial\omega}d\omega$. In our case, we get from
Eq. (\ref{stati}) that $\left\vert dn_{k}\right\rangle =\left\vert
0\right\rangle $. From Eq. (\ref{smetric}), $ds_{\mathrm{Sj\ddot{o}qvist}}%
^{2}$ reduces to%
\begin{equation}
ds_{\mathrm{Sj\ddot{o}qvist}}^{2}=\frac{1}{4}\sum_{k=0}^{1}\frac{dp_{k}^{2}%
}{p_{k}}=\frac{1}{4}\left(  \frac{dp_{0}^{2}}{p_{0}}+\frac{dp_{1}^{2}}{p_{1}%
}\right)  \text{,} \label{n2}%
\end{equation}
where the differentials $dp_{0}$ and $dp_{1}$ are given by
\begin{equation}
dp_{0}\overset{\text{def}}{=}\frac{\partial p_{0}}{\partial\beta}d\beta
+\frac{\partial p_{0}}{\partial\omega}d\omega\text{, and }dp_{1}%
\overset{\text{def}}{=}\frac{\partial p_{1}}{\partial\beta}d\beta
+\frac{\partial p_{1}}{\partial\omega}d\omega\text{,} \label{n1}%
\end{equation}
respectively. Therefore, substituting Eq. (\ref{n1}) into Eq. (\ref{n2}), we
get%
\begin{align}
ds_{\mathrm{Sj\ddot{o}qvist}}^{2}  &  =\frac{1}{4}\frac{\left(  \partial
_{\beta}p_{0}d\beta+\partial_{\omega}p_{0}d\omega\right)  ^{2}}{p_{0}}%
+\frac{1}{4}\frac{\left(  \partial_{\beta}p_{1}d\beta+\partial_{\omega}%
p_{1}d\omega\right)  ^{2}}{p_{1}}\nonumber\\
&  =\frac{\left(  \partial_{\beta}p_{0}\right)  ^{2}d\beta^{2}+\left(
\partial_{\omega}p_{0}\right)  ^{2}d\omega^{2}+2\partial_{\beta}p_{0}%
\partial_{\omega}p_{0}d\beta d\omega}{4p_{0}}+\frac{\left(  \partial_{\beta
}p_{1}\right)  ^{2}d\beta^{2}+\left(  \partial_{\omega}p_{1}\right)
^{2}d\omega^{2}+2\partial_{\beta}p_{1}\partial_{\omega}p_{1}d\beta d\omega
}{4p_{1}}\nonumber\\
&  =\left[  \frac{\left(  \partial_{\beta}p_{0}\right)  ^{2}}{4p_{0}}%
+\frac{\left(  \partial_{\beta}p_{1}\right)  ^{2}}{4p_{1}}\right]  d\beta
^{2}+\left[  \frac{\left(  \partial_{\omega}p_{0}\right)  ^{2}}{4p_{0}}%
+\frac{\left(  \partial_{\omega}p_{1}\right)  ^{2}}{4p_{1}}\right]
d\omega^{2}+\left[  \frac{2\partial_{\beta}p_{0}\partial_{\omega}p_{0}}%
{4p_{0}}+\frac{2\partial_{\beta}p_{1}\partial_{\omega}p_{1}}{4p_{1}}\right]
d\beta d\omega\text{,}%
\end{align}
that is,%
\begin{equation}
ds_{\mathrm{Sj\ddot{o}qvist}}^{2}=\left[  \frac{\left(  \partial_{\beta}%
p_{0}\right)  ^{2}}{4p_{0}}+\frac{\left(  \partial_{\beta}p_{1}\right)  ^{2}%
}{4p_{1}}\right]  d\beta^{2}+\left[  \frac{\left(  \partial_{\omega}%
p_{0}\right)  ^{2}}{4p_{0}}+\frac{\left(  \partial_{\omega}p_{1}\right)  ^{2}%
}{4p_{1}}\right]  d\omega^{2}+\left[  \frac{2\partial_{\beta}p_{0}%
\partial_{\omega}p_{0}}{4p_{0}}+\frac{2\partial_{\beta}p_{1}\partial_{\omega
}p_{1}}{4p_{1}}\right]  d\beta d\omega\text{.} \label{chi2}%
\end{equation}
From Eq. (\ref{prob}), we observe that%
\begin{align}
\partial_{\beta}p_{0}  &  =-\frac{\hslash\omega}{4}\left[  1-\tanh^{2}\left(
\beta\frac{\hslash\omega}{2}\right)  \right]  \text{, }\partial_{\omega}%
p_{0}=-\frac{\hslash\beta}{4}\left[  1-\tanh^{2}\left(  \beta\frac
{\hslash\omega}{2}\right)  \right]  \text{,}\nonumber\\
& \nonumber\\
\partial_{\beta}p_{1}  &  =\frac{\hslash\omega}{4}\left[  1-\tanh^{2}\left(
\beta\frac{\hslash\omega}{2}\right)  \right]  \text{, }\partial_{\omega}%
p_{1}=\frac{\hslash\beta}{4}\left[  1-\tanh^{2}\left(  \beta\frac
{\hslash\omega}{2}\right)  \right]  \text{.} \label{chi1}%
\end{align}
Finally, substituting Eq. (\ref{chi1}) into Eq. (\ref{chi2}), we obtain%
\begin{equation}
ds_{\mathrm{Sj\ddot{o}qvist}}^{2}=\frac{\hslash^{2}\omega^{2}}{16}\left[
1-\tanh^{2}\left(  \beta\frac{\hslash\omega}{2}\right)  \right]  d\beta
^{2}+\frac{\hslash^{2}\beta^{2}}{16}\left[  1-\tanh^{2}\left(  \beta
\frac{\hslash\omega}{2}\right)  \right]  d\omega^{2}+\frac{\hslash^{2}%
\beta\omega}{8}\left[  1-\tanh^{2}\left(  \beta\frac{\hslash\omega}{2}\right)
\right]  d\beta d\omega\text{.} \label{f0}%
\end{equation}
Using Einstein's summation convention, $ds_{\mathrm{Sj\ddot{o}qvist}}%
^{2}=g_{ij}^{\left(  \mathrm{Sj\ddot{o}qvist}\right)  }\left(  \beta\text{,
}\omega\right)  d\theta^{i}d\theta^{j}$ with $\theta^{1}\overset{\text{def}%
}{=}\beta$ and $\theta^{2}\overset{\text{def}}{=}\omega$. Finally, using Eq.
(\ref{f0}), the Sj\"{o}qvist metric metric tensor becomes%
\begin{equation}
g_{ij}^{\left(  \mathrm{Sj\ddot{o}qvist}\right)  }\left(  \beta\text{, }%
\omega\right)  =\frac{\hslash^{2}}{16}\left[  1-\tanh^{2}\left(  \beta
\frac{\hslash\omega}{2}\right)  \right]  \left(
\begin{array}
[c]{cc}%
\omega^{2} & \beta\omega\\
\beta\omega & \beta^{2}%
\end{array}
\right)  \text{,} \label{f1}%
\end{equation}
with $1\leq i$, $j\leq2$. Note that $g_{ij}^{\left(  \mathrm{Sj\ddot{o}%
qvist}\right)  }\left(  \beta\text{, }\omega\right)  $\textbf{ }in Eq.
(\ref{f2}) is equal to the classical Fisher-Rao metric since the non-classical
contribution is absent in this case. The derivation of $g_{ij}^{\left(
\mathrm{Sj\ddot{o}qvist}\right)  }\left(  \beta\text{, }\omega\right)  $ in
Eq. (\ref{f1}) ends our calculation of the Sj\"{o}qvist metric tensor for spin qubits.

Recalling the general expressions of the Bures and Sj\"{o}qvist metrics in
Eqs. (\ref{7}) and (\ref{vetta}) and, moreover, from our first set of explicit
calculations, a few remarks are in order. First, both metrics have a classical
and a non-classical contribution. Second, the classical Fisher-Rao metric
contribution is related to changes\textbf{ }$dp_{n}=\partial_{\beta}%
p_{n}d\beta+\partial_{h}p_{n}dh$\textbf{ }in the probabilities\textbf{ }%
$p_{n}\left(  \beta\text{, }h\right)  \propto e^{-\beta E_{n}\left(  h\right)
}$\textbf{ }with\textbf{ }$\left\{  E_{n}\left(  h\right)  \right\}  $\textbf{
}being the eigenvalues of the Hamiltonian. Finally, the non-classical
contribution in the two metrics is linked to changes $\left\vert
dn\right\rangle =\partial_{h}\left\vert n\right\rangle dh=\left\vert
\partial_{h}n\right\rangle dh$\textbf{ }in the eigenvectors\textbf{ }$\left\{
\left\vert n\left(  h\right)  \right\rangle \right\}  $\textbf{ }of the
Hamiltonian. In our first Hamiltonian model,\textbf{ }H$\propto\sigma_{z}%
$\textbf{ }is diagonal and, thus, its eigenvectors do not depend on any
parameter. Therefore, we found that both the Bures and Sj\"{o}qvist metrics
reduce to the classical Fisher-Rao metric. However, one expects that
if\textbf{ \ }H\textbf{ }is not proportional to the Pauli matrix
operator\textbf{ }$\sigma_{z}$\textbf{, }non-classical contributions do not
vanish any longer and the two metrics may yield different quantum (i.e.,
non-classical) metric contributions. Indeed, if one considers a spin qubit
Hamiltonian specified by a magnetic field with an orientation that is not
constrained to be along the $z$-axis, the Bures and Sj\"{o}qvist metrics
happen to be different. In particular, for a time-independent and uniform
magnetic field given by $\vec{B}=B_{x}\hat{x}+B_{z}\hat{z}$, the spin qubit
Hamiltonian becomes \textrm{H}$_{\mathrm{SQ}}\left(  \omega_{x}\text{, }%
\omega_{z}\right)  \overset{\text{def}}{=}\left(  \hslash/2\right)
(\omega_{x}\sigma_{x}+\omega_{z}\sigma_{z})$. Assuming $\omega_{x}$%
-fixed$\neq0$, tuning only the parameters $\beta$ and $\omega_{z}$, and
repeating our metric calculations, it can be shown that the Bures and
Sj\"{o}qvist metric tensor components $g_{ij}^{\mathrm{Bures}}\left(
\beta\text{, }\omega_{z}\right)  $ and $g_{ij}^{\mathrm{Sj\ddot{o}qvist}%
}\left(  \beta\text{, }\omega_{z}\right)  $ are%
\begin{equation}
g_{ij}^{\mathrm{Bures}}\left(  \beta\text{, }\omega_{z}\right)  =\frac
{\hslash^{2}}{16}\left[  1-\tanh^{2}\left(  \beta\frac{\hslash\sqrt{\omega
_{x}^{2}+\omega_{z}^{2}}}{2}\right)  \right]  \left(
\begin{array}
[c]{cc}%
\omega_{x}^{2}+\omega_{z}^{2} & \beta\omega_{z}\\
\beta\omega_{z} & \beta^{2}\frac{\omega_{z}^{2}}{\omega_{x}^{2}+\omega_{z}%
^{2}}+\frac{4}{\hslash^{2}}\frac{\omega_{x}^{2}}{\left(  \omega_{x}^{2}%
+\omega_{z}^{2}\right)  ^{2}}\frac{\tanh^{2}\left(  \beta\frac{\hslash
\sqrt{\omega_{x}^{2}+\omega_{z}^{2}}}{2}\right)  }{1-\tanh^{2}\left(
\beta\frac{\hslash\sqrt{\omega_{x}^{2}+\omega_{z}^{2}}}{2}\right)  }%
\end{array}
\right)  \text{,} \label{G1A}%
\end{equation}
and,%
\begin{equation}
g_{ij}^{\mathrm{Sj\ddot{o}qvist}}\left(  \beta\text{, }\omega_{z}\right)
=\frac{\hslash^{2}}{16}\left[  1-\tanh^{2}\left(  \beta\frac{\hslash
\sqrt{\omega_{x}^{2}+\omega_{z}^{2}}}{2}\right)  \right]  \left(
\begin{array}
[c]{cc}%
\omega_{x}^{2}+\omega_{z}^{2} & \beta\omega_{z}\\
\beta\omega_{z} & \beta^{2}\frac{\omega_{z}^{2}}{\omega_{x}^{2}+\omega_{z}%
^{2}}+\frac{4}{\hslash^{2}}\frac{\omega_{x}^{2}}{\left(  \omega_{x}^{2}%
+\omega_{z}^{2}\right)  ^{2}}\frac{1}{1-\tanh^{2}\left(  \beta\frac
{\hslash\sqrt{\omega_{x}^{2}+\omega_{z}^{2}}}{2}\right)  }%
\end{array}
\right)  \text{,} \label{G1B}%
\end{equation}
respectively.\textbf{ }For completeness, we remark that useful calculation
techniques to arrive at expressions as in Eqs. (\ref{G1A}) and (\ref{G1B})
will be performed in the next subsection where\textbf{ }\textrm{H}%
$_{\mathrm{SQ}}\left(  \omega_{x}\text{, }\omega_{z}\right)  $ will be
replaced by the superconducting flux qubit\textbf{ }Hamiltonian $\mathrm{H}%
_{\mathrm{SFQ}}\left(  \Delta\text{, }\epsilon\right)  \overset{\text{def}%
}{=}\left(  -\hslash/2\right)  \left(  \Delta\sigma_{x}+\epsilon\sigma
_{z}\right)  $. Returning to our considerations, recall that for any\textbf{
}$x\in%
\mathbb{R}
$\textbf{,} we have%
\begin{equation}
\frac{\tanh^{2}\left(  x\right)  }{1-\tanh^{2}\left(  x\right)  }=\sinh
^{2}\left(  x\right)  \text{, and }\frac{1}{1-\tanh^{2}\left(  x\right)
}=\cosh^{2}\left(  x\right)  \text{.}%
\end{equation}
\textbf{T}hen, using Eqs. (\ref{G1A}) and (\ref{G1B}), we obtain%
\begin{equation}
0\leq\frac{g_{\omega_{z}\omega_{z}}^{\mathrm{nc}\text{, \textrm{Bures}}%
}\left(  \beta\text{, }\omega_{z}\right)  }{g_{\omega_{z}\omega_{z}%
}^{\mathrm{nc}\text{, }\mathrm{Sj\ddot{o}qvist}}\left(  \beta\text{, }%
\omega_{z}\right)  }=\frac{\sinh^{2}\left(  \beta\frac{\hslash\sqrt{\omega
_{x}^{2}+\omega_{z}^{2}}}{2}\right)  }{\cosh^{2}\left(  \beta\frac
{\hslash\sqrt{\omega_{x}^{2}+\omega_{z}^{2}}}{2}\right)  }=\tanh^{2}\left(
\beta\frac{\hslash\sqrt{\omega_{x}^{2}+\omega_{z}^{2}}}{2}\right)
\leq1\text{,} \label{G1C}%
\end{equation}
with\textbf{ }$g_{\omega_{z}\omega_{z}}^{\mathrm{nc}\text{, \textrm{Bures}}%
}\left(  \beta\text{, }\omega_{z}\right)  $ and $g_{\omega_{z}\omega_{z}%
}^{\mathrm{nc}\text{, }\mathrm{Sj\ddot{o}qvist}}\left(  \beta\text{, }%
\omega_{z}\right)  $ denoting the non-classical contributions in the Bures and
Sj\"{o}qvist metric cases, respectively. From Eqs. (\ref{G1A}) and
(\ref{G1B}), we conclude that the introduction of a nonvanishing component of
the magnetic field along the $x$-direction introduces a visible
non-commutative probabilistic structure in the quantum mechanics of the system
characterized by a non-classical scenario with $\left[  \rho\text{, }%
\rho+d\rho\right]  \neq0$). In such a case, the Bures and the Sj\"{o}qvist
metrics exhibit a different behavior as evident from their nonclassical metric
tensor components (i.e., $g_{\omega_{z}\omega_{z}}^{\mathrm{nc}}\left(
\beta\text{, }\omega_{z}\right)  $) in Eq. (\ref{G1C}).

\subsection{Superconducting flux qubits}

Let us consider a system with an Hamiltonian described by $\mathrm{H}%
_{\mathrm{SFQ}}\left(  \Delta\text{, }\epsilon\right)  \overset{\text{def}%
}{=}\left(  -\hslash/2\right)  \left(  \Delta\sigma_{x}+\epsilon\sigma
_{z}\right)  $ in Eq. (\ref{superH}). The thermal state $\rho_{\mathrm{SFQ}%
}\left(  \beta\text{, }\epsilon\right)  $ corresponding to $\mathrm{H}%
_{\mathrm{SFQ}}\left(  \Delta\text{, }\epsilon\right)  $ with $\Delta$ assumed
to be constant is given by%
\begin{equation}
\rho_{\mathrm{SFQ}}\left(  \beta\text{, }\epsilon\right)  \overset{\text{def}%
}{=}\frac{e^{-\beta\mathrm{H}_{\mathrm{SFQ}}\left(  \Delta\text{, }%
\epsilon\right)  }}{\mathrm{tr}\left(  e^{-\beta\mathrm{H}_{\mathrm{SFQ}%
}\left(  \Delta\text{, }\epsilon\right)  }\right)  }\text{.} \label{anto1}%
\end{equation}
Observe that $\mathrm{H}_{\mathrm{SFQ}}\left(  \Delta\text{, }\epsilon\right)
$ is diagonalizable and can be recast as $\mathrm{H}_{\mathrm{SFQ}%
}=M_{\mathrm{H}_{\mathrm{SFQ}}}\mathrm{H}_{\mathrm{SFQ}}^{\left(
\mathrm{diagonal}\right)  }M_{\mathrm{H}_{\mathrm{SFQ}}}^{-1}$ where\textbf{
}$M_{\mathrm{H}_{\mathrm{SFQ}}}$\textbf{ }and\textbf{ }$M_{\mathrm{H}%
_{\mathrm{SFQ}}}^{-1}$\textbf{ }are the eigenvector matrix and its inverse,
respectively. Therefore, after some algebra, $\rho_{\mathrm{SFQ}}\left(
\beta\text{, }\epsilon\right)  $ in Eq. (\ref{anto1}) can be rewritten as%
\begin{align}
\rho_{\mathrm{SFQ}}\left(  \beta\text{, }\epsilon\right)   &  =\frac
{e^{-\beta\mathrm{H}_{\mathrm{SFQ}}\left(  \Delta\text{, }\epsilon\right)  }%
}{\mathrm{tr}(e^{-\beta\mathrm{H}_{\mathrm{SFQ}}\left(  \Delta\text{,
}\epsilon\right)  })}=\frac{e^{-\beta M_{\mathrm{H}_{\mathrm{SFQ}}}%
\mathrm{H}_{\mathrm{SFQ}}^{\left(  \mathrm{diagonal}\right)  }M_{\mathrm{H}%
_{\mathrm{SFQ}}}^{-1}}}{\mathrm{tr}(e^{-\beta M_{\mathrm{H}_{\mathrm{SFQ}}%
}\mathrm{H}_{\mathrm{SFQ}}^{\left(  \mathrm{diagonal}\right)  }M_{\mathrm{H}%
_{\mathrm{SFQ}}}^{-1}})}\nonumber\\
&  =\frac{M_{\mathrm{H}_{\mathrm{SFQ}}}e^{-\beta\mathrm{H}_{\mathrm{SFQ}%
}^{\left(  \mathrm{diagonal}\right)  }}M_{\mathrm{H}_{\mathrm{SFQ}}}^{-1}%
}{\mathrm{tr}(M_{\mathrm{H}_{\mathrm{SFQ}}}e^{-\beta\mathrm{H}_{\mathrm{SFQ}%
}^{\left(  \mathrm{diagonal}\right)  }}M_{\mathrm{H}_{\mathrm{SFQ}}}^{-1}%
)}\nonumber\\
&  =M_{\mathrm{H}_{\mathrm{SFQ}}}\frac{e^{-\beta\mathrm{H}_{\mathrm{SFQ}%
}^{\left(  \mathrm{diagonal}\right)  }}}{\mathrm{tr}(e^{-\beta\mathrm{H}%
_{\mathrm{SFQ}}^{\left(  \mathrm{diagonal}\right)  }})}M_{\mathrm{H}%
_{\mathrm{SFQ}}}^{-1}\text{,}%
\end{align}
that is,%
\begin{equation}
\rho_{\mathrm{SFQ}}\left(  \beta\text{, }\epsilon\right)  =M_{\mathrm{H}%
_{\mathrm{SFQ}}}\frac{e^{-\beta\mathrm{H}_{\mathrm{SFQ}}^{\left(
\mathrm{diagonal}\right)  }}}{\mathrm{tr}(e^{-\beta\mathrm{H}_{\mathrm{SFQ}%
}^{\left(  \mathrm{diagonal}\right)  }})}M_{\mathrm{H}_{\mathrm{SFQ}}}%
^{-1}=M_{\mathrm{H}_{\mathrm{SFQ}}}\rho_{\mathrm{SFQ}}^{\left(
\mathrm{diagonal}\right)  }\left(  \beta\text{, }\epsilon\right)
M_{\mathrm{H}_{\mathrm{SFQ}}}^{-1}\text{.} \label{anto0}%
\end{equation}
The quantity $\mathrm{H}_{\mathrm{SFQ}}^{\left(  \mathrm{diagonal}\right)  }$
in Eq. (\ref{anto0}) is defined as,%
\begin{equation}
\mathrm{H}_{\mathrm{SFQ}}^{\left(  \mathrm{diagonal}\right)  }%
\overset{\text{def}}{=}E_{0}\left\vert n_{1}\right\rangle \left\langle
n_{1}\right\vert +E_{1}\left\vert n_{0}\right\rangle \left\langle
n_{0}\right\vert \text{.} \label{anto2}%
\end{equation}
The the eigenvalues $E_{0}$ and $E_{1}$ are given by $E_{0}\overset{\text{def}%
}{=}-\left(  \hslash/2\right)  \nu$ and $E_{1}\overset{\text{def}}{=}+\left(
\hslash/2\right)  \nu$, respectively, with $\nu\overset{\text{def}}{=}%
\sqrt{\Delta^{2}+\epsilon^{2}}$. For later use, it is convenient to introduce
the notation $\tilde{E}_{0}\overset{\text{def}}{=}E_{1}$ and $\tilde{E}%
_{1}\overset{\text{def}}{=}E_{0}$ so that $\mathrm{H}_{\mathrm{SFQ}}^{\left(
\mathrm{diagonal}\right)  }\overset{\text{def}}{=}\tilde{E}_{0}\left\vert
n_{0}\right\rangle \left\langle n_{0}\right\vert +\tilde{E}_{1}\left\vert
n_{1}\right\rangle \left\langle n_{1}\right\vert $. The two orthonormal
eigenvectors corresponding to $E_{0}$ and $E_{1}$ are $\left\vert
n_{1}\right\rangle $ and $\left\vert n_{0}\right\rangle $, respectively. They
are given by%
\begin{equation}
\left\vert n_{0}\right\rangle \overset{\text{def}}{=}\frac{1}{\sqrt{2}}\left(
\begin{array}
[c]{c}%
\frac{\epsilon-\sqrt{\epsilon^{2}+\Delta^{2}}}{\sqrt{\epsilon^{2}+\Delta
^{2}-\epsilon\sqrt{\epsilon^{2}+\Delta^{2}}}}\\
\frac{\Delta}{\sqrt{\epsilon^{2}+\Delta^{2}-\epsilon\sqrt{\epsilon^{2}%
+\Delta^{2}}}}%
\end{array}
\right)  \text{ and, }\left\vert n_{1}\right\rangle \overset{\text{def}%
}{=}\frac{1}{\sqrt{2}}\left(
\begin{array}
[c]{c}%
\frac{\epsilon+\sqrt{\epsilon^{2}+\Delta^{2}}}{\sqrt{\epsilon^{2}+\Delta
^{2}+\epsilon\sqrt{\epsilon^{2}+\Delta^{2}}}}\\
\frac{\Delta}{\sqrt{\epsilon^{2}+\Delta^{2}+\epsilon\sqrt{\epsilon^{2}%
+\Delta^{2}}}}%
\end{array}
\right)  \text{,} \label{anto444}%
\end{equation}
respectively. A suitable choice for the eigenvector matrix $M_{\mathrm{H}%
_{\mathrm{SFQ}}}$ and its inverse $M_{\mathrm{H}_{\mathrm{SFQ}}}^{-1}$ in Eq.
(\ref{anto0}) can be expressed as%
\begin{equation}
M_{\mathrm{H}_{\mathrm{SFQ}}}\overset{\text{def}}{=}\left(
\begin{array}
[c]{cc}%
\frac{\epsilon+\nu}{\Delta} & \frac{\epsilon-\nu}{\Delta}\\
1 & 1
\end{array}
\right)  \text{ and, }M_{\mathrm{H}_{\mathrm{SFQ}}}^{-1}\overset{\text{def}%
}{=}\left(
\begin{array}
[c]{cc}%
\frac{\Delta}{2\nu} & \frac{\nu-\epsilon}{2\nu}\\
-\frac{\Delta}{2\nu} & \frac{\nu+\epsilon}{2\nu}%
\end{array}
\right)  \text{,} \label{anto44}%
\end{equation}
respectively. Using Eqs. (\ref{anto2}) and (\ref{anto44}), $\rho
_{\mathrm{SFQ}}\left(  \beta\text{, }\epsilon\right)  $ in Eq. (\ref{anto0})
becomes%
\begin{equation}
\rho_{\mathrm{SFQ}}\left(  \beta\text{, }\epsilon\right)  =\frac{1}{2}\left(
\begin{array}
[c]{cc}%
1+\frac{\epsilon}{\sqrt{\epsilon^{2}+\Delta^{2}}}\tanh\left(  \beta
\hslash\frac{\sqrt{\epsilon^{2}+\Delta^{2}}}{2}\right)  & \frac{\Delta}%
{\sqrt{\epsilon^{2}+\Delta^{2}}}\tanh\left(  \beta\hslash\frac{\sqrt
{\epsilon^{2}+\Delta^{2}}}{2}\right) \\
\frac{\Delta}{\sqrt{\epsilon^{2}+\Delta^{2}}}\tanh\left(  \beta\hslash
\frac{\sqrt{\epsilon^{2}+\Delta^{2}}}{2}\right)  & 1-\frac{\epsilon}%
{\sqrt{\epsilon^{2}+\Delta^{2}}}\tanh\left(  \beta\hslash\frac{\sqrt
{\epsilon^{2}+\Delta^{2}}}{2}\right)
\end{array}
\right)  \text{,}%
\end{equation}
that is,%
\begin{equation}
\rho_{\mathrm{SFQ}}\left(  \beta\text{, }\epsilon\right)  =\frac{1}{2}\left[
\mathrm{I}+\left(  \frac{\Delta}{\sqrt{\epsilon^{2}+\Delta^{2}}}\sigma
_{x}+\frac{\epsilon}{\sqrt{\epsilon^{2}+\Delta^{2}}}\sigma_{z}\right)
\tanh\left(  \beta\hslash\frac{\sqrt{\epsilon^{2}+\Delta^{2}}}{2}\right)
\right]  \text{.} \label{anto5}%
\end{equation}
For completeness, we note here that the spectral decomposition of
$\rho_{\mathrm{SFQ}}\left(  \beta\text{, }\epsilon\right)  =M_{\mathrm{H}%
_{\mathrm{SFQ}}}\rho_{\mathrm{SFQ}}^{\left(  \mathrm{diagonal}\right)
}\left(  \beta\text{, }\epsilon\right)  M_{\mathrm{H}_{\mathrm{SFQ}}}^{-1}$ in
Eq. (\ref{anto5}) is given by $\rho_{\mathrm{SFQ}}\overset{\text{def}}{=}%
p_{0}\left\vert n_{0}\right\rangle \left\langle n_{0}\right\vert
+p_{1}\left\vert n_{1}\right\rangle \left\langle n_{1}\right\vert $. The
probabilities $p_{0}$ and $p_{1}$ are%
\begin{equation}
p_{0}\overset{\text{def}}{=}\frac{e^{-\beta\tilde{E}_{0}}}{\mathcal{Z}}%
=\frac{1}{2}\left[  1-\tanh\left(  \beta\hslash\frac{\sqrt{\epsilon^{2}%
+\Delta^{2}}}{2}\right)  \right]  \text{, and }p_{1}\overset{\text{def}%
}{=}\frac{e^{-\beta\tilde{E}_{1}}}{\mathcal{Z}}=\frac{1}{2}\left[
1+\tanh\left(  \beta\hslash\frac{\sqrt{\epsilon^{2}+\Delta^{2}}}{2}\right)
\right]  \text{,} \label{sorry}%
\end{equation}
respectively, with $\mathcal{Z}\overset{\text{def}}{\mathcal{=}}e^{-\beta
E_{0}}+e^{-\beta E_{1}}=e^{-\beta\tilde{E}_{0}}+e^{-\beta\tilde{E}_{1}}%
=2\cosh(\beta\hslash\frac{\sqrt{\epsilon^{2}+\Delta^{2}}}{2})$ denoting the
partition function of the system. In what follows, we shall use $\rho
_{\mathrm{SFQ}}\left(  \beta\text{, }\epsilon\right)  $ in Eq. (\ref{anto5})
to calculate the Bures and the Sj\"{o}qvist metrics.

\subsubsection{The Bures metric}

For simplicity of notation, we replace $\mathrm{H}_{\mathrm{SFQ}}$ with
$\mathrm{H}$ in the forthcoming calculation. We begin by noting that, in our
case, the general expression of the Bures metric $ds_{\mathrm{Bures}}^{2}$ in
Eq. (\ref{general}) becomes%
\begin{align}
ds_{\mathrm{Bures}}^{2}  &  =\frac{1}{4}\left[  \left\langle \mathrm{H}%
^{2}\right\rangle -\left\langle \mathrm{H}\right\rangle ^{2}\right]
d\beta^{2}\nonumber\\
&  +\frac{1}{4}\left\{  \beta^{2}\left\{  \left\langle \left[  \left(
\partial_{\epsilon}\mathrm{H}\right)  _{d}\right]  ^{2}\right\rangle
-\left\langle \left(  \partial_{\epsilon}\mathrm{H}\right)  _{d}\right\rangle
^{2}\right\}  +2\sum_{n\neq m}\left\vert \frac{\left\langle n|\partial
_{\epsilon}\mathrm{H}|m\right\rangle }{\tilde{E}_{n}-\tilde{E}_{m}}\right\vert
^{2}\frac{\left(  e^{-\beta\tilde{E}_{n}}-e^{-\beta\tilde{E}_{m}}\right)
^{2}}{\mathcal{Z}\cdot\left(  e^{-\beta\tilde{E}_{n}}+e^{-\beta\tilde{E}_{m}%
}\right)  }\right\}  d\epsilon^{2}+\nonumber\\
&  +\frac{1}{4}\left\{  2\beta\left[  \left\langle \mathrm{H}\left(
\partial_{\epsilon}\mathrm{H}\right)  _{d}\right\rangle -\left\langle
\mathrm{H}\right\rangle \left\langle \left(  \partial_{\epsilon}%
\mathrm{H}\right)  _{d}\right\rangle \right]  \right\}  d\beta d\epsilon
\text{.} \label{zarrillo}%
\end{align}
As previously pointed out in this manuscript, $ds_{\mathrm{Bures}}^{2}%
$\textbf{ }is the sum of two contributions, the classical Fisher-Rao
information metric contribution and the non-classical metric contribution
described in the summation term in the right-hand-side of Eq. (\ref{zarrillo}%
). In what follows, we shall the that the presence of nonvanishing
terms\textbf{ }$\left\vert \left\langle n|\partial_{\epsilon}\mathrm{H}%
|m\right\rangle \right\vert ^{2}$ leads to the existence of a non-classical
contribution in\textbf{ }$ds_{\mathrm{Bures}}^{2}$. Following our previous
line of reasoning, we partition our calculation in three parts. In particular,
since $ds_{\mathrm{Bures}}^{2}=g_{\beta\beta}\left(  \beta\text{, }%
\epsilon\right)  d\beta^{2}+g_{\epsilon\epsilon}\left(  \beta\text{, }%
\epsilon\right)  d\epsilon^{2}+2g_{\beta\epsilon}\left(  \beta\text{,
}\epsilon\right)  d\beta d\epsilon$, we focus on computing $g_{\beta\beta
}\left(  \beta\text{, }\epsilon\right)  $, $2g_{\beta\epsilon}\left(
\beta\text{, }\epsilon\right)  $, and $g_{\epsilon\epsilon}\left(
\beta\text{, }\epsilon\right)  $.

\subsubsection{First sub-calculation}

We proceed here with the first sub-calculation. We recall that $g_{\beta\beta
}\left(  \beta\text{, }\epsilon\right)  d\beta^{2}=(1/4)\left[  \left\langle
\mathrm{H}^{2}\right\rangle -\left\langle \mathrm{H}\right\rangle ^{2}\right]
d\beta^{2}$. Note that $\left\langle \mathrm{H}^{2}\right\rangle $ and
$\left\langle \mathrm{H}\right\rangle ^{2}$ are given by,%
\begin{equation}
\left\langle \mathrm{H}^{2}\right\rangle =\mathrm{tr}\left(  \mathrm{H}%
^{2}\rho\right)  =\frac{\hslash^{2}}{4}\left(  \epsilon^{2}+\Delta^{2}\right)
\text{,} \label{jo2}%
\end{equation}
and,%
\begin{equation}
\left\langle \mathrm{H}\right\rangle ^{2}=\left[  \mathrm{tr}\left(
\mathrm{H}\rho\right)  \right]  ^{2}=\frac{\hslash^{2}}{4}\left(  \epsilon
^{2}+\Delta^{2}\right)  \tanh^{2}\left(  \beta\hslash\frac{\sqrt{\epsilon
^{2}+\Delta^{2}}}{2}\right)  \text{,} \label{jo3}%
\end{equation}
respectively. Therefore, using Eqs. (\ref{jo2}) and (\ref{jo3}),
$g_{\beta\beta}\left(  \beta\text{, }\epsilon\right)  d\beta^{2}$ becomes%
\begin{equation}
g_{\beta\beta}\left(  \beta\text{, }\epsilon\right)  d\beta^{2}=\frac
{\hslash^{2}}{16}\left(  \epsilon^{2}+\Delta^{2}\right)  \left[  1-\tanh
^{2}\left(  \beta\hslash\frac{\sqrt{\epsilon^{2}+\Delta^{2}}}{2}\right)
\right]  d\beta^{2}\text{.} \label{J1}%
\end{equation}
Our fist sub-calculation ends with the derivation of Eq. (\ref{J1}).

\subsubsection{Second sub-calculation}

In our second calculation, we focus on calculating the term $2g_{\beta
\epsilon}\left(  \beta\text{, }\epsilon\right)  d\beta d\epsilon$ defined as
\begin{equation}
2g_{\beta\epsilon}\left(  \beta\text{, }\epsilon\right)  d\beta d\epsilon
\overset{\text{def}}{=}\frac{1}{4}\left\{  2\beta\left[  \left\langle
\mathrm{H}\left(  \partial_{\epsilon}\mathrm{H}\right)  _{d}\right\rangle
-\left\langle \mathrm{H}\right\rangle \left\langle \left(  \partial_{\epsilon
}\mathrm{H}\right)  _{d}\right\rangle \right]  \right\}  d\beta d\epsilon
\text{.} \label{jo4}%
\end{equation}
Note that $\left\langle \mathrm{H}\left(  \partial_{\epsilon}\mathrm{H}%
\right)  _{d}\right\rangle $ can be recast as%
\begin{align}
\left\langle \mathrm{H}\left(  \partial_{\epsilon}\mathrm{H}\right)
_{d}\right\rangle  &  =\sum_{i=0}^{1}p_{i}\tilde{E}_{i}\partial_{\epsilon
}\tilde{E}_{i}=p_{0}\tilde{E}_{0}\partial_{\epsilon}\tilde{E}_{0}+p_{1}%
\tilde{E}_{1}\partial_{\epsilon}\tilde{E}_{1}\nonumber\\
&  =p_{0}\left(  \frac{\hslash}{2}\sqrt{\epsilon^{2}+\Delta^{2}}\right)
\partial_{\epsilon}\left(  \frac{\hslash}{2}\sqrt{\epsilon^{2}+\Delta^{2}%
}\right)  +p_{1}\left(  -\frac{\hslash}{2}\sqrt{\epsilon^{2}+\Delta^{2}%
}\right)  \partial_{\epsilon}\left(  -\frac{\hslash}{2}\sqrt{\epsilon
^{2}+\Delta^{2}}\right) \nonumber\\
&  =\left(  p_{0}+p_{1}\right)  \left(  \frac{\hslash}{2}\sqrt{\epsilon
^{2}+\Delta^{2}}\right)  \partial_{\epsilon}\left(  \frac{\hslash}{2}%
\sqrt{\epsilon^{2}+\Delta^{2}}\right) \nonumber\\
&  =\left(  \frac{\hslash}{2}\sqrt{\epsilon^{2}+\Delta^{2}}\right)
\partial_{\epsilon}\left(  \frac{\hslash}{2}\sqrt{\epsilon^{2}+\Delta^{2}%
}\right) \nonumber\\
&  =\frac{\hslash^{2}}{4}\epsilon\text{,}%
\end{align}
that is,%
\begin{equation}
\left\langle \mathrm{H}\left(  \partial_{\epsilon}\mathrm{H}\right)
_{d}\right\rangle =\frac{\hslash^{2}}{4}\epsilon\text{.} \label{mas1}%
\end{equation}
We also note that the expectation value $\left\langle \mathrm{H}\right\rangle
$ of the Hamiltonian $\mathrm{H}$ equals%
\begin{equation}
\left\langle \mathrm{H}\right\rangle =-\frac{\hslash}{2}\sqrt{\epsilon
^{2}+\Delta^{2}}\tanh\left(  \beta\hslash\frac{\sqrt{\epsilon^{2}+\Delta^{2}}%
}{2}\right)  \text{.} \label{mas2}%
\end{equation}
Finally, the quantity $\left\langle \left(  \partial_{\epsilon}\mathrm{H}%
\right)  _{d}\right\rangle $ can be rewritten as
\begin{align}
\left\langle \left(  \partial_{\epsilon}\mathrm{H}\right)  _{d}\right\rangle
&  =\sum_{i=0}^{1}p_{i}\partial_{\epsilon}\tilde{E}_{i}=p_{0}\partial
_{\epsilon}\tilde{E}_{0}+p_{1}\partial_{\epsilon}\tilde{E}_{1}\nonumber\\
&  =p_{0}\partial_{\epsilon}\left(  \frac{\hslash}{2}\sqrt{\epsilon^{2}%
+\Delta^{2}}\right)  +p_{1}\partial_{\epsilon}\left(  -\frac{\hslash}{2}%
\sqrt{\epsilon^{2}+\Delta^{2}}\right) \nonumber\\
&  =\frac{e^{-\beta\hslash\frac{\sqrt{\epsilon^{2}+\Delta^{2}}}{2}}%
}{\mathcal{Z}}\left(  \frac{\hslash}{2}\right)  \frac{\epsilon}{\sqrt
{\epsilon^{2}+\Delta^{2}}}+\frac{e^{\beta\hslash\frac{\sqrt{\epsilon
^{2}+\Delta^{2}}}{2}}}{\mathcal{Z}}\left(  -\frac{\hslash}{2}\right)
\frac{\epsilon}{\sqrt{\epsilon^{2}+\Delta^{2}}}\nonumber\\
&  =-\frac{\hslash}{2}\frac{2\sinh\left(  \beta\hslash\frac{\sqrt{\epsilon
^{2}+\Delta^{2}}}{2}\right)  }{2\cosh\left(  \beta\hslash\frac{\sqrt
{\epsilon^{2}+\Delta^{2}}}{2}\right)  }\frac{\epsilon}{\sqrt{\epsilon
^{2}+\Delta^{2}}}\nonumber\\
&  =-\frac{\hslash}{2}\frac{\epsilon}{\sqrt{\epsilon^{2}+\Delta^{2}}}%
\tanh\left(  \beta\hslash\frac{\sqrt{\epsilon^{2}+\Delta^{2}}}{2}\right)
\text{,}%
\end{align}
that is,%
\begin{equation}
\left\langle \left(  \partial_{\epsilon}\mathrm{H}\right)  _{d}\right\rangle
=-\frac{\hslash}{2}\frac{\epsilon}{\sqrt{\epsilon^{2}+\Delta^{2}}}\tanh\left(
\beta\hslash\frac{\sqrt{\epsilon^{2}+\Delta^{2}}}{2}\right)  \text{.}
\label{mas3}%
\end{equation}
Finally, using Eqs. (\ref{mas1}), (\ref{mas2}), and (\ref{mas3}),
$2g_{\beta\epsilon}\left(  \beta\text{, }\epsilon\right)  d\beta d\epsilon$ in
Eq. (\ref{jo4}) becomes%
\begin{align}
2g_{\beta\epsilon}\left(  \beta\text{, }\epsilon\right)  d\beta d\epsilon &
=\frac{1}{4}\left\{  2\beta\left[
\begin{array}
[c]{c}%
\frac{\hslash^{2}}{4}\epsilon+\frac{\hslash}{2}\sqrt{\epsilon^{2}+\Delta^{2}%
}\tanh\left(  \beta\hslash\frac{\sqrt{\epsilon^{2}+\Delta^{2}}}{2}\right)
\cdot\\
\left(  -\frac{\hslash}{2}\frac{\epsilon}{\sqrt{\epsilon^{2}+\Delta^{2}}}%
\tanh\left(  \beta\hslash\frac{\sqrt{\epsilon^{2}+\Delta^{2}}}{2}\right)
\right)
\end{array}
\right]  \right\}  d\beta d\epsilon\nonumber\\
&  =\frac{1}{4}\left\{  2\beta\left[  \frac{\hslash^{2}}{4}\epsilon
-\frac{\hslash^{2}}{4}\epsilon\tanh^{2}\left(  \beta\hslash\frac
{\sqrt{\epsilon^{2}+\Delta^{2}}}{2}\right)  \right]  \right\}  d\beta
d\epsilon\nonumber\\
&  =\frac{\hslash^{2}}{8}\beta\epsilon\left[  1-\tanh^{2}\left(  \beta
\hslash\frac{\sqrt{\epsilon^{2}+\Delta^{2}}}{2}\right)  \right]  d\beta
d\epsilon\text{,}%
\end{align}
that is,%
\begin{equation}
2g_{\beta\epsilon}\left(  \beta\text{, }\epsilon\right)  d\beta d\epsilon
=\frac{\hslash^{2}}{8}\beta\epsilon\left[  1-\tanh^{2}\left(  \beta
\hslash\frac{\sqrt{\epsilon^{2}+\Delta^{2}}}{2}\right)  \right]  d\beta
d\epsilon\text{.} \label{J2}%
\end{equation}
Our second sub-calculation ends with the derivation of Eq. (\ref{J2}).

\subsubsection{Third sub-calculation}

In what follows, we focus on the calculation of $g_{\epsilon\epsilon}\left(
\beta\text{, }\epsilon\right)  d\epsilon^{2}$,%
\begin{equation}
\ g_{\epsilon\epsilon}\left(  \beta\text{, }\epsilon\right)  d\epsilon
^{2}\ \overset{\text{def}}{=}\frac{1}{4}\left\{  \beta^{2}\left\{
\left\langle \left[  \left(  \partial_{\epsilon}\mathrm{H}\right)
_{d}\right]  ^{2}\right\rangle -\left\langle \left(  \partial_{\epsilon
}\mathrm{H}\right)  _{d}\right\rangle ^{2}\right\}  +2\sum_{n\neq m}\left\vert
\frac{\left\langle n|\partial_{\epsilon}\mathrm{H}|m\right\rangle }{\tilde
{E}_{n}-\tilde{E}_{m}}\right\vert ^{2}\frac{\left(  e^{-\beta\tilde{E}_{n}%
}-e^{-\beta\tilde{E}_{m}}\right)  ^{2}}{\mathcal{Z}\cdot\left(  e^{-\beta
\tilde{E}_{n}}+e^{-\beta\tilde{E}_{m}}\right)  }\right\}  d\epsilon
^{2}\text{.} \label{zar1}%
\end{equation}
Let us recall that $\left\langle \left(  \partial_{\epsilon}\mathrm{H}\right)
_{d}\right\rangle $ is given in Eq. (\ref{mas3}). Therefore, we get%
\begin{equation}
\left\langle \left(  \partial_{\epsilon}\mathrm{H}\right)  _{d}\right\rangle
^{2}=\frac{\hslash^{2}}{4}\frac{\epsilon^{2}}{\epsilon^{2}+\Delta^{2}}%
\tanh^{2}\left(  \beta\hslash\frac{\sqrt{\epsilon^{2}+\Delta^{2}}}{2}\right)
\text{.} \label{zar2}%
\end{equation}
Moreover, $\left\langle \left(  \partial_{\epsilon}\mathrm{H}\right)  _{d}%
^{2}\right\rangle $ is given by%
\begin{align}
\left\langle \left(  \partial_{\epsilon}\mathrm{H}\right)  _{d}^{2}%
\right\rangle  &  =\sum_{i=0}^{1}p_{i}\left(  \partial_{\epsilon}\tilde{E}%
_{i}\right)  ^{2}=p_{0}\left(  \partial_{\epsilon}\tilde{E}_{0}\right)
^{2}+p_{1}\left(  \partial_{\epsilon}\tilde{E}_{1}\right)  ^{2}\nonumber\\
&  =p_{0}\frac{\hslash^{2}}{4}\frac{\epsilon^{2}}{\epsilon^{2}+\Delta^{2}%
}+p_{1}\frac{\hslash^{2}}{4}\frac{\epsilon^{2}}{\epsilon^{2}+\Delta^{2}%
}\nonumber\\
&  =\left(  p_{0}+p_{1}\right)  \frac{\hslash^{2}}{4}\frac{\epsilon^{2}%
}{\epsilon^{2}+\Delta^{2}}\nonumber\\
&  =\frac{\hslash^{2}}{4}\frac{\epsilon^{2}}{\epsilon^{2}+\Delta^{2}}\text{,}%
\end{align}
that is,%
\begin{equation}
\left\langle \left(  \partial_{\epsilon}\mathrm{H}\right)  _{d}^{2}%
\right\rangle =\frac{\hslash^{2}}{4}\frac{\epsilon^{2}}{\epsilon^{2}%
+\Delta^{2}}\text{.} \label{zar3}%
\end{equation}
Finally, let us focus on the term in Eq. (\ref{zar1}) given by%
\begin{align}
2\sum_{n\neq m}\left\vert \frac{\left\langle n|\partial_{\epsilon}%
\mathrm{H}|m\right\rangle }{\tilde{E}_{n}-\tilde{E}_{m}}\right\vert ^{2}%
\frac{\left(  e^{-\beta\tilde{E}_{n}}-e^{-\beta\tilde{E}_{m}}\right)  ^{2}%
}{\mathcal{Z}\cdot\left(  e^{-\beta\tilde{E}_{n}}+e^{-\beta\tilde{E}_{m}%
}\right)  }  &  =\frac{2}{\mathcal{Z}}\left\vert \frac{\left\langle
n_{0}|\partial_{\epsilon}\mathrm{H}|n_{1}\right\rangle }{\tilde{E}_{0}%
-\tilde{E}_{1}}\right\vert ^{2}\frac{\left(  e^{-\beta\tilde{E}_{0}}%
-e^{-\beta\tilde{E}_{1}}\right)  ^{2}}{\left(  e^{-\beta\tilde{E}_{0}%
}+e^{-\beta\tilde{E}_{1}}\right)  }+\nonumber\\
&  +\frac{2}{\mathcal{Z}}\left\vert \frac{\left\langle n_{1}|\partial
_{\epsilon}\mathrm{H}|n_{0}\right\rangle }{\tilde{E}_{1}-\tilde{E}_{0}%
}\right\vert ^{2}\frac{\left(  e^{-\beta\tilde{E}_{1}}-e^{-\beta\tilde{E}_{0}%
}\right)  ^{2}}{\left(  e^{-\beta\tilde{E}_{0}}+e^{-\beta\tilde{E}_{1}%
}\right)  }\nonumber\\
&  =\frac{2}{\mathcal{Z}}\frac{\left(  e^{-\beta\tilde{E}_{0}}-e^{-\beta
\tilde{E}_{1}}\right)  ^{2}}{\left(  e^{-\beta\tilde{E}_{0}}+e^{-\beta
\tilde{E}_{1}}\right)  }\frac{\left(  \left\vert \left\langle n_{0}%
|\partial_{\epsilon}\mathrm{H}|n_{1}\right\rangle \right\vert ^{2}+\left\vert
\left\langle n_{1}|\partial_{\epsilon}\mathrm{H}|n_{0}\right\rangle
\right\vert ^{2}\right)  }{\left\vert \tilde{E}_{0}-\tilde{E}_{1}\right\vert
^{2}}\nonumber\\
&  =2\frac{\left(  e^{-\beta\tilde{E}_{0}}-e^{-\beta\tilde{E}_{1}}\right)
^{2}}{\left(  e^{-\beta\tilde{E}_{0}}+e^{-\beta\tilde{E}_{1}}\right)  ^{2}%
}\frac{\frac{\hslash^{2}}{4}\frac{\Delta^{2}}{\Delta^{2}+\epsilon^{2}}%
+\frac{\hslash^{2}}{4}\frac{\Delta^{2}}{\Delta^{2}+\epsilon^{2}}}{\hslash
^{2}\left(  \epsilon^{2}+\Delta^{2}\right)  }\nonumber\\
&  =\frac{\Delta^{2}}{\left(  \Delta^{2}+\epsilon^{2}\right)  ^{2}}\tanh
^{2}\left(  \beta\hslash\frac{\sqrt{\epsilon^{2}+\Delta^{2}}}{2}\right)
\text{,} \label{alsing}%
\end{align}
that is,%
\begin{equation}
2\sum_{n\neq m}\left\vert \frac{\left\langle n|\partial_{\epsilon}%
\mathrm{H}|m\right\rangle }{\tilde{E}_{n}-\tilde{E}_{m}}\right\vert ^{2}%
\frac{\left(  e^{-\beta\tilde{E}_{n}}-e^{-\beta\tilde{E}_{m}}\right)  ^{2}%
}{\mathcal{Z}\cdot\left(  e^{-\beta\tilde{E}_{n}}+e^{-\beta\tilde{E}_{m}%
}\right)  }d\epsilon^{2}=\frac{\Delta^{2}}{\left(  \Delta^{2}+\epsilon
^{2}\right)  ^{2}}\tanh^{2}\left(  \beta\hslash\frac{\sqrt{\epsilon^{2}%
+\Delta^{2}}}{2}\right)  d\epsilon^{2}\text{.} \label{zar4}%
\end{equation}
For clarity, note that $\partial_{\epsilon}\mathrm{H}$ in Eq. (\ref{zar4})
equals $\left(  -\hslash/2\right)  \sigma_{z}$ in the standard computational
basis $\left\{  \left\vert 0\right\rangle \text{, }\left\vert 1\right\rangle
\right\}  $. Therefore, combining Eqs. (\ref{zar2}), (\ref{zar3}), and
(\ref{zar4}) we get that $g_{\epsilon\epsilon}\left(  \beta\text{, }%
\epsilon\right)  d\epsilon^{2}$ in Eq. (\ref{zar1}) equals%
\begin{equation}
g_{\epsilon\epsilon}\left(  \beta\text{, }\epsilon\right)  d\epsilon^{2}%
=\frac{\hslash^{2}}{16}\left\{  \frac{4\Delta^{2}}{\hslash^{2}\left(
\Delta^{2}+\epsilon^{2}\right)  ^{2}}\tanh^{2}\left(  \beta\hslash\frac
{\sqrt{\epsilon^{2}+\Delta^{2}}}{2}\right)  +\frac{\beta^{2}\epsilon^{2}%
}{\epsilon^{2}+\Delta^{2}}\left[  1-\tanh^{2}\left(  \beta\hslash\frac
{\sqrt{\epsilon^{2}+\Delta^{2}}}{2}\right)  \right]  \right\}  d\epsilon
^{2}\text{.} \label{J3}%
\end{equation}
Then, using Eqs. (\ref{J1}), (\ref{J2}), and (\ref{J3}), $ds_{\mathrm{Bures}%
}^{2}$ in Eq. (\ref{zarrillo}) becomes%
\begin{align}
ds_{\mathrm{Bures}}^{2}  &  =\frac{\hslash^{2}}{16}\left(  \epsilon^{2}%
+\Delta^{2}\right)  \left[  1-\tanh^{2}\left(  \beta\hslash\frac
{\sqrt{\epsilon^{2}+\Delta^{2}}}{2}\right)  \right]  d\beta^{2}+\nonumber\\
&  +\frac{\hslash^{2}}{8}\beta\epsilon\left[  1-\tanh^{2}\left(  \beta
\hslash\frac{\sqrt{\epsilon^{2}+\Delta^{2}}}{2}\right)  \right]  d\beta
d\epsilon+\nonumber\\
&  +\frac{\hslash^{2}}{16}\left\{  \frac{4\Delta^{2}}{\hslash^{2}\left(
\Delta^{2}+\epsilon^{2}\right)  ^{2}}\tanh^{2}\left(  \beta\hslash\frac
{\sqrt{\epsilon^{2}+\Delta^{2}}}{2}\right)  +\frac{\beta^{2}\epsilon^{2}%
}{\epsilon^{2}+\Delta^{2}}\left[  1-\tanh^{2}\left(  \beta\hslash\frac
{\sqrt{\epsilon^{2}+\Delta^{2}}}{2}\right)  \right]  \right\}  d\epsilon
^{2}\text{.} \label{f4}%
\end{align}
Finally, using Eq. (\ref{f4}), the Bures metric tensor in the case of a
superconducting flux qubit becomes%
\begin{equation}
g_{ij}^{\left(  \mathrm{Bures}\right)  }\left(  \beta\text{, }\epsilon\right)
=\frac{\hslash^{2}}{16}\left[  1-\tanh^{2}\left(  \beta\hslash\frac
{\sqrt{\epsilon^{2}+\Delta^{2}}}{2}\right)  \right]  \left(
\begin{array}
[c]{cc}%
\epsilon^{2}+\Delta^{2} & \beta\epsilon\\
\beta\epsilon & \frac{\beta^{2}\epsilon^{2}}{\epsilon^{2}+\Delta^{2}}%
+\frac{4\Delta^{2}}{\hslash^{2}\left(  \Delta^{2}+\epsilon^{2}\right)  ^{2}%
}\frac{\tanh^{2}\left(  \beta\hslash\frac{\sqrt{\epsilon^{2}+\Delta^{2}}}%
{2}\right)  }{1-\tanh^{2}\left(  \beta\hslash\frac{\sqrt{\epsilon^{2}%
+\Delta^{2}}}{2}\right)  }%
\end{array}
\right)  \text{,} \label{chetu1}%
\end{equation}
with $1\leq i$, $j\leq2$. The derivation of Eqs. (\ref{f4}) and (\ref{chetu1})
completes our calculation of the Bures metric structure for a superconducting
flux qubit.

\subsubsection{The Sj\"{o}qvist metric}

Let us observe that the Sj\"{o}qvist metric in Eq. (\ref{vetta}) can be
rewritten in our case as%
\begin{equation}
ds_{\mathrm{Sj\ddot{o}qvist}}^{2}\overset{\text{def}}{=}\frac{1}{4}\sum
_{k=0}^{1}\frac{dp_{k}^{2}}{p_{k}}+\sum_{k=0}^{1}p_{k}ds_{k}^{2}\text{,}
\label{cami0}%
\end{equation}
where $ds_{k}^{2}\overset{\text{def}}{=}\left[  \left\langle dn_{k}|\left(
\mathrm{I}-\left\vert n_{k}\right\rangle \left\langle n_{k}\right\vert
\right)  |dn_{k}\right\rangle \right]  $ and $\left\langle n_{k}\left\vert
n_{k^{\prime}}\right.  \right\rangle =\delta_{kk^{\prime}}$. From Eq.
(\ref{anto444}), the states $\left\vert dn_{0}\right\rangle $ and $\left\vert
dn_{1}\right\rangle $ become
\begin{equation}
\left\vert dn_{0}\right\rangle \overset{\text{def}}{=}\frac{1}{\sqrt{2}%
}\left(
\begin{array}
[c]{c}%
\frac{1}{2}\frac{\Delta^{2}}{\sqrt{\epsilon^{2}+\Delta^{2}}}\frac
{\sqrt{\epsilon^{2}+\Delta^{2}}-\epsilon}{\left(  \epsilon^{2}+\Delta
^{2}-\epsilon\sqrt{\epsilon^{2}+\Delta^{2}}\right)  ^{3/2}}d\epsilon\\
\frac{1}{2}\frac{\Delta^{2}}{\sqrt{\epsilon^{2}+\Delta^{2}}}\frac{\left(
\sqrt{\epsilon^{2}+\Delta^{2}}-\epsilon\right)  ^{2}}{\left(  \epsilon
^{2}+\Delta^{2}-\epsilon\sqrt{\epsilon^{2}+\Delta^{2}}\right)  ^{3/2}%
}d\epsilon
\end{array}
\right)  \text{, and }\left\vert dn_{1}\right\rangle \overset{\text{def}%
}{=}\frac{1}{\sqrt{2}}\left(
\begin{array}
[c]{c}%
\frac{1}{2}\frac{\Delta^{2}}{\sqrt{\epsilon^{2}+\Delta^{2}}}\frac
{\epsilon+\sqrt{\epsilon^{2}+\Delta^{2}}}{\left(  \epsilon^{2}+\Delta
^{2}+\epsilon\sqrt{\epsilon^{2}+\Delta^{2}}\right)  ^{3/2}}d\epsilon\\
-\frac{1}{2}\frac{\Delta^{2}}{\sqrt{\epsilon^{2}+\Delta^{2}}}\frac{\left(
\epsilon+\sqrt{\epsilon^{2}+\Delta^{2}}\right)  ^{2}}{\left(  \epsilon
^{2}+\Delta^{2}+\epsilon\sqrt{\epsilon^{2}+\Delta^{2}}\right)  ^{3/2}%
}d\epsilon
\end{array}
\right)  \text{,} \label{cami1}%
\end{equation}
respectively. Eqs. (\ref{anto444}) and (\ref{cami1}) will be used to calculate
the nonclassical contribution that appears in the Sj\"{o}qvist metric in\ Eq.
(\ref{cami0}). In what follows, however, let us consider the classical
contribution $\left[  ds_{\mathrm{Sj\ddot{o}qvist}}^{2}\right]  ^{\left(
\mathrm{classical}\right)  }$ in Eq. (\ref{cami0}). We note that $\left[
ds_{\mathrm{Sj\ddot{o}qvist}}^{2}\right]  ^{\left(  \mathrm{classical}\right)
}$ equals%
\begin{align}
\frac{1}{4}\sum_{k=0}^{1}\frac{dp_{k}^{2}}{p_{k}}  &  =\frac{1}{4}\frac
{dp_{0}^{2}}{p_{0}}+\frac{1}{4}\frac{dp_{1}^{2}}{p_{1}}\nonumber\\
&  =\left[  \frac{\left(  \partial_{\beta}p_{0}\right)  ^{2}}{4p_{0}}%
+\frac{\left(  \partial_{\beta}p_{1}\right)  ^{2}}{4p_{1}}\right]  d\beta
^{2}+\left[  \frac{\left(  \partial_{\epsilon}p_{0}\right)  ^{2}}{4p_{0}%
}+\frac{\left(  \partial_{\epsilon}p_{1}\right)  ^{2}}{4p_{1}}\right]
d\epsilon^{2}+\left[  \frac{2\partial_{\beta}p_{0}\partial_{\epsilon}p_{0}%
}{4p_{0}}+\frac{2\partial_{\beta}p_{1}\partial_{\epsilon}p_{1}}{4p_{1}%
}\right]  d\beta d\epsilon\text{.} \label{q2}%
\end{align}
Using Eq. (\ref{sorry}), $\left[  ds_{\mathrm{Sj\ddot{o}qvist}}^{2}\right]
^{\left(  \mathrm{classical}\right)  }$ in Eq. (\ref{q2}) reduces to%
\begin{align}
\left[  ds_{\mathrm{Sj\ddot{o}qvist}}^{2}\right]  ^{\left(  \mathrm{classical}%
\right)  }  &  =\frac{1}{4}\sum_{k=0}^{1}\frac{dp_{k}^{2}}{p_{k}}\nonumber\\
&  =\frac{\hslash^{2}}{16}\left(  \epsilon^{2}+\Delta^{2}\right)  \left[
1-\tanh^{2}\left(  \beta\hslash\frac{\sqrt{\epsilon^{2}+\Delta^{2}}}%
{2}\right)  \right]  d\beta^{2}+\nonumber\\
&  +\frac{\hslash^{2}}{16}\frac{\beta^{2}\epsilon^{2}}{\epsilon^{2}+\Delta
^{2}}\left[  1-\tanh^{2}\left(  \beta\hslash\frac{\sqrt{\epsilon^{2}%
+\Delta^{2}}}{2}\right)  \right]  d\epsilon^{2}+\nonumber\\
&  +\frac{\hslash^{2}}{8}\beta\epsilon\left[  1-\tanh^{2}\left(  \beta
\hslash\frac{\sqrt{\epsilon^{2}+\Delta^{2}}}{2}\right)  \right]  d\beta
d\epsilon\text{.} \label{classical}%
\end{align}
We can now return our focus on the nonclassical contribution $\left[
ds_{\mathrm{Sj\ddot{o}qvist}}^{2}\right]  ^{\left(  \text{\textrm{quantum}%
}\right)  }$ that specifies the Sj\"{o}qvist metric. We have%
\begin{align}
\left[  ds_{\mathrm{Sj\ddot{o}qvist}}^{2}\right]  ^{\left(
\text{\textrm{quantum}}\right)  }  &  =\sum_{k=0}^{1}p_{k}\left\langle
dn_{k}|\left(  \mathrm{I}-\left\vert n_{k}\right\rangle \left\langle
n_{k}\right\vert \right)  |dn_{k}\right\rangle \nonumber\\
&  =p_{0}\left\langle dn_{0}|dn_{0}\right\rangle -p_{0}\left\vert \left\langle
dn_{0}|n_{0}\right\rangle \right\vert ^{2}+p_{1}\left\langle dn_{1}%
|dn_{1}\right\rangle -p_{1}\left\vert \left\langle dn_{1}|n_{1}\right\rangle
\right\vert ^{2}\text{.}%
\end{align}
A simple check allows us to verify that $\left\langle dn_{0}|n_{0}%
\right\rangle =0$ and $\left\langle dn_{1}|n_{1}\right\rangle =0$. Therefore,
$\left[  ds_{\mathrm{Sj\ddot{o}qvist}}^{2}\right]  ^{\left(
\text{\textrm{quantum}}\right)  }$ becomes
\begin{align}
\left[  ds_{\mathrm{Sj\ddot{o}qvist}}^{2}\right]  ^{\left(
\text{\textrm{quantum}}\right)  }  &  =p_{0}\left\langle dn_{0}|dn_{0}%
\right\rangle +p_{1}\left\langle dn_{1}|dn_{1}\right\rangle \nonumber\\
&  =p_{0}\frac{1}{8}\frac{\Delta^{2}}{\Delta^{2}+\epsilon^{2}}\left(
\epsilon-\sqrt{\Delta^{2}+\epsilon^{2}}\right)  ^{2}\frac{2\Delta
^{2}+2\epsilon^{2}-2\epsilon\sqrt{\Delta^{2}+\epsilon^{2}}}{\left(  \Delta
^{2}+\epsilon^{2}-\epsilon\sqrt{\Delta^{2}+\epsilon^{2}}\right)  ^{3}%
}d\epsilon^{2}+\nonumber\\
&  +p_{1}\frac{1}{8}\frac{\Delta^{2}}{\Delta^{2}+\epsilon^{2}}\left(
\epsilon+\sqrt{\Delta^{2}+\epsilon^{2}}\right)  ^{2}\frac{2\Delta
^{2}+2\epsilon^{2}+2\epsilon\sqrt{\Delta^{2}+\epsilon^{2}}}{\left(  \Delta
^{2}+\epsilon^{2}+\epsilon\sqrt{\Delta^{2}+\epsilon^{2}}\right)  ^{3}%
}d\epsilon^{2}\nonumber\\
&  =\frac{1}{4}\frac{\Delta^{2}}{\left(  \Delta^{2}+\epsilon^{2}\right)  ^{2}%
}d\epsilon^{2}\nonumber\\
&  =\frac{\hslash^{2}}{16}\frac{4\Delta^{2}}{\hslash^{2}\left(  \Delta
^{2}+\epsilon^{2}\right)  ^{2}}d\epsilon^{2}\text{,}%
\end{align}
that is,%
\begin{equation}
\left[  ds_{\mathrm{Sj\ddot{o}qvist}}^{2}\right]  ^{\left(
\text{\textrm{quantum}}\right)  }=\frac{\hslash^{2}}{16}\frac{4\Delta^{2}%
}{\hslash^{2}\left(  \Delta^{2}+\epsilon^{2}\right)  ^{2}}d\epsilon
^{2}\text{.} \label{quantum}%
\end{equation}
Finally, combining Eqs. (\ref{classical}) and (\ref{quantum}), the
Sj\"{o}qvist metric $ds_{\mathrm{Sj\ddot{o}qvist}}^{2}$ becomes%
\begin{align}
ds_{\mathrm{Sj\ddot{o}qvist}}^{2}  &  =\frac{\hslash^{2}}{16}\left(
\epsilon^{2}+\Delta^{2}\right)  \left[  1-\tanh^{2}\left(  \beta\hslash
\frac{\sqrt{\epsilon^{2}+\Delta^{2}}}{2}\right)  \right]  d\beta
^{2}+\nonumber\\
&  +\frac{\hslash^{2}}{8}\beta\epsilon\left[  1-\tanh^{2}\left(  \beta
\hslash\frac{\sqrt{\epsilon^{2}+\Delta^{2}}}{2}\right)  \right]  d\beta
d\epsilon+\nonumber\\
&  +\frac{\hslash^{2}}{16}\left\{  \frac{4\Delta^{2}}{\hslash^{2}\left(
\Delta^{2}+\epsilon^{2}\right)  ^{2}}+\frac{\beta^{2}\epsilon^{2}}%
{\epsilon^{2}+\Delta^{2}}\left[  1-\tanh^{2}\left(  \beta\hslash\frac
{\sqrt{\epsilon^{2}+\Delta^{2}}}{2}\right)  \right]  \right\}  d\epsilon
^{2}\text{.} \label{f3}%
\end{align}
The metric tensor $g_{ij}^{\left(  \mathrm{Sj\ddot{o}qvist}\right)  }\left(
\beta\text{, }\epsilon\right)  $ from Eq. (\ref{f3}) is given by
\begin{equation}
g_{ij}^{\left(  \mathrm{Sj\ddot{o}qvist}\right)  }\left(  \beta\text{,
}\epsilon\right)  =\frac{\hslash^{2}}{16}\left[  1-\tanh^{2}\left(
\beta\hslash\frac{\sqrt{\epsilon^{2}+\Delta^{2}}}{2}\right)  \right]  \left(
\begin{array}
[c]{cc}%
\epsilon^{2}+\Delta^{2} & \beta\epsilon\\
\beta\epsilon & \frac{\beta^{2}\epsilon^{2}}{\epsilon^{2}+\Delta^{2}}%
+\frac{4\Delta^{2}}{\hslash^{2}\left(  \Delta^{2}+\epsilon^{2}\right)  ^{2}%
}\frac{1}{1-\tanh^{2}\left(  \beta\hslash\frac{\sqrt{\epsilon^{2}+\Delta^{2}}%
}{2}\right)  }%
\end{array}
\right)  \text{,} \label{chetu}%
\end{equation}
with $1\leq i$, $j\leq2$. The derivation of Eqs. (\ref{f3}) and (\ref{chetu})
completes our calculation of the Sj\"{o}qvist metric structure for
superconducting flux qubits.

\section{Considerations from the comparative analysis}

In this section, we discuss the outcomes of the comparative analysis carried
out in Section V concerning the Bures and Sj\"{o}qvist metrics for spin qubits
and superconducting flux qubits Hamiltonian models presented in Section IV.

\subsection{The asymptotic limit of $\beta$ approaching infinity}

We begin by discussing the asymptotic limit of $\beta$ approaching infinity.
In \ the case of a spin qubit with Hamiltonian \textrm{H}$_{\mathrm{SQ}%
}\left(  \omega\right)  $ in Eq. (\ref{spinH}), the density matrix
$\rho_{\mathrm{SQ}}\left(  \beta\text{, }\omega\right)  $ in Eq. (\ref{ro1})
approaches $\rho_{\mathrm{SQ}}^{\beta\rightarrow\infty}\left(  \omega\right)
\overset{\text{def}}{=}\left\vert 1\right\rangle \left\langle 1\right\vert $
as $\beta\rightarrow\infty$. Observe that $\left\vert 1\right\rangle $ denotes
here the ground state, the state of lowest energy $-\hslash\omega/2$. Since
$\rho_{\mathrm{SQ}}^{\beta\rightarrow\infty}\left(  \omega\right)  $ is a
constant in $\omega$, the Bures and Sj\"{o}qvist metrics in Eqs. (\ref{07})
and (\ref{f0}), respectively, both vanish in this limiting scenario. In this
regard, the case of the superconducting flux qubit specified by the
Hamiltonian \textrm{H}$_{\mathrm{SFQ}}\left(  \Delta\text{, }\epsilon\right)
$ in Eq. (\ref{superH}) is more interesting. Indeed, in this case the density
matrix $\rho_{\mathrm{SFQ}}\left(  \beta\text{, }\epsilon\right)  $ in Eq.
(\ref{anto5}) approaches $\rho_{\mathrm{SFQ}}^{\beta\rightarrow\infty}\left(
\epsilon\right)  $ when $\beta$ approaches infinity. The quantity
$\rho_{\mathrm{SFQ}}^{\beta\rightarrow\infty}\left(  \epsilon\right)  $
represents a pure (ground) state of lowest energy $\left(  -\hslash/2\right)
\sqrt{\Delta^{2}+\epsilon^{2}}$ and is given by%
\begin{equation}
\rho_{\mathrm{SFQ}}^{\beta\rightarrow\infty}\left(  \epsilon\right)  =\frac
{1}{2}\left(
\begin{array}
[c]{cc}%
1+\frac{\epsilon}{\sqrt{\epsilon^{2}+\Delta^{2}}} & \frac{\Delta}%
{\sqrt{\epsilon^{2}+\Delta^{2}}}\\
\frac{\Delta}{\sqrt{\epsilon^{2}+\Delta^{2}}} & 1-\frac{\epsilon}%
{\sqrt{\epsilon^{2}+\Delta^{2}}}%
\end{array}
\right)  \text{,}%
\end{equation}
with $\rho_{\mathrm{SFQ}}^{\beta\rightarrow\infty}\left(  \epsilon\right)
=(\rho_{\mathrm{SFQ}}^{\beta\rightarrow\infty}\left(  \epsilon\right)  )^{2}$
and \textrm{tr}$\left(  \rho_{\mathrm{SFQ}}^{\beta\rightarrow\infty}\left(
\epsilon\right)  \right)  =1$. Furthermore, when $\beta\rightarrow\infty$, the
Bures and Sj\"{o}qvist metrics in Eqs. (\ref{f4}) and (\ref{f3}),
respectively, reduce to the same expression%
\begin{equation}
ds_{\mathrm{Bures}}^{2}\overset{\beta\rightarrow\infty}{\rightarrow}\frac
{1}{4}\frac{\Delta^{2}}{\left(  \Delta^{2}+\epsilon^{2}\right)  ^{2}}%
d\epsilon^{2}\text{, and }ds_{\mathrm{Sj\ddot{o}qvist}}^{2}\overset{\beta
\rightarrow\infty}{\rightarrow}\frac{1}{4}\frac{\Delta^{2}}{\left(  \Delta
^{2}+\epsilon^{2}\right)  ^{2}}d\epsilon^{2}\text{.} \label{betalim}%
\end{equation}
The limiting expressions assumed by the Bures and Sj\"{o}qvist metrics in Eq.
(\ref{betalim}) are, modulo an unimportant constant factor, the Fubini-Study
metric $ds_{\mathrm{FS}}^{2}$ for pure states. Indeed, we have%
\begin{equation}
ds_{\mathrm{FS}}^{2}\overset{\text{def}}{=}\mathrm{tr}\left[  \left(
\frac{\partial\rho_{\mathrm{SFQ}}^{\beta\rightarrow\infty}\left(
\epsilon\right)  }{\partial\epsilon}\right)  ^{2}\right]  d\epsilon^{2}%
=\frac{1}{2}\frac{\Delta^{2}}{\left(  \Delta^{2}+\epsilon^{2}\right)  ^{2}%
}d\epsilon^{2}\text{.} \label{FSlimit}%
\end{equation}
In the next subsection, we discuss the discrepancy in the Bures (Eqs.
(\ref{07}) and (\ref{f4})) and Sj\"{o}qvist (Eqs. (\ref{f0}) and (\ref{f3}))
metrics emerging from the different nature of the nonclassical contributions
in the two metrics.

\subsection{The metrics discrepancy}

\begin{figure}[t]
\centering
\includegraphics[width=0.5\textwidth] {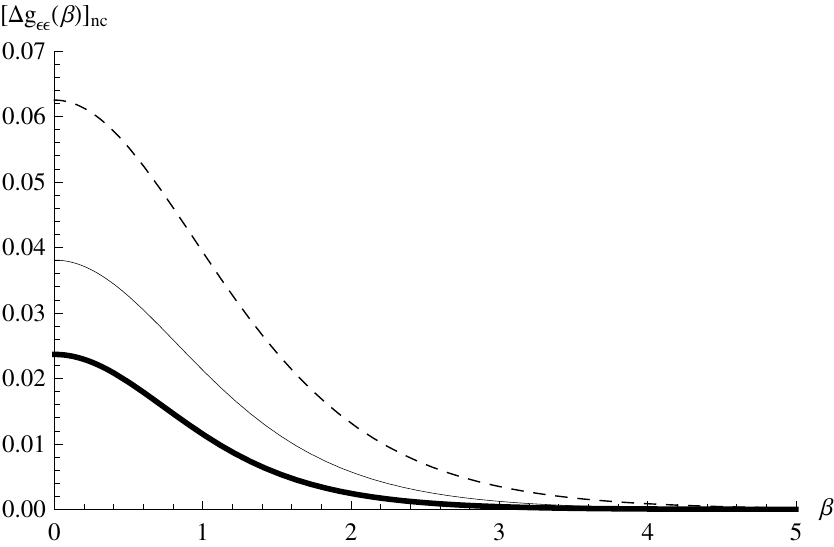}\caption{Plots of the metric
discrepancy $\Delta g_{\epsilon\epsilon}^{\mathrm{nc}}\left(  \beta\text{,
}\epsilon\right)  $ in Eq. (\ref{discrepancy}) versus $\beta$ for $\epsilon=1$
(dashed), $\epsilon=1.25$ (thin solid), and $\epsilon=1.5$ (thick solid). For
simplicity, we set $\hslash=1$ and $\Delta=1$ in each plot. Observe that
$\Delta g_{\epsilon\epsilon}^{\mathrm{nc}}\left(  \beta\text{, }%
\epsilon\right)  $ is nonvanishing because of the difference between the
nonclassical parts $g_{\epsilon\epsilon}^{\mathrm{nc}}\left(  \beta\text{,
}\epsilon\right)  $ of the metric tensor component $g_{\epsilon\epsilon
}\left(  \beta\text{, }\epsilon\right)  \protect\overset{\text{def}%
}{=}g_{\epsilon\epsilon}^{\mathrm{c}}\left(  \beta\text{, }\epsilon\right)
+g_{\epsilon\epsilon}^{\mathrm{nc}}\left(  \beta\text{, }\epsilon\right)  $ in
the Bures and Sj\"{o}qvist metric scenarios. Finally, note that the metric
discrepancy vanishes in the zero-temperature limit, i.e., in the limit of
$\beta$ approaching infinity.}%
\end{figure}

We begin by noting that in the case of the spin qubit Hamiltonian model in Eq.
(\ref{spinH}), there is no discrepancy since the Bures and the Sj\"{o}qvist
metrics in Eqs. (\ref{07}) and (\ref{f0}), respectively, coincide. Indeed, in
this case, both metrics reduce to the classical Fisher-Rao information metric.
The nonclassical/quantum terms vanish in both metrics due to the commutativity
of $\rho_{\mathrm{SQ}}\left(  \beta\text{, }\omega\right)  $ and $\left(
\rho_{\mathrm{SQ}}+d\rho_{\mathrm{SQ}}\right)  \left(  \beta\text{, }%
\omega\right)  $, with $\rho_{\mathrm{SQ}}\left(  \beta\text{, }\omega\right)
$ in Eq. (\ref{ro1}). In the case of the superconducting flux qubit
Hamiltonian model in Eq. (\ref{superH}), instead, the nonclassical/quantum
terms vanish in neither the Bures nor the Sj\"{o}qvist metrics due to the
non-commutativity of $\rho_{\mathrm{SFQ}}\left(  \beta\text{, }\epsilon
\right)  $ and $\left(  \rho_{\mathrm{SFQ}}+d\rho_{\mathrm{SFQ}}\right)
\left(  \beta\text{, }\epsilon\right)  $, with $\rho_{\mathrm{SQ}}\left(
\beta\text{, }\epsilon\right)  $ in Eq. (\ref{anto5}). However, these
nonclassical contributions differ in the two metrics and this leads to a
discrepancy in the Bures and Sj\"{o}qvist metrics in Eqs. (\ref{f4}) and
(\ref{f3}), respectively. More specifically, we have
\begin{equation}
ds_{\mathrm{Sj\ddot{o}qvist}}^{2}-ds_{\mathrm{Bures}}^{2}=\Delta
g_{\epsilon\epsilon}^{\mathrm{nc}}\left(  \beta\text{, }\epsilon\right)
d\epsilon^{2}\geq0\text{,} \label{dis1}%
\end{equation}
for any $\beta$ and $\epsilon$. Note that $\Delta g_{\epsilon\epsilon
}^{\mathrm{nc}}\left(  \beta\text{, }\epsilon\right)  \overset{\text{def}%
}{=}g_{\epsilon\epsilon}^{\mathrm{nc,}\text{ }\mathrm{Sj\ddot{o}qvist}}\left(
\beta\text{, }\epsilon\right)  -g_{\epsilon\epsilon}^{\mathrm{nc,}\text{
}\mathrm{Bures}}\left(  \beta\text{, }\epsilon\right)  $ is the difference
between the nonclassical (\textrm{nc}) contributions in the metric components
$g_{\epsilon\epsilon}\left(  \beta\text{, }\epsilon\right)
\overset{\text{def}}{=}g_{\epsilon\epsilon}^{\mathrm{c}}\left(  \beta\text{,
}\epsilon\right)  +g_{\epsilon\epsilon}^{\mathrm{nc}}\left(  \beta\text{,
}\epsilon\right)  $ and is given by
\begin{equation}
\Delta g_{\epsilon\epsilon}^{\mathrm{nc}}\left(  \beta\text{, }\epsilon
\right)  \overset{\text{def}}{=}\frac{1}{4}\frac{\Delta^{2}}{\left(
\Delta^{2}+\epsilon^{2}\right)  ^{2}}\left[  1-\tanh^{2}\left(  \beta
\hslash\frac{\sqrt{\epsilon^{2}+\Delta^{2}}}{2}\right)  \right]  \text{,}
\label{discrepancy}%
\end{equation}
with $0\leq$ $\tanh^{2}\left(  x\right)  \leq1$ for any $x\in%
\mathbb{R}
$. To be crystal clear, it is useful to view the metric tensor $g_{ij}\left(
\beta\text{, }\epsilon\right)  $ with $1\leq i$, $j\leq2$ (i.e., $\theta
^{1}\overset{\text{def}}{=}\beta$ and $\theta^{2}\overset{\text{def}%
}{=}\epsilon$) recast as
\begin{equation}
g_{ij}\left(  \beta\text{, }\epsilon\right)  =\left(
\begin{array}
[c]{cc}%
g_{\beta\beta}\left(  \beta\text{, }\epsilon\right)  & g_{\beta\epsilon
}\left(  \beta\text{, }\epsilon\right) \\
g_{\epsilon\beta}\left(  \beta\text{, }\epsilon\right)  & g_{\epsilon\epsilon
}^{\mathrm{c}}\left(  \beta\text{, }\epsilon\right)  +g_{\epsilon\epsilon
}^{\mathrm{nc}}\left(  \beta\text{, }\epsilon\right)
\end{array}
\right)  \text{.}%
\end{equation}
The discrepancy between the Bures and Sj\"{o}qvist metrics arises only because
$g_{\epsilon\epsilon}^{\mathrm{nc,}\text{ }\mathrm{Sj\ddot{o}qvist}}\left(
\beta\text{, }\epsilon\right)  \neq g_{\epsilon\epsilon}^{\mathrm{nc,}\text{
}\mathrm{Bures}}\left(  \beta\text{, }\epsilon\right)  $. However, the metric
discrepancy $\Delta g_{\epsilon\epsilon}^{\mathrm{nc}}\left(  \beta\text{,
}\epsilon\right)  $ in Eq. (\ref{discrepancy}) vanishes in the asymptotic
limit of $\beta$ approaching infinity. In Fig. $1$, we plot the discrepancy
between the Bures and the Sj\"{o}qvist metrics for the superconducting flux
qubit Hamiltonian model. In Table I, instead, we summarize the links between
the Bures and the Sj\"{o}qvist metrics.\begin{table}[t]
\centering
\begin{tabular}
[c]{c|c|c|c}\hline\hline
\textbf{Description of quantum states} & \textbf{Quantum states} &
\textbf{Bures metric} & \textbf{Sj\"{o}qvist metric}\\\hline
Pure & $\rho=\rho^{2}$ & Fubini-Study metric & Fubini-Study metric\\
Mixed, classical scenario & $\rho\neq\rho^{2}$, $\left[  \rho\text{, }%
\rho+d\rho\right]  =0$ & Fisher-Rao metric & Fisher-Rao metric\\
Mixed, nonclassical scenario & $\rho\neq\rho^{2}$, $\left[  \rho\text{, }%
\rho+d\rho\right]  \neq0$ & $ds_{\mathrm{Bures}}^{2}\neq ds_{\mathrm{Sj\ddot
{o}qvist}}^{2}$ & $ds_{\mathrm{Sj\ddot{o}qvist}}^{2}\neq ds_{\mathrm{Bures}%
}^{2}$\\\hline
\end{tabular}
\caption{Relation between the Bures and the Sj\"{o}qvist metrics\textbf{. }The
Bures and the Sj\"{o}qvist metrics are identical when considering pure quantum
states $\left(  \rho=\rho^{2}\right)  $ or mixed quantum states $\left(
\rho\neq\rho^{2}\right)  $ for which the non-commutative probabilistic
structure underlying quantum theory is not visible (i.e., in the classical
scenario with $\left[  \rho\text{, }\rho+d\rho\right]  =0$). In particular, in
the former and latter cases, they becomes the Fubini-Study and the Fisher-Rao
information metrics, respectively. However, the Bures and the Sj\"{o}qvist
metrics are generally distinct when considering mixed quantum states $\left(
\rho\neq\rho^{2}\right)  $ for which the non-commutative probabilistic
structure of quantum mechanics is visible (i.e., in the nonclassical scenario
with $\left[  \rho\text{, }\rho+d\rho\right]  \neq0$).}%
\end{table}

\section{Conclusive Remarks}

In this paper, building on our recent scientific effort in Ref.
\cite{cafaroprd22}, we presented an explicit analysis of the Bures and
Sj\"{o}qvist metrics over the manifolds of thermal states for the spin qubit
(Eq. (\ref{spinH})) and the superconducting flux qubit Hamiltonian (Eq.
(\ref{superH})) models. We observed that while both metrics (Eqs. (\ref{07})
and (\ref{f0})) reduce to the Fubini-Study metric in the (zero-temperature)
asymptotic limiting case of the inverse temperature $\beta$ approaching
infinity for both Hamiltonian models, the two metrics are generally different.
We observed this different behavior in the case of the superconducting flux
Hamiltonian model. In general, we note that the two metrics (Eqs. (\ref{f4})
and (\ref{f3})) seem to differ when nonclassical behavior is present since
they quantify noncommutativity of neighboring mixed quantum states in
different manners (Eqs. (\ref{dis1}) and (\ref{discrepancy})). In summary, we
reach (see Table I) the conclusion that for pure quantum states $\left(
\rho=\rho^{2}\right)  $ and for mixed quantum states $\left(  \rho\neq\rho
^{2}\right)  $ for which the non-commutative probabilistic structure
underlying quantum theory is not visible (i.e., in the classical scenario with
$\left[  \rho\text{, }\rho+d\rho\right]  =0$), the Bures and the Sj\"{o}qvist
metrics are the same (Eqs. (\ref{f2}) and (\ref{f1})). Indeed, in the former
and latter cases, they reduce to the Fubini-Study and Fisher-Rao information
metrics, respectively. Instead, when investigating mixed quantum states
$\left(  \rho\neq\rho^{2}\right)  $ for which the non-commutative
probabilistic structure of quantum mechanics is visible (i.e., in the
non-classical scenario with $\left[  \rho\text{, }\rho+d\rho\right]  \neq0$),
the Bures and the Sj\"{o}qvist metrics exhibit a different behavior (Eqs.
(\ref{G1A}) and (\ref{G1B}); Eqs. (\ref{chetu1}) and (\ref{chetu})).

Our main conclusions can be outlined as follows:

\begin{enumerate}
\item[{[i]}] We presented an explicit derivation of Bures metric for arbitrary
density matrices in H\"{u}bner's form (Eq. (\ref{buri14})) and in Zanardi's
form (Eq. (\ref{7})). Moreover, we presented a clear derivation of Zanardi's
form of the Bures metric suitable for the special class of thermal states (Eq.
(\ref{general})). Finally, we reported an explicit derivation of the
Sj\"{o}qvist metric for nondegenerate density matrices (Eq. (\ref{vetta})).

\item[{[ii]}] Using our explicit derivations outlined in [i], we performed
detailed analytical calculations yielding the expressions of the Bures (Eqs.
(\ref{07}) and (\ref{f4})) and Sj\"{o}qvist (Eqs. (\ref{f0}) and (\ref{f3}))
metrics on manifolds of thermal states (Eqs. (\ref{ro1}) and (\ref{anto5}))
that correspond to a spin qubit (Eq. (\ref{spinH})) and a superconducting flux
qubit (Eq. (\ref{superH})) Hamiltonian models.

\item[{[iii]}] In the absence of nonclassical features in which the
neighboring density matrices $\rho$ and $d\rho$ commute, the Bures and the
Sj\"{o}qvist metrics lead to and identical metric expression exemplified by
the classical Fisher-Rao metric tensor. We have explicitly verified this
similarity in the case of a manifold of thermal states for spin qubits in the
presence of a constant magnetic field along the quantization $z$-axis.

\item[{[iv]}] In general, the Bures and the Sj\"{o}qvist metrics are expected
to yield different expressions. Indeed, the Bures and Sj\"{o}qvist metrics
seem to quantify the noncommutativity of neighboring mixed states $\rho$ and
$d\rho$ in different manners, in general. We have explicitly verified this
difference in the case of a manifold of thermal states for superconducting
flux qubits (see Fig. $2$).

\item[{[v]}] In the asymptotic limit of $\beta\rightarrow\infty$, both Bures
and Sj\"{o}qvist metric tensors reduce to the same limiting value (Eq.
(\ref{betalim})) specified by, modulo an unimportant constant factor, the
Fubini-Study metric tensor (Eq. (\ref{FSlimit})) for the zero-temperature pure states.

\item[{[vi]}] In the superconducting flux qubit Hamiltonian model considered
here, we observe that the difference $ds_{\mathrm{Sj\ddot{o}qvist}}%
^{2}-ds_{\mathrm{Bures}}^{2}$ is a positive quantity that depends solely on
the difference between the nonclassical nature of the metric tensor component
$g_{\epsilon\epsilon}\left(  \beta\text{, }\epsilon\right)  $. Indeed, we have
shown that $ds_{\mathrm{Sj\ddot{o}qvist}}^{2}-ds_{\mathrm{Bures}}^{2}=\Delta
g_{\epsilon\epsilon}^{\mathrm{nc}}\left(  \beta\text{, }\epsilon\right)
d\epsilon^{2}\geq0$ (Eqs. (\ref{dis1}) and (\ref{discrepancy})).

\item[{[vii]}] The existence of nonclassical contributions in the Bures and
Sj\"{o}qvist metrics is related to the presence of non-vanishing quadratic
terms like\textbf{ }$\left\vert \left\langle n\left\vert \partial
_{h}\mathrm{H}\right\vert m\right\rangle \right\vert ^{2}$\textbf{
}and\textbf{ }$\left\vert \left\langle dn\left\vert n\right.  \right\rangle
\right\vert ^{2}$\textbf{,} respectively. The former term is related to
modulus squared of suitable quantum overlaps defined in terms of parametric
variations in the Hamiltonian of the system. The latter term, instead, is
specified by the modulus squared of suitable quantum overlaps characterized by
parametric variations of the eigenstates of the Hamiltonian of the system. It
is not unreasonable to expect a formal connection between these two types of
terms causing the noncommutativity between\textbf{ }$\rho$ and $\rho+d\rho$
(see, for instance, Eq. (15.30) in Ref. \cite{karol06}) and find a deeper
relation between the Bures and Sj\"{o}qvist metrics for the class of thermal
quantum states. Indeed, for a more quantitative discussion on the link between
these two terms, see Ref. \cite{alsing23}.

\item[{[viii]}] The differential\textbf{ }$d\rho\left(  \beta\text{,
}h\right)  \overset{\text{def}}{=}\partial_{\beta}\rho d\beta+\partial_{h}\rho
dh$\textbf{ }depends both on eigenvalues and eigenvectors parametric
variations.\textbf{ }However, the noncommutativity between\textbf{ }$\rho
$\textbf{ }and\textbf{ }$d\rho$\textbf{ }is related to that part of\textbf{
}$d\rho$\textbf{ }that emerges from the eigenvectors parametric variations.
These changes, in turn, can be related to the existence of a nonvanishing
commutator between the Hamiltonian of the system and the density matrix
specifying the thermal state. Indeed, in the two main examples studied in this
paper, we have\textbf{ }$\left[  \mathrm{H}_{\mathrm{SQ}}\left(
\omega\right)  \text{, }\rho_{\mathrm{SQ}}\left(  \beta\text{, }\omega\right)
\right]  =0$\textbf{ }and\textbf{ }$\left[  \mathrm{H}_{\mathrm{SFQ}}\left(
\Delta\text{, }\epsilon\right)  \text{, }\rho_{\mathrm{SFQ}}\left(
\beta\text{, }\epsilon\right)  \right]  \neq0$\textbf{, }respectively. In the
former case, unlike the latter case, there is no contribution to\textbf{
}$d\rho$\textbf{ }arising from a variation in the eigenvectors of the Hamiltonian.
\end{enumerate}

For the set of pure states, the scenario is rather unambiguous. The
Fubini--Study metric represents the only natural option for a measure that
characterizes \textquotedblleft random states\textquotedblright.
Alternatively, for mixed-state density matrices, the geometry of the state
space is more complicated \cite{karol06,brody19}. There is a collection of
distinct metrics that can be used, each of them with different physical
inspiration, benefits and disadvantages that can rest on the peculiar
application one \ might be interested in examining. Specifically, both simple
and complicated geometric quantities (i.e., for instance, path, path length,
volume, curvature, and complexity) seem to depend on the measure selected on
the space of mixed states that specify the quantum system being investigated
\cite{karol99,cafaroprd22}. Therefore, our work in this paper can be
particularly important in offering an explicit comparative study between the
(emerging) Sj\"{o}qvist interferometric geometry and the (established) Bures
geometry for mixed quantum states. Gladly, the importance of the existence of
this kind of comparative investigation was lately emphasized in Refs.
\cite{mera22} and \cite{cafaroprd22} too.

From a mathematics standpoint, it would be interesting to formally prove (or,
alternatively, disprove with an explicit counterexample) the monotonicity
\cite{petz96a,petz99} of the Sj\"{o}qvist metric in an arbitrarily
finite-dimensional space of mixed quantum states. From a physics perspective
that relies on empirical evidence, instead, it would be very important to
understand what the physical significance of employing either metric is
\cite{mera22,cafaroprd22}.

In conclusion, despite its relatively limited scope, we hope this work will
inspire either young or senior scientists to pursue ways to deepen our
understanding (both mathematically and physically) of this fascinating
connection among information geometry, statistical physics, and quantum
mechanics \cite{cafaropre20,gassner21,hasegawa21,miller20,cc,saito20,ito20}.

\begin{acknowledgments}
C.C. is grateful to the United States Air Force Research Laboratory (AFRL)
Summer Faculty Fellowship Program for providing support for this work. C. C.
acknowledges helpful discussions with Orlando Luongo,\ Cosmo Lupo, Stefano
Mancini, and Hernando Quevedo. P.M.A. acknowledges support from the Air Force
Office of Scientific Research (AFOSR). Any opinions, findings and conclusions
or recommendations expressed in this material are those of the author(s) and
do not necessarily reflect the views of the Air Force Research Laboratory (AFRL).
\end{acknowledgments}

\end{document}